\documentclass[a4paper,11pt]{article}
\usepackage{jheppub}
\usepackage{epsfig}
\usepackage[T1]{fontenc} 
\usepackage{rotating}

\usepackage{hyperref}
\usepackage{amssymb,amsmath}
\usepackage[caption=false]{subfig}

\usepackage{breqn}

\newcommand\gA{  f  }

\newcommand\eps{\epsilon}
\newcommand\cN{  {\cal N}  }

\newcommand{\bea}{\begin{eqnarray}}
\newcommand{\eea}{\end{eqnarray}}

\newcommand{\nn}{\nonumber}

\newcommand\bu{{\beta}_u}
\newcommand\bv{{\beta}_v}
\newcommand\buv{{\beta}_{uv}}

\renewcommand\u{u}
\renewcommand\v{v}
\newcommand\oneplusu{1+u}

\newcommand\Li{{\rm Li}}
\def\Log#1{\log\left({#1}\right)}
\def\sb#1{{#1}}
\def\GA#1{G_{{#1}}} 
\def\GG#1#2#3{G_{{#1},{#2},{#3}}}  
\def\GC#1#2#3#4{G_{{#1},{#2},{#3},{#4}}}  
\def\cd#1#2{(#1#2)}   
\newcommand\vf{\vec{f}}
\newcommand\vg{\vec{g}}
\def\lll#1{\multicolumn{1}{|c}{{#1}}}
\def\rr#1{\multicolumn{1}{c|}{{#1}}}

\def\be{\begin{equation}}
\def\ee{\end{equation}}

\allowdisplaybreaks[4]

\title{
Iterative structure of finite loop integrals
}

\author[a,b]{Simon Caron-Huot}
\author[a]{Johannes M.\ Henn}
\affiliation[a]{Institute for Advanced Study, Princeton, NJ 08540, USA}
\affiliation[b]{Niels Bohr International Academy and Discovery Center, Blegdamsvej 17, 
Copenhagen 2100, Denmark}
\emailAdd {schuot@nbi.dk}
\emailAdd {jmhenn@ias.edu}

\abstract{
In this paper we develop further and refine the method of differential
equations for computing Feynman integrals. In particular, we show 
that an additional iterative structure emerges for finite loop integrals.
As a concrete non-trivial example we study planar master integrals 
of light-by-light scattering to three loops, 
and derive analytic results  for all values of the Mandelstam variables $s$ and $t$ and the mass $m$.
We start with a recent proposal for defining a basis of loop integrals
having uniform transcendental weight properties and use this approach to compute all planar two-loop
master integrals in dimensional regularization. 
We then show how this approach can be further
simplified when computing finite loop integrals. 
This allows us to discuss precisely  the subset
of integrals that are relevant to the problem. 
We find that this  leads to a block triangular structure of the differential equations, where the blocks correspond to integrals of different weight. We explain how this block triangular form is found in an algorithmic way.
Another advantage of working in four dimensions is that integrals of different loop orders are interconnected and can be seamlessly discussed within the same formalism. 
We use this method to compute all finite master integrals needed up to three loops.
Finally, we remark that all integrals have simple Mandelstam representations.
}


\begin{document}

\maketitle
\flushbottom

\section{Introduction}  

In perturbative quantum field theory, beyond the leading order interesting quantities such 
as scattering amplitudes or correlation functions involve Feynman integrals. 
They are multivalued functions of the kinematical invariants and masses. 
One finds that one-loop integrals in four dimensions can be expressed in terms of logarithms and dilogarithms.
At higher orders, generalizations to other special functions are typically required.
Understanding the structure of loop integrals, in particular precisely which special functions are needed, 
and their analytic computation is an important problem.

In the literature on Feynman integrals one finds various generalizations of the logarithms and dilogarithms encountered at one loop order. Most of them fall into the rather general class of multiple polylogarithms \cite{Goncharov:1998kja,arXiv:math/0606419},
also referred to as Goncharov polylogarithms or hyperlogarithms.\footnote{Elliptic functions also appear in certain Feynman integrals, but will not be needed in the present paper.} As we will see, our method will naturally suggest to work with
a slightly more general class of functions, Chen iterated integrals. 
They enjoy beautiful mathematical properties that we are going to benefit from.

A variety of different approaches for computing Feynman integrals can be found in the literature, and often there are interrelations between them.
Here we wish to focus on the method of differential equations 
\cite{Kotikov:1990kg,Remiddi:1997ny,Gehrmann:1999as,Argeri:2007up}, 
which has provided many results in the past, in particular for virtual corrections to scattering processes. 
Recently, a proposal has appeared to render this technique more systematic.
The suggestion is that a standard form for the differential equations can be obtained, 
by choosing a particularly convenient basis \cite{Henn:2013pwa}. 
The choice of the latter is guided by the properties of the functions expected to appear in the answer, and in particular their transcendental weight properties.

Following the approach of \cite{Henn:2013pwa} 
we write a system of first-order differential
equations for the required integrals, and bring the latter into a simple form by choosing a convenient basis.
Denoting the set of basis integrals by $\vec{f}$, we seek to express their differential as
\begin{align}\label{diffeqsimpleDdim}
d \, \vf(s,t,m^2;\eps) = \eps\, [d \,\tilde{A}(s,t,m^2)] \, \vf(s,t,m^2;\eps) \,,
\end{align}
where the differential acts on the kinematical variables $s,t,m^2$. 
The special feature of eq. (\ref{diffeqsimpleDdim}) is that its dependence on $\eps$
is simple, namely the matrix $\tilde{A}$ is independent of $\eps$.
Moreover, it contains only logarithms of combinations of the kinematical variables,
in agreement with the expected singularities of Feynman integrals.
More precisely,
\begin{align}\label{defalphabet}
\tilde{A} = \sum_{k} a_{k} \, d\log(\alpha_{k}) \,,
\end{align}
where $a_{k}$ are certain constant matrices, and the $\alpha_{k}$ 
are algebraic functions of $s,t,m^2$.
They are called {\it letters}, and the set set of letters appearing
in $\tilde{A}$ specifies the {\it alphabet} required for the integrals $\vec{f}$.
Once the form (\ref{diffeqsimpleDdim}) is reached, one can immediately write 
down a solution to eq. (\ref{diffeqsimpleDdim})
in terms of Chen iterated integrals \cite{Chen1997}, to all orders in the $\eps$ expansion.
Depending on the properties of the alphabet $\{ \alpha_{k} \}$, one may choose
a representation in terms of more familiar functions.

The method was successfully applied to many integrals, including 
\cite{Henn:2013tua,Henn:2013woa,Henn:2013nsa,Henn:2014lfa}.
Although the conjecture in \cite{Henn:2013pwa} is still unproven, 
that the special form (\ref{diffeqsimpleDdim}) can always be achieved,
this was verified in all cases.

A convenient feature of (\ref{diffeqsimpleDdim}) is that the 
equations give all orders in the $\epsilon$ expansion.
This feature will not, however, be so desirable for us here, for several reasons.
First of all, since all the integrals we will consider are already convergent in $D=4$, the $\epsilon$ expansion
represents unneeded information. Possibly the differential equation would become even simpler
if it did not need to carry this information. 
Second, for particular applications, one may have a symmetry enhancement at $D=4$.
In particular, the integrals we will be interested in 
enjoy (dual) conformal symmetry, in the sense of ref. \cite{Alday:2009zm}, which is broken away from $D=4$.
A similar situation occurs for example for correlation functions in conformal field theories.
Clearly, it is desirable to set up the computation in a way which preserves the symmetries.
Third, in integer dimensions integrals with different loop orders can be related to each other 
\cite{Drummond:2006rz,Drummond:2010cz,Dixon:2011ng},
and this allows us to discuss them in a seamless way.

We will thus develop a simplified version of the approach of \cite{Henn:2013pwa}, applicable directly in $D=4$ dimensions.
In a first step, we will assume that, by using suitable integral identities that are valid among (convergent) integrals defined in strictly $D=4$,
a basis of ``master integrals'' has been identified such that the derivatives of all integrals with respect to external parameters can be expressed within the basis.
We will then seek a change a basis where the differential takes the special form
\begin{align}\label{diffeqsimple}
d \, \vg(s,t,m^2) = (dA) \, \vg(s,t,m^2) \,,
\end{align}
where the matrix $A$ will be required to be \emph{block triangular}.
By this we mean that there should exist a grading (``weight'') such that the derivative of a weight $n$ integral is expressed
in terms of only weight $(n-1)$ integrals. In particular, all diagonal elements of $A$ vanish.
The weight will turn out to correspond the so-called transcendental weight, although at this stage we are merely
discussing the matrix structure of $A$.

Just as in the $D$-dimensional case, one can immediately read off from eq. (\ref{diffeqsimple}), 
and more precisely from the entries of the matrix $A$, what class of functions the answer will be 
expressed in. It defines the integration kernels that can appear in the iterated integrals that
the solution is expressed in.
Moreover, the block triangular structure of $A$ implies that the integrals can be organized according
to their (transcendental) weight, and can be computed in an iterative way,
as will be explained in detail in section \ref{sec:deoneloop}.

In this paper we will compute families of one, two, and three-loop three-scale integrals defined in $D=4$ 
and giving the complete light-by-light scattering in planar $\mathcal{N}=4$ super Yang-Mills \cite{chpaper2}.
Anticipating other applications, we will also compute the one and two loop families at general value of $\eps$.
This set of integrals contains a number of two-scale integrals that appear in
massive form factor calculations. They were computed previously (in a different basis) to some order in $\eps$ in ref. \cite{Anastasiou:2006hc}.
We confirm these results. 
Moreover, the formulation we give can be trivially expanded to any order in $\eps$, where
the result is given by a homogeneous expression in terms of harmonic polylogarithms.

This paper is organized as follows. In section \ref{sec:definitions}, we give definitions of the 
loop integrals that appear in the paper.
Then, in section \ref{sec:deoneloop}, we briefly review the differential equations method for loop integrals,
and discuss simplifications in the four-dimensional limit. We use the one-loop
integrals as a pedagogical example. 
In section \ref{sec:defourdim}, we explain
how to systematically set up the differential equations directly in four dimensions,
and present an algorithm for putting the latter into a canonical block-triangular form.
We give the differential equations at two and three loops and discuss the
iterative structure of the analytic solution.
In section \ref{sec:mandelstam}, we discuss the analytic properties of the functions to three loops
and show that they satisfy a Mandelstam representation.
We discuss checks of our results in section \ref{sec:checks}.
We conclude in section \ref{sec:discussion}.
There are three appendices.
In appendix \ref{app:twoloop} we apply the differential equation method to the set of two-loop master integrals in $D=4-2\eps$
dimensions and compute them using the method of ref. \cite{Henn:2013pwa}.
Additional material on our method for writing down identities and differential equations for four-dimensional loop
integrals can be found in appendix \ref{app:embedding}.
Appendix \ref{app:polylogs} contains expressions for the one- and two-loop box integrals in terms of multiple polylogarithms.
Appendix \ref{app:de} contains the differential equations up to three loops.
We supply several ancillary electronic files together with the arXiv submission of this paper.

%
%
%
%
%

\section{Definitions and kinematical preliminaries}
\label{sec:definitions}

\begin{figure}[t] 
\captionsetup[subfigure]{labelformat=empty}
\begin{center}
\subfloat[(a)]{\includegraphics[width=0.4\textwidth]{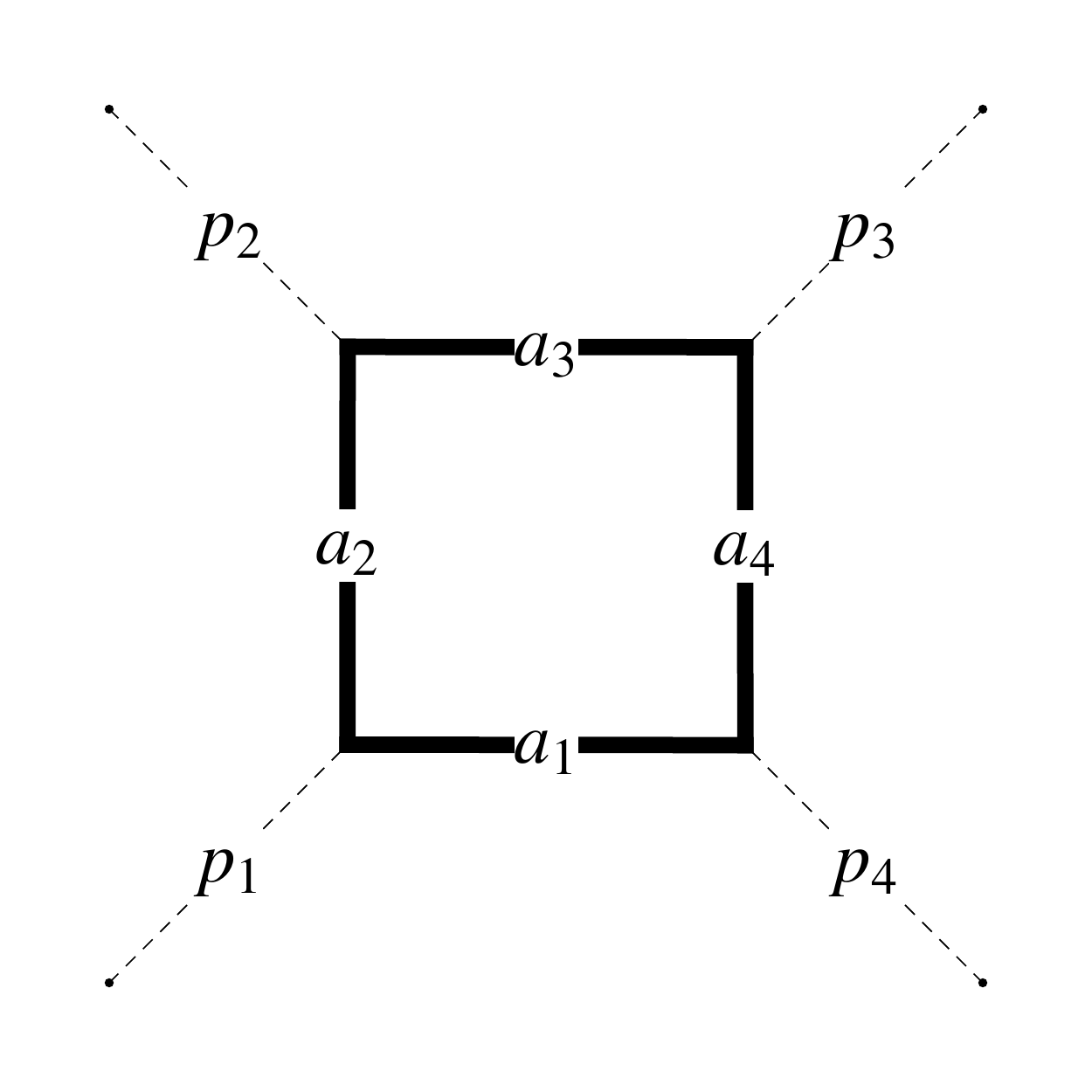}}
\subfloat[(b)]{\includegraphics[width=0.4\textwidth]{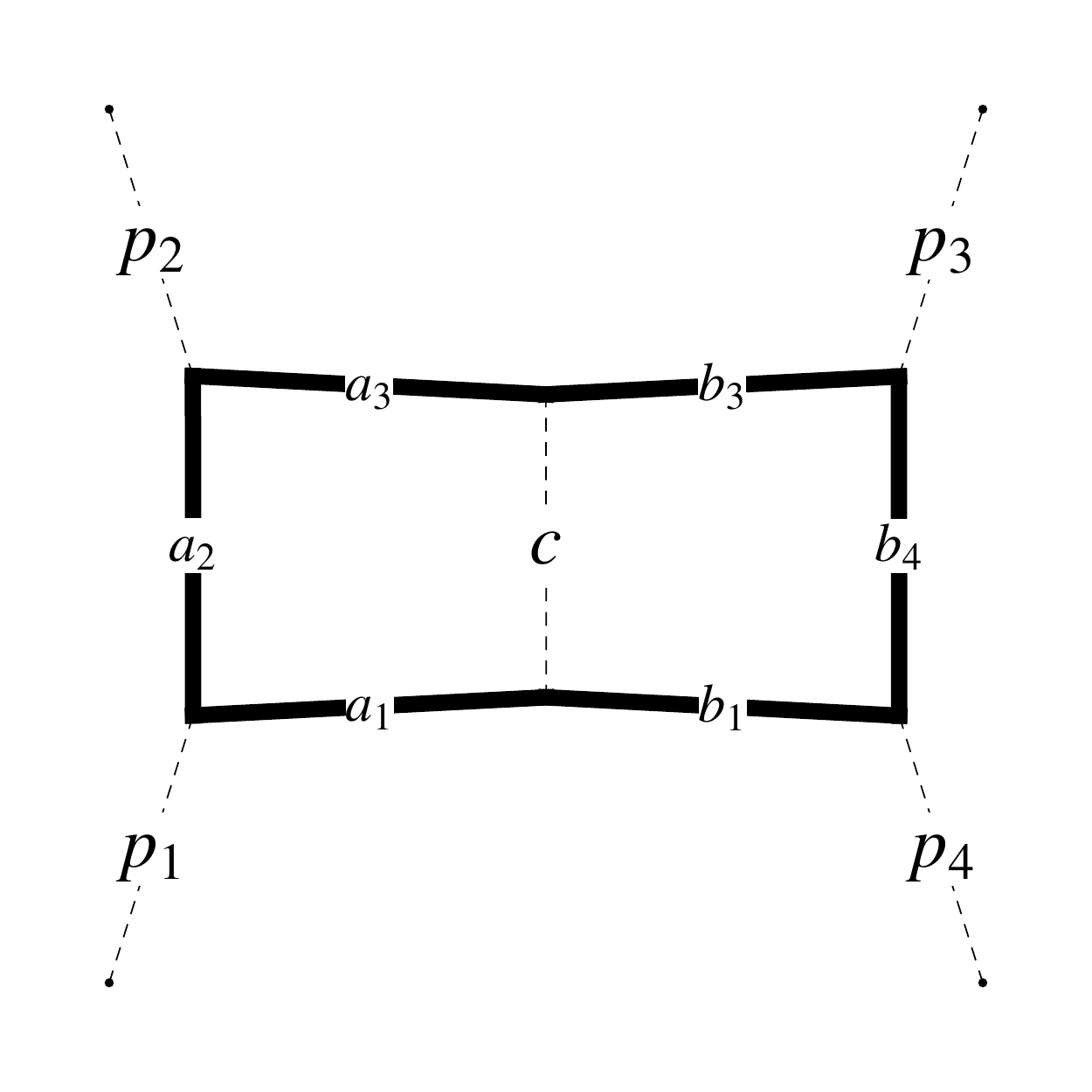}}
\caption{Families of massive one- and two-loop integrals for light-by-light scattering.
Possible irreducible numerator factors at two loops are not shown. }
\label{fig:massivefamily}
\end{center}
\end{figure}

Let us introduce the integral families that we are going to discuss in this paper.
As was already mentioned, the motivation for these particular integrals is that they
can be used to study light-by-light scattering in planar $\mathcal{N}=4$ super 
Yang-Mills \cite{chpaper2}.

\begin{figure}[t] 
\captionsetup[subfigure]{labelformat=empty}
\begin{center}
\subfloat[(a)]{\includegraphics[width=0.4\textwidth]{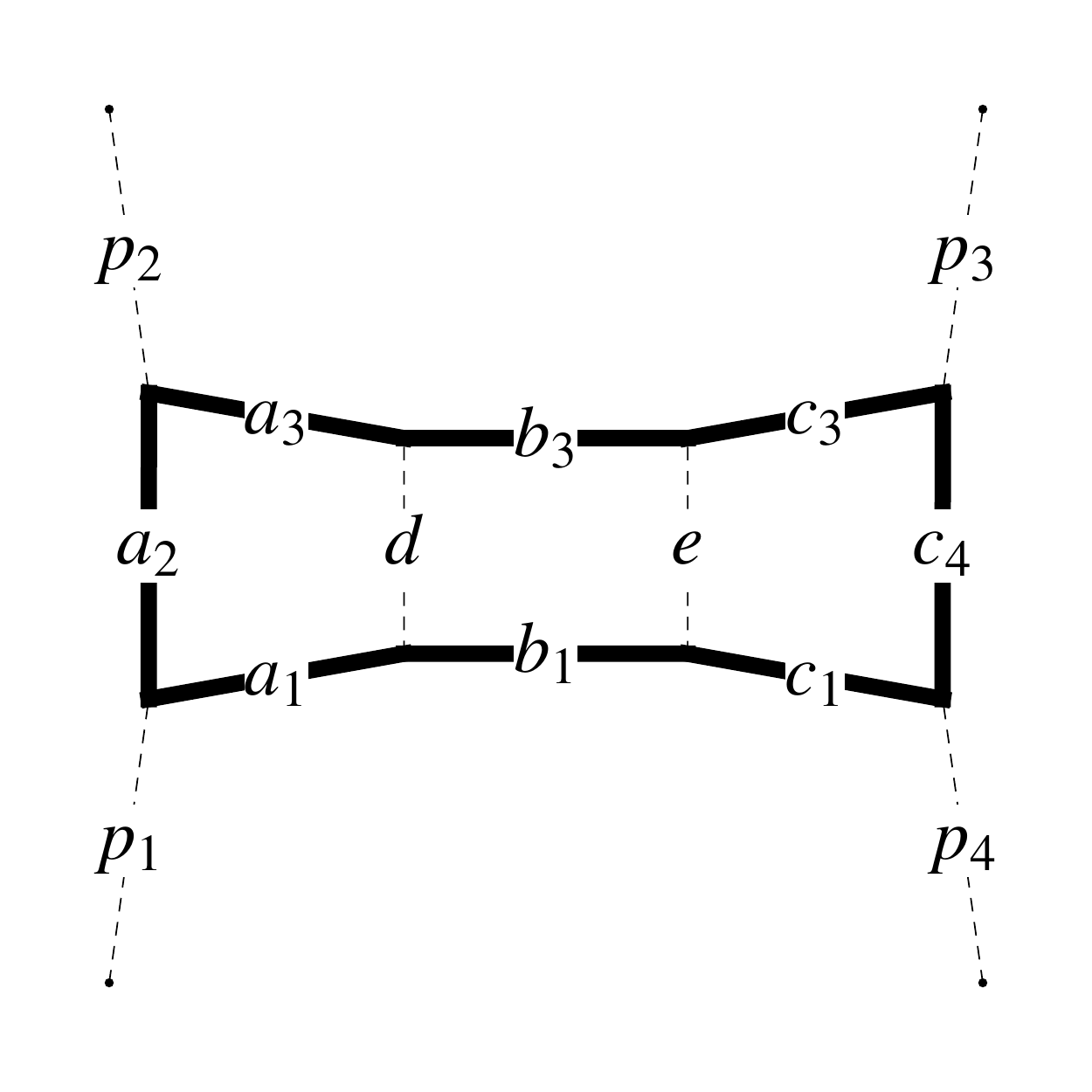}}
\subfloat[(b)]{\includegraphics[width=0.4\textwidth]{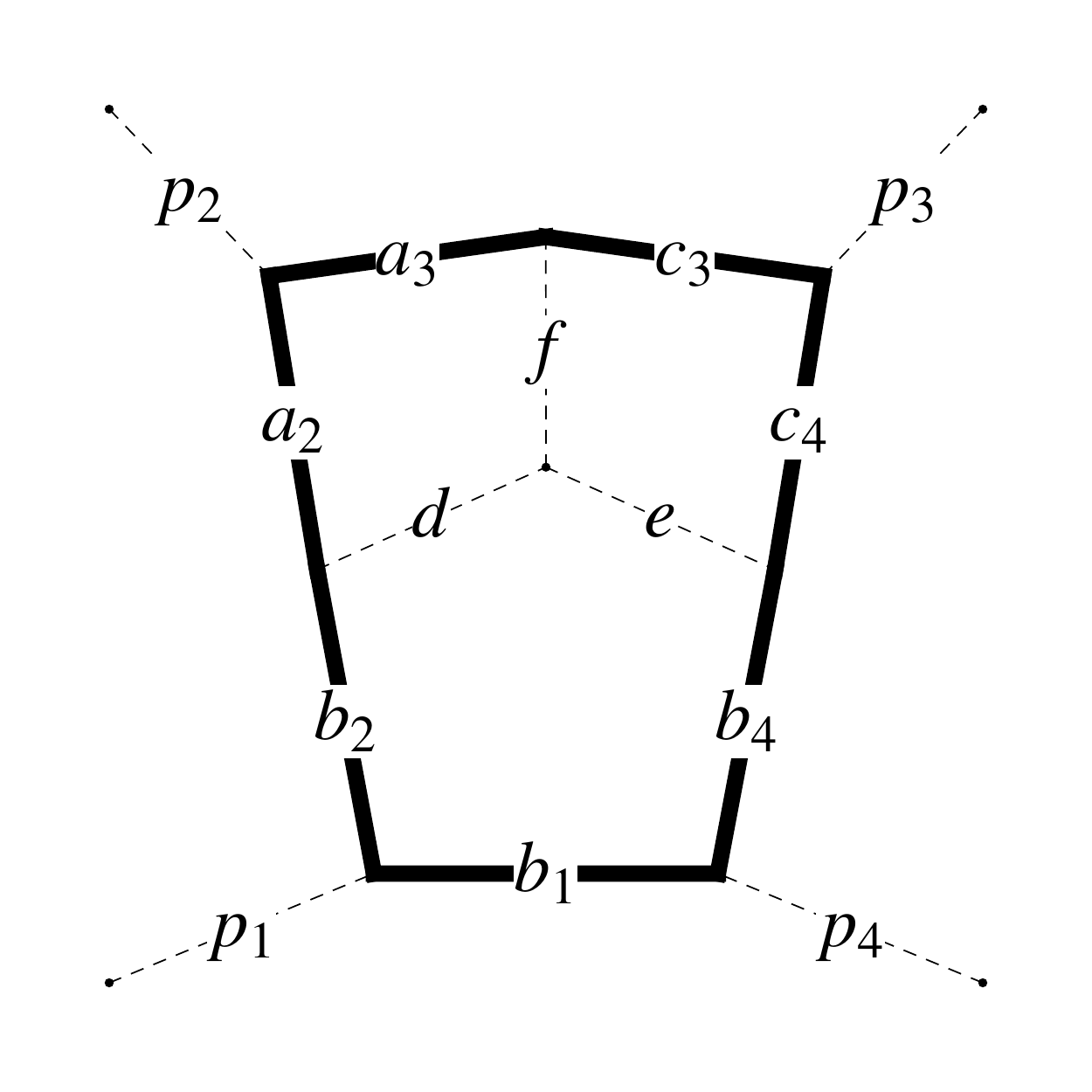}}
\caption{Families of massive three-loop integrals for light-by-light scattering.
Possible irreducible numerator factors are not shown. }
\label{fig:massivefamily3loop}
\end{center}
\end{figure}

We define the inverse propagators
\begin{align}
 D_1(k) =& -k^2+m^2,\quad
&  D_3(k) =& -(k+p_1+p_2)^2+m^2,\nn\\
 D_2(k) =& -(k+p_1)^2+m^2,\quad
& D_4(k) =& -(k-p_4)^2+m^2. \label{inverse_propagators}
\end{align}
Note that we changed the convention for the metric w.r.t. \cite{Alday:2009zm} to the mostly minus metric ($+-\ldots-$).
The integrals we discuss include the one-loop family
\begin{align}
\GA{a_1, \ldots a_4} \label{def-1loopfamily}
 =& \int \frac{ d^{D}k_1}{i \pi^{D/2}} \frac{1}{D_1(k_1)^{a_1}D_2(k_1)^{a_2}D_3(k_1)^{a_3}D_4(k_1)^{a_4}}\,,
\end{align}
where the $a_{i}$ can take any integer values.
At two loops we discuss the family
\begin{align}\label{def-2loopfamily}
\GG{a_1,\ldots, a_4}{b_1,\ldots, b_4}{c}
 =& \int \frac{ d^{D}k_1 d^{D}k_2}{(i \pi^{D/2})^2} 
\frac{D_4(k_1)^{-a_4}}{D_1(k_1)^{a_1}D_2(k_1)^{a_2}D_3(k_1)^{a_3}}
 \nonumber\\
&\hspace{-2cm}  \times \, 
\frac{D_2(k_2)^{-b_2}}{D_1(k_2)^{b_1}D_3(k_2)^{b_3}D_4(k_2)^{b_4}}\frac{1}{[-(k_1 - k_2 )^2 ]^{c} } \,,
\end{align}
where $a_{4}  \le 0, b_{2} \le 0$ represent possible numerator factors. See Fig. \ref{fig:massivefamily}(a) 
and Fig. \ref{fig:massivefamily}(b) for the one- and two-loop cases, respectively.
Finally, defining
\begin{align}\label{def-3loopfamily}
\GC{a_1,\ldots, a_4}{b_1,\ldots, b_4}{c_1,\ldots, c_4}{d,e,f}
 =& \int \frac{ d^{D}k_1 d^{D}k_2d^{D}k_3}{(i \pi^{D/2})^3} 
\frac{D_1(k_1)^{-a_1}D_4(k_1)^{-a_4}}{D_2(k_1)^{a_2}D_3(k_1)^{a_3}} 
\frac{D_1(k_3)^{-c_1}D_2(k_3)^{-c_2}}{D_3(k_3)^{c_3}D_4(k_3)^{c_4} } 
 \nonumber\\
&\hspace{-4cm}  \times \, 
\frac{D_3(k_2)^{-b_3}}{ D_1(k_2)^{b_1}D_2(k_2)^{b_2}D_4(k_2)^{b_4} }
\frac{1}{[-(k_1 - k_2 )^2 ]^{d}[-(k_2 - k_3 )^2 ]^{e}[-(k_1 - k_3 )^2 ]^{f}}\,,
\end{align}
the two three-loop families we discuss are the following.
For the first family, $a_4\le 0,b_2\le0,b_4\le 0,c_2\le 0,f\le0$ represent possible numerator factors,
and the second family is defined in the same way, but instead with $a_1\le0,a_4\leq0,b_3\le0,c_1\le 0,c_2\le 0$. 
They are represented in Fig.~\ref{fig:massivefamily3loop}(a) and Fig.~\ref{fig:massivefamily3loop}(b),
respectively.

Each $G$ is a function of the dimension $D$, the Mandelstam variables $s= (p_1+p_2)^2$ and $t = (p_2 + p_3)^2$,
and the internal mass $m$. The on-shell conditions for the external legs read $p_{i}^2 = 0$ for $i=1,2,3,4$. 

A one-loop integral in the above family will be said to be dual conformal invariant, or DCI for short, if
\begin{align}\label{DCI}
 a_1+a_2+a_3+a_4=&4\,. 
\end{align}
The name is motivated by a natural generalization to a multiple mass case, 
where one can define dual conformal transformations under which the integral is covariant, see  \cite{Alday:2009zm}
and the review \cite{Henn:2011xk}. 
We use a similar definition at higher loops, equating to 4 the sum of indices associated with each loop variable.
For example, a two-loop integral is DCI if $(a_1+a_2+a_3+a_4+c)=4$ and $(b_1+b_2+b_3+b_4+c)=4$.

In the following, we will define all integrals in the Euclidean region $s/m^2<0,t/m^2<0$, where all functions are real-valued. 
One then defines the functions elsewhere by analytic continuation, using the Feynman prescription.
This implies giving the kinematical variables a small imaginary part, according to
$m^2 \rightarrow m^2 -i0, s \rightarrow s+i0, t \rightarrow t+i0$.

Let us have a first look at these integrals, and take $I_{1}$ as an example.
It is given analytically by (the form below is due to \cite{Davydychev:1993ut}),
\begin{align}\label{I1exact}
I_{1}=& \frac{2}{\buv} \Big\{ 
2 \log^2\left( \frac{\buv + \bu}{\buv + \bv} \right) + 
\log\left( \frac{\buv - \bu}{\buv + \bu} \right)  \log \left( \frac{\buv - \bv}{\buv + \bv} \right)
- \frac{\pi^2}{2} \nonumber\\
& + \sum_{i=1,2} \Big[ 
2 \, \Li_{2} \left( \frac{\beta_{i} - 1}{\buv + \beta_{i}} \right)
-2 \, \Li_{2} \left(- \frac{\buv - \beta_{i}}{\beta_{i} + 1} \right)
- \, \log^2 \left(\frac{\beta_{i} +1}{\buv + \beta_{i}} \right)
\Big]
 \Big\} \,.
\end{align}
Here we introduced dimensionless variables\footnote{From the context there should be
no confusion between the 
ratio $u$ and the Mandelstam invariant $u=-s-t$.
Also note that our normalization of $u$ and $v$ differs by a factor 4 from those in ref.~\cite{Alday:2009zm}.} 
\begin{align}
u = \frac{4m^2}{-s}\,,\qquad \v=\frac{4m^2}{-t}\,,
\end{align}
and the following abbreviations,
\begin{align}
\bu=\sqrt{1+u}\,,\qquad \bv=\sqrt{1+v}\,,\qquad \buv=\sqrt{1+u+v}\,.
\end{align}
The functions appearing in eq. (\ref{I1exact}) are examples of polylogarithms. For these
and more general classes of integral functions that we will discuss one can define a  
``symbol'' \cite{Goncharov:1998kja,arXiv:math/0606419,Goncharov:2010jf}.
Roughly speaking, the symbol contains information about the integration kernels leading to
those functions, while forgetting about boundary constants at each integration step.
We note that the symbol of the above formula is very simple, and visibly more compact compared to eq. (\ref{I1exact}),
\begin{align}\label{I1exactsymbol}
{\mathcal{S}}\left[\buv \, I_{1}\right] = 
2  \left[ \frac{ \bu -1}{\bu +1} \otimes \frac{ \buv - \bu}{\buv+\bu}
+ \frac{ \bv -1}{\bv +1} \otimes \frac{ \buv - \bv}{\buv+\bv} \right] \,.
\end{align}
This foreshadows a simple structure of the integral under the action of differential operators.
In the next two sections, we will see how this structure arises in a systematic way.
It will however not be necessary to restrict the analysis to the level of the symbol, rather
the observations will apply to the functions directly.
Indeed, we will be able to write compact formulas in terms of 
iterated integrals that make the simplicity manifest, while at
the same time keeping track of the integration constants.

In particular, the complete information 
specifying the multi-loop integrals we will discuss will be contained in
simple formulas similar to eq. (\ref{I1exactsymbol}). 
In the next two sections, we will first see how to reproduce this formula from 
differential equations, and then proceed to compute the required
integrals at two and three loops.

%
%
%
%
%

\section{Differential equation at one loop: $4$ versus $D$ dimensions}
\label{sec:deoneloop}

Computing loop integrals via differential equations is by now a fairly standard procedure \cite{Kotikov:1990kg,Remiddi:1997ny,Gehrmann:1999as,Argeri:2007up}, so we will only briefly outline the main steps.
For a given class of integrals under consideration, defined by the propagator structure, 
one defines integral families for any possible powers of the propagators. 
E.g. at one loop, the family we consider is defined by eq. (\ref{def-1loopfamily}), see also Fig.~\ref{fig:massivefamily}(a), for any integer values of the parameters $a_{i}$.
The integrals in a given family are in general not independent. 
They satisfy integration-by-parts (IBP) identities \cite{Chetyrkin:1981qh}. The latter are linear
in the integrals. 
In practice, this means that one can relate the calculation of any given integral to a finite 
number of master integrals. The algebraic manipulations needed are straightforward but cumbersome,
so that one usually performs them using computer algebra codes. We have used FIRE \cite{Smirnov:2008iw,Smirnov:2013dia}, 
as well as an own implementation, to be described below. 
Other public computer codes include \cite{vonManteuffel:2012np,Lee:2012cn}. 
The independent integrals obtained in this way are usually referred to as
master integrals. The basis choice is not unique.
We have found \cite{Lee:2013hzt} helpful to verify the number of master integrals needed in a given sector.

For a given set of master integrals, one sets up differential equations in the kinematical invariants. The derivative w.r.t. $m^2$ is straightforward to implement. 
The derivatives w.r.t. $s$ and $t$ can be written down via the chain rule, in terms of derivatives w.r.t. the external momenta, e.g.\footnote{Note that the operator on the r.h.s. of eq. (\ref{differs}) is defined to act on expressions such as eq. (\ref{def-1loopfamily}), where $p_{3}^\mu$ has been eliminated via momentum conservation, and it is chosen such that it commutes with the on-shell conditions $p_{1}^2 = 0$ and $(p_1 + p_2 + p_4 )^2=0$.}
\begin{align}\label{differs}
 \frac{\partial}{\partial s} = \frac{1}{2} \left[  \frac{1}{s} p_{1}^\mu - \frac{2 s +t}{s (s+t)} p_{2}^\mu + \frac{1}{s+t} p_{4}^{\mu} \right] \frac{\partial}{\partial p_{2}^{\mu}} \,.
\end{align}
When acting with the operator (\ref{differs}) on integrals such as (\ref{def-1loopfamily}), it is clear that the resulting terms will still be within the same family of integrals. Therefore, any derivative of a master integral can be expressed in terms of a linear combination of master integrals. This implies that the master integrals satisfy a linear system of first-order equations. Let us discuss this system in the one-loop case.

It is well known that the set of master integrals at one-loop, in the presence of internal masses, consists of the box, triangles, bubbles and tadpole topologies.
Thanks to symmetries, for the present configuration this amounts to six master integrals.
As we will see, the corresponding $6 \times 6$ system of differential equations can be put in the form of eq.~(\ref{diffeqsimpleDdim}) by 
choosing the following basis of master integrals,
\begin{equation}
\label{UTbasis1loop}
\begin{aligned}
f_1=& 2c\, m^{2} \GA{0,0,0,3}\\
f_2=& -\epsilon c\, \sqrt{s(s-4m^2)}\GA{1,0,2,0}\\
f_3=& -\epsilon c\,  \sqrt{t(t-4m^2)}\GA{0,1,0,2}\\
f_4=& \epsilon^2 c\,s \GA{1,1,1,0}\\
f_5=& \epsilon^2 c\,t \GA{1,1,0,1}\\
f_6=& \frac12\epsilon^2 c\, \sqrt{st(st-4m^2(s+t))} \GA{1,1,1,1}\,,
\end{aligned}
\end{equation}
where $c=m^{2\epsilon}/\Gamma(1+\epsilon)$ is a convenient overall factor.

We start by setting up the system of first-order differential equations for the basis functions $f_{i}$.
which takes the form, in our basis choice,\footnote{Note that the notation for $A$ and $\tilde{A}$ in refs. \cite{Henn:2013pwa,Henn:2013tua,Henn:2013woa} is slightly different from the one used here.}
\begin{align}\label{partialeqs}
\partial_s  \vf = \eps \, \tilde A_{s} \vf  \qquad  
\partial_t  \vf = \eps \, \tilde A_{t} \vf  \,,  \qquad
\partial_{m^2}  \vf = \eps \, \tilde A_{m^2} \vf \,.
\end{align}
The crucial simplification of our basis choice lies in the fact that the partial derivative matrices $\tilde A_{s}, \tilde  A_{t}$ and $\tilde  A_{m^2}$ are
independent of $\eps$. Moreover, as we will see presently, they can be expressed in terms of derivatives of logarithms only.

The $f_{i}$ are normalized to be dimensionless functions and therefore satisfy 
\begin{align}
(s \partial_s + t \partial_t  + m^2 \partial_{m^2} ) \vf = 0\,,
\end{align}
which is a useful consistency check.
Other consistency checks are obtained from requiring that partial derivatives commute, which implies
\begin{align}\label{integrabilityconditions}
[ \tilde A_{i} , \tilde A_{j} ] = 0  \,, \qquad \partial_{i} \tilde A_{j} - \partial_{j} \tilde A_{i} = 0\,,
\end{align}
where $i,j$ can be $s,t$, or $m^2$.

The structure of the differential equations, and hence that of the $f_{i}$, is greatly clarified by 
combining equations (\ref{partialeqs}) by writing them in differential form
\begin{align}\label{canonicalequationssection3}
d \, \vec{f} = \eps \, (d \tilde{A}) \, \vec{f} \,.
\end{align}
This is the form quoted in the introduction.
Here $\tilde{A}$ was obtained by integrating the partial differential equations
\begin{align}
\partial_s \tilde{A} = \tilde A_s \,,\qquad
\partial_t \tilde{A} = \tilde A_t \,,\qquad 
\partial_{m^2} \tilde{A} = \tilde A_{m^2} \,. 
\end{align}
This integration is very similar to the one that is needed to integrate $\vf$ (at the first order in the $\eps$ expansion).
Performing it at this stage allows us to completely elucidate the properties of the functions $\vf$.
In fact, we find that $\tilde{A}$ can be written in terms of logarithms only,
\begin{align}\label{Amatrix1loop}
 \tilde A=& \left(\begin{array}{*{6}{c}}
0&0&0&0&0&0\\ 
\cline{1-1}
\multicolumn{1}{|c|}{\Log{\frac{\bu-1}{\bu+1}}}&\log{\frac{\u}{1+\u}}&0&0&0&0\\
\multicolumn{1}{|c|}{\Log{\frac{\bv-1}{\bv+1}}}&0&\log{\frac{\v}{1+\v}}&0&0&0\\
\cline{1-1}\cline{2-3}
0&\multicolumn{1}{|c}{-\Log{\frac{\bu-1}{\bu+1}}}&\multicolumn{1}{c|}{0}&0&0&0\\
0&\multicolumn{1}{|c}{0}&\multicolumn{1}{c|}{-\Log{\frac{\bv-1}{\bv+1}}}&0&0&0\\
0&\multicolumn{1}{|c}{\Log{\frac{\buv-\bu}{\buv+\bu}}}&\multicolumn{1}{c|}{\Log{\frac{\buv-\bv}{\buv+\bv}}}&\Log{\frac{\buv-1}{\buv+1}}&\Log{\frac{\buv-1}{\buv+1}}&\Log{\frac{\u+\v}{1+\u+\v}}\\ \cline{2-3}
\end{array}\right)\,.
\end{align}
We have `boxed' the elements which survive in the $\eps\to0$ limit, to be discussed presently.

To fully define the answer we need to specify a boundary condition.
For the integrals under consideration a particularly natural boundary point is $m=\infty$, where
\begin{align}\label{boundaryoneloop}
 \vec{h}(\eps) \equiv  \lim_{m\to\infty} \vf(s,t,m^2;\eps) = \delta_{i,1}\,.
\end{align}
Before commenting on the solution to the differential equations, let us first discuss
a simplification for finite integrals.

We will be particularly interested in the $\eps\to 0$ limit. Since all the above integrals
are both ultraviolet and infrared convergent (thanks to the internal masses, and thanks to the choice of representatives we made),
it is natural to remove the powers of $\eps$ in eq.~(\ref{UTbasis1loop}) in the limit.
We thus let
\begin{align}\label{fromftog}
 (g_1,g_2,g_3,\tilde g_4,\tilde g_5,g_6) = \lim_{\epsilon\to 0} \,
  (f_1,\frac1\epsilon f_2, \frac1\epsilon f_3,\frac1{\epsilon^2}f_4,\frac1{\epsilon^2} f_5,\frac1{\epsilon^2} f_6).
\end{align}
The powers of $\epsilon$ reflect the transcendental weights of the functions $g_i$.
We have placed `tilde' on $\tilde g_4$ and $\tilde g_5$ to distinguish them from two-loop integrals $g_4$ and $g_5$ which will be introduced in the next section.
The integrals $g_1,g_2,g_3$ and $g_6$ will carry the same meaning throughout this paper.

The differential equation for the $g_i$ takes the similar canonical form (\ref{diffeqsimple}),
with the $A$-matrix obtained from $\tilde A$ in eq.~(\ref{Amatrix1loop}) by retaining only the boxed elements.

\begin{figure}[t] 
\captionsetup[subfigure]{labelformat=empty}
\begin{center}
{\includegraphics[width=1.0\textwidth]{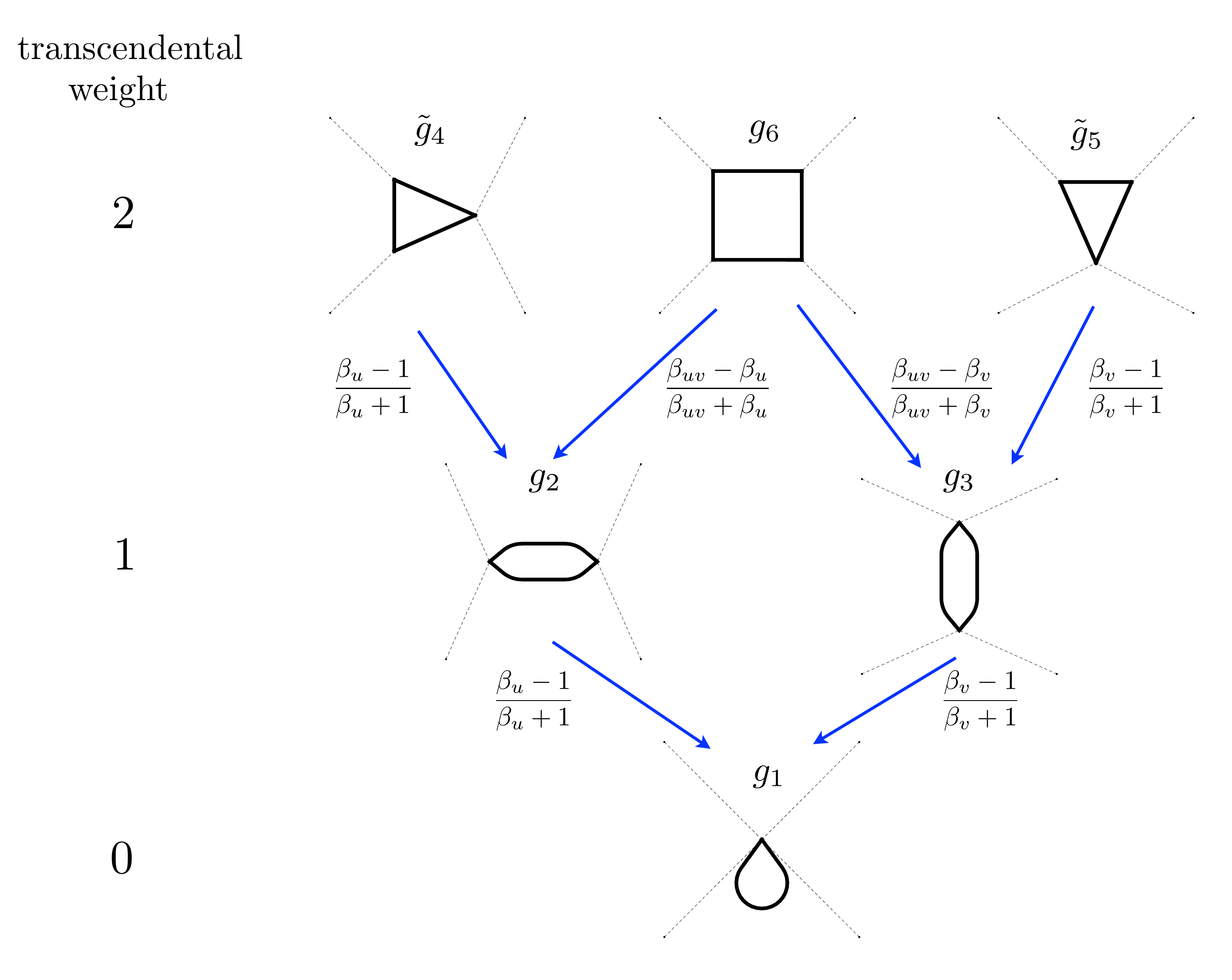}}
\caption{Hierarchy of one-loop functions. 
The integrals are classified according to their (transcendental) weight, shown in the leftmost column. 
Each arrow corresponds to one non-zero element of the derivative matrix $A$, cf. eq. (\ref{Amatrix1loop}). 
The fact that arrows only link integrals in adjacent rows
is the statement that the matrix is {\it block triangular}. 
Solid and dashed lines denote massive and massless propagators, respectively.}
\label{fig:blocktriangular}
\end{center}
\end{figure}

Two observations will be important for us.
\begin{itemize}
\item The integrals $g_i$ have uniform transcendental weights $(0,1,1,2,2,2)$, respectively. This follows immediately from the block-triangular structure
of the $A$-matrix: the derivatives of $g_6$ are expressed in terms of $g_2,g_3$ only, while the derivatives of the latter are expressed in terms of $g_1$ only.
Another way of seeing this is to note that the functions $f_{i}$ have uniform weight zero, and the rescaling by powers of $\eps$, which can be assigned weight $-1$, leads to the weight of the $g_{i}$ given above.
\item The box $g_{6}$ and triangles $\tilde{g}_{4}, \tilde{g}_{5}$ are fully decoupled in $D=4$.
\end{itemize}
The block triangular structure of the matrix and its implication for the weight of the functions is visualized in a different way in Fig.~\ref{fig:blocktriangular}. There one can also see the decoupling of the box and triangle integrals.
The decoupling of the triangles implies that, if we are interested only in evaluating the box integral in $D=4$, we could consistently
work with the truncated basis $g_1,g_2,g_3$ and $g_6$.  
Specifically, the system of differential equations in this case reduces to, cf. eq. (\ref{Amatrix1loop}),
\begin{align}\label{DE1looptruncatedDeq4}
d \left(\begin{array}{*{1}{c}}
g_1\\
g_2\\
g_3\\
g_6
\end{array}\right)
 =& d \, \left(\begin{array}{*{4}{c}}
0&0&0&0\\ 
\cline{1-1}
\multicolumn{1}{|c|}{\Log{\frac{\bu-1}{\bu+1}}}&0&0&0\\
\multicolumn{1}{|c|}{\Log{\frac{\bv-1}{\bv+1}}}&0&0&0\\
\cline{1-1}\cline{2-3}
0&\multicolumn{1}{|c}{\Log{\frac{\buv-\bu}{\buv+\bu}}}&\multicolumn{1}{c|}{\Log{\frac{\buv-\bv}{\buv+\bv}}}&0\\ \cline{2-3}
\end{array}\right)
\left(\begin{array}{*{1}{c}}
g_1\\
g_2\\
g_3\\
g_6
\end{array}\right)
\,,
\end{align}
in agreement with the general form of eq. (\ref{diffeqsimple}).
Such truncations will be exploited extensively in the next section.

Let us now discuss the general solution in $D=4-2\eps$ dimensions, and then come back to the simplifications as 
$\eps \to 0$.
With the differential equations in the form (\ref{canonicalequationssection3}) we can immediately write down the analytic answer in terms of Chen iterated integrals \cite{Chen1997}. 
We have
\begin{align}\label{path_ordered}
\vf(s,t,m^2;\eps) = {\mathbb P} e^{\eps \int_{{\gamma}} d\, \tilde{A}} \, \vec{h}(\eps)\,.
\end{align}
Here the integration contour ${\gamma}$ is a path in the space of kinematical variables, which begins at a base point, in our case $m\to \infty$, where we have the simple boundary condition (\ref{boundaryoneloop}). 

Let us be more specific about the notation, following closely the recent lecture notes \cite{iterated1,iterated2} on iterated integrals. We denote by $\mathcal{M}$ the space of kinematical variables, here $(u,v) \in \mathbb{R}^2$, 
and let $\omega_i$ be some differential one-forms (corresponding to entries of $d\, \tilde{A}$ or $d\, A$).
Moreover, define the pull-back of the differential forms to the unit interval $[0,1]$ via
\begin{align}
\gamma^{*}(\omega_{i}) = f_{i}(t) dt\,.
\end{align}
Then, an ordinary line integral is given by
\begin{align}
\int_{\gamma} \omega_1 = \int_{[0,1]}\gamma^{\star}(\omega_1) = \int_0^1 f_{1}(t_1) t_1 \,.
\end{align}
Then, the iterated integral of $\omega_1, \ldots \omega_n$ along $\gamma$ is defined by
\begin{align}\label{iterateddef3}
\int_{\gamma} \omega_1 \ldots \omega_n = \int_{0\le t_1 \le \ldots \le t_n \le 1} f_{1}(t_1) dt_1 \ldots f_{n}(t_n) dt_{n} \,.
\end{align}
Iterated integrals have many nice properties, see \cite{iterated1,iterated2}. 
Moreover, the iterated integrals appearing in eq. (\ref{path_ordered}) are homotopy invariant (on $\mathcal{M}$ with the set of singularities removed).
This property makes them very flexible, and we will see later that how to rewrite them in terms of more familiar functions, if desired.
We will also discuss expansions in limits and their numerical evaluation.

Expanding eq. (\ref{path_ordered}) to second order in $\eps$ and noting that the box integral $I_1$ in eq.~(\ref{I1exact}) is simply $\frac{1}{\eps^2}\frac{2}{\buv} f_6$,
it is trivial to reproduce its symbol given in eq. (\ref{I1exactsymbol}) starting from the $\tilde{A}$-matrix.
With a little bit more work, it is also possible to verify that $I_1$ indeed obeys the claimed differential equation and boundary condition.

More generally, one may expand the solution to any desired order in $\eps$. With the normalization of the master integrals of eq. (\ref{UTbasis1loop}), the term multiplying $\eps^k$ will be a $\mathbb{Q}$-linear combination of $k$-fold iterated integrals, i.e. integrals of weight $k$. Such functions are sometimes called {\it pure} functions.

Let us now discuss the simplifications that occur for the four-dimensional basis (\ref{DE1looptruncatedDeq4}).
The solution can be again be written as in eq. (\ref{path_ordered}), with the difference that $\eps$ is removed, and that only
a finite number of terms have to be kept in the exponential. 

The block diagonal form of $d\,A$ implies the particularly simple iterative structure of the integrals
that we already alluded to earlier when discussing the symbol. 
The iterative solution starts with the weight zero function $1$, represented by the tadpole integral $g_1$.
To obtain the weight one integrals, we simply have to follow the arrows shown in Fig.~\ref{fig:blocktriangular} back up (starting from the tadpole). To each arrow corresponds an element of the $d\,A$-matrix, and the contribution is given by the integral over the latter. For example,
\begin{align}
g_2 = \int_{{\gamma}} d \, \log   \frac{ \bu -1}{\bu +1}    \,.
\end{align}
As was explained before, the integration is along a contour $\gamma$ in $(u,v)$ space, 
starting at the boundary point $u,v \to \infty$ corresponding to $m^2 \to \infty$. 
Of course the above integral just evaluates to $\log  \frac{ \bu -1}{\bu +1}$, but this notion will be important when generalizing to higher weight functions.
Next, the formulas for the weight two functions are obtained by summing over all paths from the tadpole to the integral under consideration. 
The contribution of each path is an iterated integral over the differential forms corresponding to each arrow.
In this way, one arrives at
\begin{align}\label{I1exactfunction}
g_6 = \int_{{\gamma}} d \, \log   \frac{ \bu -1}{\bu +1} \,  d \log  \frac{ \buv - \bu}{\buv+\bu}
+ \int_{{\gamma}} d \, \log \frac{ \bv -1}{\bv +1} \, d \, \log  \frac{ \buv - \bv}{\buv+\bv}  \,.
\end{align}
Note that only the sum of the two iterated integrals in eq. (\ref{I1exactfunction}) is homotopy invariant, not the individual terms.
We can compare this to eq.  (\ref{I1exactsymbol}), keeping in mind that $g_{6} = \lim_{\eps \to 0} \frac{\beta_{uv}}{2} I_{1}$.
As was already mentioned, the difference is that, unlike the symbol,
eq. (\ref{I1exactfunction}) completely determines the answer.

We see that the solution is specified by the set of differential forms $\omega_{i}$ allowed to appear in the iterated integrals. This {\it alphabet} can be read off directly from the $d\,\tilde{A}$ or $d\, A$ matrices.
Of course, for multivalued functions depending on many kinematical variables, one can expect this alphabet, which is related to possible singularities of the functions,
to be relatively complicated. As we will see, one can often choose specific contours $\gamma$ that express the Chen iterated integrals in terms of more common multiple polylogarithms \cite{Goncharov:1998kja,arXiv:math/0606419}, at the cost of giving up the homotopy invariance and the compactness of the expressions. An example is given in appendix \ref{app:polylogs}. The general formulas are also very useful for discussing analytic continuation and for studying simplifying limits, where again one often finds expressions in terms of multiple polylogarithms. Many examples can be found in \cite{chpaper2}.
Discontinuities of the functions are also very transparent, see section \ref{sec:mandelstam}.

In this section we have explained how to obtain a simplified version of the differential equations in the $\eps \to 0$ limit, 
starting from the $D=4-2\eps$ dimensional case. We saw that this implied that we could work with a smaller set of integrals
which satisfy block-triangular systems of differential equations.
In the next section, we will explain how to arrive at such results working directly in four dimensions. 

%
%
%
%
%

\section{Differential equation in $D=4$: two and three loops}
\label{sec:defourdim}

In $D=4$ a natural subset of integrals to consider are the dual conformal ones, cf. eq (\ref{DCI}).
As we will see shortly, the derivatives of dual conformal integrals with respect to kinematic variables are themselves dual conformal.
Therefore, one expects to be able to find a closed set of differential equations among dual conformal integrals.
This has several advantages over the $D$-dimensional approach.\footnote{
We note that the idea of using IBP relations directly in four dimensions has appeared previously in refs. \cite{Smirnov:1996yi,Kravtsova:1997hk}, although not in the present context of dual conformal integrals.}   
   
By restricting to a subset of integrals closed under differential equations, the number of integrals
we need to consider in total decreases. This, and the possibility of setting $D=4$ from the start results in much faster computer implementations of the algebra needed to derive the differential equations. 
This allowed us to extend the analysis to the three-loop level.

In this section we describe the method.
The first step, to generate tables of integration-by-part
identities among dual conformal integrals, is discussed in subsection \ref{sec:IBPdci}.
An interesting feature of working in $D=4$ dimensions is that the identities sometimes contain contact terms,
allowing different loop orders to be seamlessly merged.
Once \emph{some} basis of dual conformal master integrals has been found, our next step was to find a change of basis that puts the differential equation into the canonical \emph{block triangular} form.
It turns out that in $D=4$ this problem is much simpler compared to the $\eps \neq 0$ case, and
can be solved systematically. We detail the procedure in subsection \ref{sec:canonicalform}. 
We conclude this section with the main result for the two- and three-loop integrals.

\subsection{Integration by part identities among dual conformal integrals}
\label{sec:IBPdci}

It is natural to expect that among all
integration-by-parts identities, there exists a \emph{non-empty}
subset of identities, which hold exactly in $D=4$, and which relate only dual conformal
integrals to one another.   Identifying this subset
starting from a list of $D$-dimensional reduction identities among non-conformal
integrals might seem like a daunting task.
It turns out that this admits
a rather elegant solution in terms of the \emph{embedding formalism}.

The embedding formalism is ideally suited to this problem, because
writing a non-conformal expression in this formalism requires a deliberate effort.
The main idea is to write propagators as \emph{linear} functions
of a $(D{+}2)$-vector $Y_i\simeq (k_i^\mu,-k_i^2,1)$, where the
$Y_i$'s obey certain equivalence relations.  The formalism is
described in appendix \ref{app:embedding}.
As an example, the one-loop integral of eq. (\ref{def-1loopfamily}) is rewritten in the form 
\be
 \GA{a_1,\ldots,a_4} = \int_{Y_1}
 \frac{\cd{Y_1}{I}^{\sum_i
     a_i-D}}{\cd{Y_1}{X_1}^{a_1}\cd{Y_1}{X_2}^{a_2}\cd{Y_1}{X_3}^{a_3}\cd{Y_1}{X_4}^{a_4}}\,. \nonumber
\ee
The numerator is necessary to ensure that the integrand has uniform homogeneity degree $D$ under rescaling of $Y_1$.
The important feature here is that dual conformal integrals are precisely those for which the
``point at infinity'' $I$ is absent.
By simply never introducing this point, one automatically stays within the class of conformal integrals.

Since the $(D{+}2)$-vectors $Y_i$ are a trivial change of
variable away from the original loop momenta $k_i$, it is straightforward to
carry through all the usual operations (generation of integration by
parts identities, derivatives with respect to external variables,
etc.) in terms of these variables.  The upshot is that we automatically generate
identities which involve only dual conformal integrals.

In order not to distract from the main course of this paper, these
technical matters are discussed extensively in appendix
\ref{app:embedding}.
Here we record only the most salient features in which $D=4$ differs from $D\neq 4$:
\begin{itemize}
\item Integration-by-part identities in $D=4$ sometimes contain contact terms, which appear through the identity (\ref{delta_function_term})
\be
 \frac{\partial}{\partial k_1^\mu} \frac{(k_1-k_2)^\mu}{(k_1-k_2)^4} =  2\pi^2 i \delta^4(k_1-k_2)\,. \nonumber
\ee
This identity effectively relates integrals with different loop
orders, since the $\delta$-function removes one integration variable.
\item When working without a regulator it is important to restrict
  attention to \emph{convergent} integrals, which converge
 in all integration regions.  To this aim we imposed a simple set
 of \emph{sufficient} conditions on the set of integrals we
 have considered, also listed in appendix \ref{app:embedding}.
\item Derivatives of dual conformal integrals with respect to kinematic variables are themselves
  dual conformal, see e.g. eq.~(\ref{DCIderivative}).
\end{itemize}

By generating identities among dual-conformal integrals in $D=4$ and following
standard integral reduction algorithms, we have reduced all integrals
which appearing in derivatives  to a \emph{minimal basis}.  Before we
describe this basis, it is convenient to perform a certain basis change.

\subsection{Algorithm for obtaining the differential in canonical form}
\label{sec:canonicalform}

Perhaps the most striking feature of working in $D=4$ is that the $A$
matrix, in a suitable basis, is expected to admit a block-triangular form. This reflects the
existence of master integrals which have uniform transcendental
weights. \emph{Assuming} that such a basis exists, a constructive algorithm can find it!

The method is actually rather straightforward: first we put the system
into triangular form, by finding the proper normalization of (suitable combinations of) the master integrals.
This already introduces a first approximation to the notion of weight.
Then we recursively shift integrals by lower weight ones, until all differentials can be written in ``d-$\log$'' form; the matrix will then automatically be block-triangular.\footnote{A related proposal for obtaining the canonical form of the differential equations of \cite{Henn:2013pwa} in the $D$-dimensional case has been put forward in ref. \cite{Argeri:2014qva}.}

We explain the three steps of our method, illustrating them with simple examples.

\subsubsection*{Step 1. \emph{Obtaining a triangular form}. }

 We decompose the integrals into two sets: correctly ``normalised'' and ``remaining'' ones.
We assume that the constant integral ($g_1$) is part of the basis; by definition it is correctly normalised.
We then successively normalise the remaining integrals as follows.

First we find the smallest subset of remaining integrals, whose derivatives involve only themselves and normalised ones.
As a trivial example, in the two-loop case after we had normalised the integrals $g_1,g_2,g_3,g_4$ (see below for their definitions),
the smallest subset consisted of just one element:
\begin{align}
  \frac{d}{ds} \GG{0,1,2,0}{2,0,1,0}{1} =& -2\frac{s-3m^2}{s(s-4m^2)} \GG{0,1,2,0}{2,0,1,0}{1}\,\, ({\rm mod}\, g_1,g_2,g_3,g_4).
\end{align}
We integrate this equation, working modulo already normalised integrals:
\begin{align}
   \frac{d}{ds} \left(s m^2\sqrt{s(s-4m^2)} \GG{0,1,2,0}{2,0,1,0}{1}\right) =&0 \,\, ({\rm mod}\, g_1,g_2,g_3,g_4).
\end{align}
Since the right-hand side involves only other, already normalised integrals, we declare the parenthesis to be correctly normalised and we add it to the list.
We then repeat the same procedure for the remaining integrals.

Sometimes  the smallest closed set of remaining integrals contains more than one element.
This happened to us at three loops; for example at some point we had the ``entangled'' pair of integrals
\begin{align}
\quad \vec{h} =& \big(\GC{0,1,2,0}{2,0,0,0}{2,0,1,0}{1,1,0},\GC{0,2,1,0}{2,0,0,0}{2,0,1,0}{1,1,0}\big) \nn
\end{align}
with the differential equation
\begin{align}
\frac{d}{ds} \vec{h} =& \vec{h}M\,\,({\rm mod\,normalised}), \qquad M=\frac{m^4}{s}\left(\begin{array}{c@{\,\,\,\,}c} \frac{2}{4m^2-s}&-2\\ \frac{10}{4m^2-s}&-4 \end{array}\right).
\end{align}
The right-hand side can be removed by considering the ansatz $I=\vec{h}.(c_1,c_2)$ and solving
the differential equation $\frac{d}{ds} (c_1,c_2)= -M(c_1,c_2)$.\footnote{For this purpose we found the {\tt Mathematica} built-in command {\tt DSolve[]} to be adequate.}
The solution gives that
\begin{align}
 \frac{d}{ds} \left(\begin{array}{l}
 s^2 \left[(4m^2-s)(2m^2-s)h_1  + (10m^4-10m^2s+s^2)h_2\right]\\
  s\sqrt{s(s-4m^2)}\left[(2m^4-4m^2s+s^2)h_1 +(2m^4-6m^2s+2s^2)h_2\right]
 \end{array}\right) =& 0 \quad({\rm mod\,normalised}). 
\end{align}
These two combinations can now be added to the list of normalised integrals, 
and the procedure repeated for the next subset.\footnote{If our integral were to have produced elliptic functions, it is at this step that complete elliptic integrals would first have appeared. This did not happen in our problem.}
It is noteworthy that even at three loops we never had to integrate anything more complicated than a $2\times 2$ matrix.\footnote{This is in contrast to the $D$-dimensional version, see appendix \ref{app:twoloop}, where already at two loops we encountered several sectors with three coupled integrals. More complicated cases are known in the literature.}

This procedure successfully puts the differential equation into strict triangular form,
which is already a tremendous simplification. We stress the critical role of $D=4$ for this to be possible (see for example appendix \ref{app:twoloop} for a discussion of the $D\neq 4$ case).
The integrals do not, yet, however, have uniform transcendental weight.

\subsubsection*{Step 2.  \emph{Obtaining a block-triangular form}.}

We start from the triangular form of the differential equation and
again split the integrals into two sets, ``UT'' (for uniform transcendental degree, or weight) and ``remaining'' ones.
The UT integrals are assigned definite weights. At the beginning only the constant integral $g_1$ is UT, and its weight is defined to be 0.

We pick one of the remaining integrals whose derivatives involve only integrals in the UT set. Such integrals always exist, since the system is already strictly triangular.
We must figure out the weight of that integral. To do this we integrate the coefficients of the maximal weight UT integrals on the right-hand-side of the differential equation and ask whether they contain transcendental parts or not. If they do, the weight of the integral will be one more than the weight on the right-hand side. If they do not, we remove the highest weight on the right-hand-side by a judicious redefinition, and repeat.

Let us illustrate this with a simple example. After $g_1,g_2,g_3,g_4$ were put in UT form, we had the equation
\begin{align}
 \frac{d}{ds} \left( s m^2\sqrt{s(s-4m^2)} \GG{0,1,2,0}{2,0,1,0}{1}\right) =&  \frac{-4m^4 g_4+2m^2sg_1}{s\sqrt{s(s-4m^2)}} + \frac{8m^4g_2}{s(s-4m^2)}.
\end{align}
The presence of $g_4$ on the right-hand side, which has weight 2, would suggest that the parenthesis has weight 3. However, when we integrate the coefficient of $g_4$,
we find that it contains no logarithm:
\begin{align}
 \int ds \frac{-4m^4}{s\sqrt{s(s-4m^2)}} = 2m^2\sqrt{(s-4m^2)/s}+ c_1,
\end{align}
where $c_1$ is an ($s$-independent) integration constant. This shows that $g_4$ can be removed by a simple redefinition of the integral.
After we do this, we are led to integrate the coefficient of $g_2$, which is now the highest weight integral on the right-hand side (it has weight 1).
This time we find that its primitive contains a logarithm:
this \emph{cannot} be removed by a simple redefinition. Including an integration constant for $g_2$, and setting
$c_1=0$ since there is no use for it anymore (the integral under consideration turned out to have the same weight as $g_4$),  we now have
\begin{equation}
\label{UTintegralexample}
\begin{aligned}
 \frac{d}{ds}& \left[s m^2\sqrt{s(s-4m^2)} \GG{0,1,2,0}{2,0,1,0}{1} - 2m^2\sqrt{(s-4m^2)/s}\,g_4+ c_2 g_2 \right]
\\ &=
 2g_2 \left[  \frac{d}{ds}\log(s-4m^2)\right] + (4+2c_2)g_1\left[\frac{d}{ds} \log \big(\sqrt{-s}+\sqrt{4m^2-s}\big)\right]\,. 
\end{aligned}
\end{equation}
Finally we set $c_2=-2$ to remove the last term, which would have the wrong weight.
The parenthesis is now a UT integral of weight 2, one more than the weight of $g_2$. We add it to the list and repeat the procedure for the next integral.

The general procedure is as follows.  Once we have identified the weight of the integral, we have two tasks: i) we must remove non-logarithmic terms in the highest-weight part of the derivatives ii) we must remove all terms involving lower-weight integrals (such as $g_1$ in the example).
Both of these tasks are carried out order by order in decreasing weight.  Non-logarithmic terms at high weights are removed by shifting the definition of the integral (like we did for $g_4$ above), while at lower weights it is also necessary to adjust integration constants from the previous weight.
We believe that this procedure will always successfully find a UT basis \emph{if it exists}.

\subsubsection*{Step 3. \emph{Optimizing the solution}.}

The UT master integrals generated by the above procedure can be quite lengthy and far from optimal.
This is especially so for the highest weight UT integrals, which can end up expressed as combinations of all the other integrals in the original basis.

The good news is that once \emph{some} UT basis has been found, it is very easy to find other, more compact, ones.

Returning to the example in eq.~(\ref{UTintegralexample}), it is natural to consider other candidate master integrals for the same topology:
\begin{align}
 \GG{0,1,2,0}{2,0,1,0}{1}\,, \quad
 \GG{0,2,1,0}{2,0,1,0}{1}\,, \quad
 \GG{0,1,2,0}{1,0,2,0}{1} \quad\mbox{and}\quad
 \GG{0,2,1,0}{1,0,2,0}{1}\,.\nn
\end{align}
By construction, each of these may be reduced to linear combinations of UT master integrals.
One can then search, among these integrals, for combinations which reduce directly to the parenthesis in eq.~(\ref{UTintegralexample}).
That way we found the rather compact combination recorded as $g_5$ in eq.~(\ref{newform1loop}) below.

Contrary to the previous steps, there is of course no systematic procedure for finding ``the nicest possible'' representative.
In most cases, we could find satisfactory combinations by trying a rather obvious set of candidates.

\subsection{Result at two and three loops: UT basis and differential equation}
\begin{figure}[t] 
\captionsetup[subfigure]{labelformat=empty}
\begin{center}
{\includegraphics[width=0.84\textwidth]{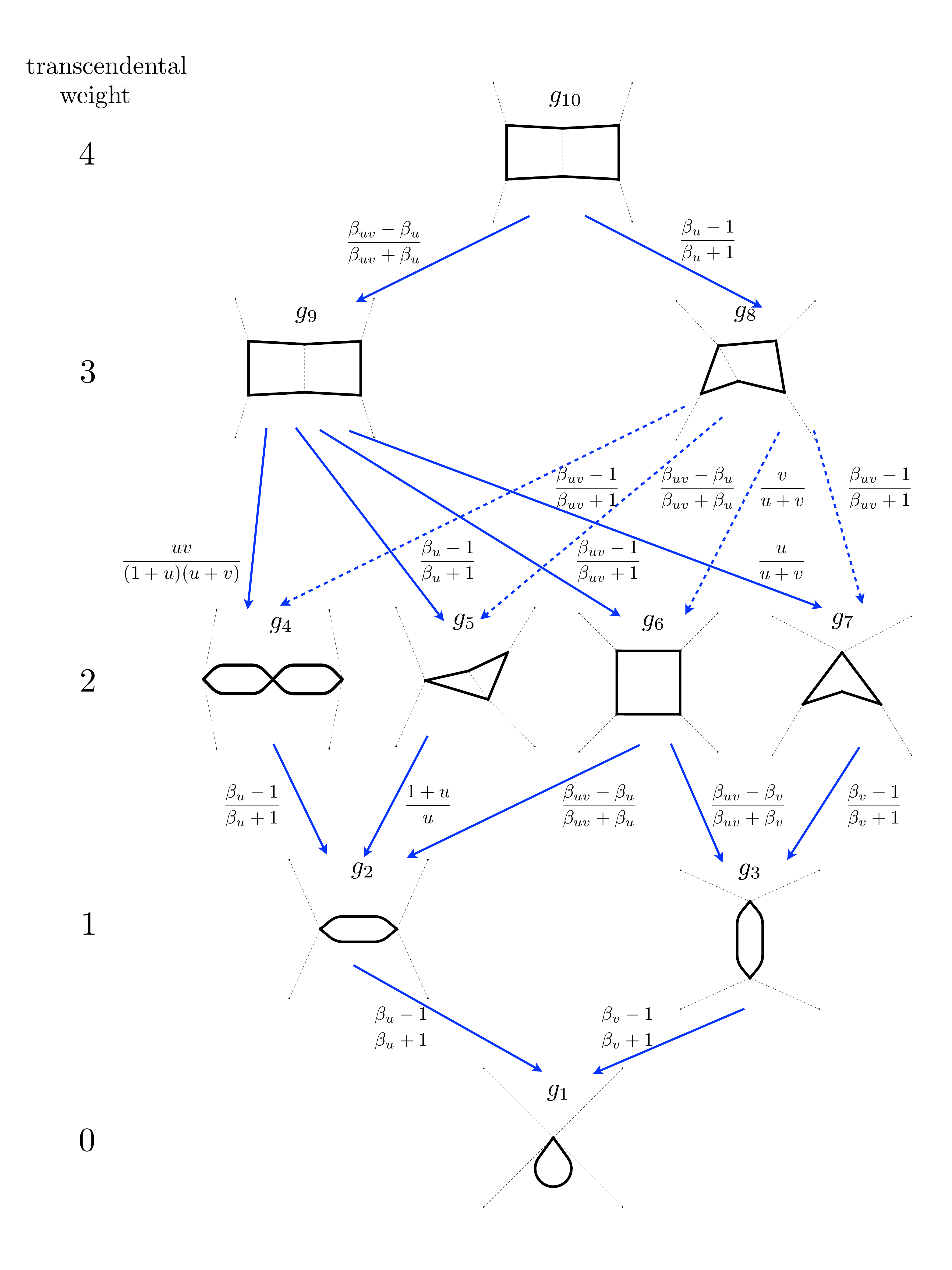}}
\vspace{0cm}
\caption{Hierarchy of master integrals up to two loops. 
The integrals are classified according to their (transcendental) weight, shown in the leftmost column. 
Each arrow corresponds to one non-zero element of the derivative matrix $A$, cf. eq. (\ref{Amat2loop}). 
The fact that arrows only link integrals in adjacent rows
is the statement that the matrix is {\it block triangular}. 
The result for an integral can immediately be written down by
summing over all paths leading up from the tadpole integral $g_{1}=1$. 
Each path gives a contribution to an iterated integral, with the 
integration kernels being specified by the `letters' written next
to the corresponding arrows.
Solid and dashed lines denote massive and massless propagators, respectively.
Note that the pictures are intended to give an idea of how the integrals look like, but omit details such as
e.g. numerator factors.}
\label{fig:blocktriangular2loop}
\end{center}
\end{figure}

The complete basis of four-dimensional DCI integrals for the family (\ref{def-2loopfamily})
is given as follows. First we have the four one-loop integrals
\begin{equation}
\label{newform1loop}
\begin{aligned}
g_1 =& 6 m^4 \GA{4,0,0,0} \\
g_2 =& -\frac12 \sqrt{s(s-4m^2)} \left[ (2m^2-s)\GA{2,0,2,0}+4m^2\GA{1,0,3,0}\right]\\
g_3 =& -\frac12 \sqrt{t(t-4m^2)} \left[ (2m^2-t)\GA{0,2,0,2}+4m^2\GA{0,1,0,3}\right]\\
g_6 =& \frac12\sqrt{st(st-4m^2(s+t))}\GA{1,1,1,1}\,.
\end{aligned}
\end{equation}
These integrals are equal to the corresponding ones defined in eq.~(\ref{UTbasis1loop}), but the present form makes the dual conformal symmetry manifest.
In addition we have the two-loop integrals
\begin{equation}
\begin{aligned}
g_4 =& \frac12 (g_2)^2\\
g_5 =& \frac12 sm^2\sqrt{s(s-4m^2)} \left[ \GG{0,1,2,0}{2,0,1,0}{1}+\GG{0,2,1,0}{2,0,1,0}{1}\right]\\ 
g_7=& -tm^4\left[ \GG{0,1,2,0}{0,0,2,1}{1}+2\GG{0,2,1,0}{0,0,1,2}{1}+3\GG{0,1,2,0}{0,0,1,2}{1}\right]\\
g_8=&-\frac12 m^2 \sqrt{st(st-4m^2(s+t))} \left[\GG{0,1,2,0}{1,0,1,1}{1}+\GG{0,2,1,0}{1,0,1,1}{1}\right]\\
g_9=&-\frac14 s m^2 \left[s \GG{1,1,1,0}{1,-1,1,2}{1}+2t \GG{1,1,1,0}{0,0,1,2}{1}\right]\\
g_{10}=&\frac18 \sqrt{s(s-4m^2)}\sqrt{st(st-4m^2(s+t))} \GG{1,1,1,0}{1,0,1,1}{1}\,.
\end{aligned}
\end{equation}
All integrals are convergent. It is easy to see that they obey the boundary condition
\begin{align}
 \lim_{m\to \infty} g_i = \delta_{i,1}. \label{bdy2loop}
\end{align}
They obey the differential equation in the canonical form (\ref{diffeqsimple}), 
with the $A$-matrix
\begin{align}
A=& \left(\begin{array}{cccccccccc}
0&0&0&0&0&0&0&0&0&0\\ \cline{1-1}
\multicolumn{1}{|c|}{\sb{\frac{\bu-1}{\bu+1}}}&0&0&0&0&0&0&0&0&0\\
\multicolumn{1}{|c|}{\sb{\frac{\bv-1}{\bv+1}}}&0&0&0&0&0&0&0&0&0\\ \cline{1-3}
0&\lll{\sb{\frac{\bu-1}{\bu+1}}}&\rr{0}&0&0&0&0&0&0&0\\
0&\lll{\sb{\frac{\oneplusu}{\u}}}&\rr{0}&0&0&0&0&0&0&0\\
0&\lll{\sb{\frac{\buv-\bu}{\buv+\bu}}}&\rr{\sb{\frac{\buv-\bv}{\buv+\bv}}}&0&0&0&0&0&0&0\\ 
0&\lll{0}&\rr{\sb{\frac{\bv-1}{\bv+1}}}&0&0&0&0&0&0&0\\ \cline{2-7}
0&0&0&\lll{\sb{\frac{\buv-1}{\buv+1}}}&\sb{\frac{\buv-\bu}{\buv+\bu}}&\sb{\frac{\v}{\u+\v}}&\rr{\sb{\frac{\buv-1}{\buv+1}}}&0&0&0\\
0&0&0&\lll{\sb{\frac{\u \v}{(\oneplusu)(\u+\v)}}}&\sb{\frac{\bu-1}{\bu+1}}&\sb{\frac{\buv-1}{\buv+1}}&\rr{\sb{\frac{\u}{\u+\v}}}&0&0&0\\ \cline{4-9}
0&0&0&0&0&0&0&\lll{\sb{\frac{\bu-1}{\bu+1}}}&\rr{\sb{\frac{\buv-\bu}{\buv+\bu}}}&0\\ \cline{8-9}
\end{array}\right)\,.
\label{Amat2loop}
\end{align}
We have abbreviated each nonzero entry of the matrix by omitting the
logarithm; for example $A_{2,1}\equiv \log {\frac{\bu-1}{\bu+1}}$.

The differential equation (\ref{diffeqsimple}), with the $A$-matrix (\ref{Amat2loop}) and boundary condition (\ref{bdy2loop}),
completely determines the integral.
The block-triangular nature of the matrix, highlighted by its presentation, reflects the uniform transcendental weights of the integrals.
The weights can be read off from the matrix structure: $g_1$; $g_{2,3}$; $g_{4,5,6,7}$; $g_{8,9}$ and $g_{10}$ have uniform weights 0,1,2,3 and 4.

As in the preceding section, the differential equation admits a natural graphical representation, see Fig.~\ref{fig:blocktriangular2loop}. The solution to the differential equation can be read off directly from this figure, as was explained in section \ref{sec:deoneloop}.

\subsection{Numerical evaluation}
\label{sec:N}

We would like to stress that, for practical purposes, Fig.~\ref{fig:blocktriangular2loop} contains \emph{all} relevant information about the double box integral.

Consider, for example, its numerical evaluation.
As discussed in the end of section \ref{sec:deoneloop},
each path between $g_{10}$ and $g_1$ translates to an iterated integral:
\begin{align}
g_{10}(u,v)=& \int_0^1 
\partial_{t_4}\log\frac{\bu(t_4)-1}{\bu(t_4)+1}\,\,
\partial_{t_3}\log\frac{\buv(t_3)-1}{\buv(t_3)+1}\,\,
\partial_{t_2}\log\frac{\bv(t_2)-1}{\bv(t_2)+1}\,\,
\partial_{t_1}\log\frac{\bv(t_1)-1}{\bv(t_1)+1}
\nn\\
&+ \mbox{9 other terms.} \label{fourfold}
\end{align}
The integration variables are ordered: $1{>}t_4{>}t_3{>}t_2{>}t_1{>}0$. The shown term corresponds to the rightmost path in Fig.~\ref{fig:blocktriangular2loop}.
In the end $g_{10}$ is independent of the choice of path going to $(u,v)$ starting from infinity,
but one simple choice is $(u(t),v(t))=(u/t,v/t)$, $t\in [0,1]$.
There are in total 10 paths in the figure, and the sum of the 10 associated terms represents $g_{10}$ exactly.
Alternatively, the 10 terms can be generated by solving formally the differential equation $dg_i=(dA)_{ij}g_j$, with the $A$-matrix given
in eq.~(\ref{Amat2loop}) and boundary condition $\lim_{u,v\to \infty}g_i=\delta_{i1}$.

Although it can be evaluated numerically, the four-fold integral (\ref{fourfold}) is obviously not optimal. Many integrals can be done analytically.
For example, the $t_1$ and $t_4$ integrations obviously only produce logarithms, so at most two integrations really need to be done numerically.
In fact, since analytic expressions for the weight 2 functions $g_{4,5,6,7}$ are easy to obtain,
the $t_2$ integration can be also done analytically.
In this way Fig.~\ref{fig:blocktriangular2loop} immediately produces a one-fold integral representation for the double box:
\begin{align}
 g_{10}(u,v)=&\int_0^1 dt\,\log \frac{(\buv-\bu)(\buv(t)+\bu(t))}{(\buv+\bu)(\buv(t)-\bu(t))}
  \left[g_4(u)\frac{d}{dt} \log \left(\frac{uv}{(1+u)(u+v)}\right)
 \right.\nn\\ & \hspace{0.5cm}\left.
 +g_5(u)\frac{d}{dt} \log \left(\frac{\bu-1}{\bu+1}\right)
 +g_6(u,v)\frac{d}{dt} \log \left(\frac{\buv-1}{\buv+1}\right)
 +g_7(v)\frac{d}{dt} \log \left(\frac{u}{u+v}\right)\right]
\nn\\
 & \!\!\!\!\!+\int_0^1 dt\,\log \frac{(\bu-1)(\bu(t)+1)}{(\bu+1)(\bu(t)-1)}
 \left[g_4(u)\frac{d}{dt} \log \left(\frac{\buv-1}{\buv+1}\right)
 \right.\nn\\ & \hspace{0.5cm}\left.
  +g_5(u)\frac{d}{dt} \log \left(\frac{\buv-\bu}{\buv+\bu}\right)
 +g_6(u,v)\frac{d}{dt} \log \left(\frac{v}{u+v}\right)
 +g_7(v)\frac{d}{dt} \log \left(\frac{\buv-1}{\buv+1}\right)\right]\,,
\label{twoloopN}
\end{align}
where everything in the square brackets is evaluated at $(u(t),v(t))$.
In addition to $g_6=\frac{\buv}{2} I_1$ which was given previously in eq.~(\ref{I1exact}), the functions are ($x\equiv\frac{\bu-1}{\bu+1}$):
\begin{equation}
 g_4(u)=\frac12(\log x)^2\,,\qquad
 \quad g_5(u)= 2\Li_2(-x)+2\log x\log(1+x)-\frac12(\log x)^2+\zeta_2\,,
\end{equation}
and $g_7=g_4(v)$.
We found the representation (\ref{twoloopN}) to be very well suited for numerical evaluation, with the number of digits of accuracy increasing linearly
with computing time (when implemented in {\tt Mathematica}).

It is also possible to convert Fig.~\ref{fig:blocktriangular2loop} automatically into expressions involving Goncharov polylogarithms.
This exploits a judicious parametrization of the kinematic variables, as discussed in appendix \ref{app:polylogs}.
These can then be evaluated numerically, using, for example \cite{Bauer:2000cp,Vollinga:2004sn}.
In practice, we found the representation (\ref{twoloopN}) to be both more direct to implement, easier to analytically continue and generally faster.

Before closing this section, we wish to emphasize that Fig.~\ref{fig:blocktriangular2loop} is also a useful starting point for expanding the integrals in various
kinematical limits.
Often, the alphabet simplifies in these cases, and a representation in terms of simpler standard functions can be given.
For example, let us consider the Regge limit, $s \rightarrow \infty$, $t$ fixed. Choosing the parametrization
$-t/m^2 = x/(1-x)^2$ for convenience, after a little algebra one sees that the alphabet required for 
all functions including the three-loop ones simplifies to 
\begin{align}
\{ x\,, 1+x\,, 1-x \}\,. \label{alphabet_Regge}
\end{align}
This implies that the Regge expansion of all integrals, to any order in $1/s$, can be given in terms of powers of $s$ and $\log s$, multiplied by harmonic polylogarithms \cite{Remiddi:1999ew} of argument $x$.
Such expansions, as well as their physical interpretation, will be discussed in greater detail in ref. \cite{chpaper2}.

\subsection{Result at three loops}

A remarkable feature of the preceding computation is the surprisingly small
amount of CPU power and memory usage that it required, as well as the fact that all manipulations are essentially
algorithmic and with little need for trial-and-error.  Furthermore the final form of the differential equation is
pleasingly compact.  These findings encouraged us to go to three loops.
Indeed, the algorithms described above carried seamlessly through
the three-loop integrals of the considered family, if only that more
CPU power was required, as detailed in appendix \ref{app:embedding}.

For the two families in Fig.~\ref{def-3loopfamily} combined (and including all one- and
two- loop sub-topologies), our final result for the UT basis contains
48 master integrals, 38 of them being relevant to the computation of
the three-loop light-by-light amplitude \cite{chpaper2} (i.e. the remaining 10 integrals decouple from the problem). 
In particular, the triple ladder integral, and the `tennis court' integral with a single numerator are part of our basis,
\begin{align}\label{defint37}
g_{37} =& - \frac{1}{16} (4 m^2 - s) s \sqrt{s t (-4 s m^2 - 4 t m^2 + s t)} \, \GC{1,1,1,0}{1,0,1,0}{1,0,1,1}{1,1,0} \,,
 \\
g_{38} =&  - \frac{1}{8} \sqrt{-(4 m^2 - t) t} \sqrt{s t (-4 s m^2 - 4 t m^2 + s t)}  \GC{0,1,1,0}{1,1,-1,1}{0,0,1,1}{1,1,1}\,. \label{defint38}
\end{align}
The system of differential equation, which takes a hierarchical form as found previously at one- and two- loops,
is given for the 38 first integrals in appendix \ref{app:de}. The full system, including the definitions of all the integrals in the basis,
is attached as an ancillary file with the arXiv submission of this article.
As before, this results contains all the information needed for efficient numerical evaluation of the integrals,
the study of their limits, analytic continuation, etc.
Here we simply note that the alphabet at three loops is not the same as at two loops.
In addition to the entries in eq.~(\ref{Amat2loop}) and their $u\leftrightarrow v$ symmetric values, the $A$-matrix, and hence the alphabet,
contains the new entries
\be\label{letters3loop}
dA \supset \left\{ d\log \left( u^2-4 v \right) \,, d\log \left( v^2-4 u \right) \,, d \log \left( \frac{2-2 \beta_{uv} + u}{2+2 \beta_{uv} + u} \right) \,,   d \log \left( \frac{2-2 \beta_{uv} + v}{2+2 \beta_{uv} + v} \right)  \right\}
\ee
These new letters reveal the existence of a new scale in the problem, $s\sim t^2/m^2$, where the three-loop
amplitude exhibits nontrivial features.
While the physical significance of this scale remains to be clarified, this suggests that other scales (and other letters)
will continue to appear at each loop oder.

%
%
%
%
%

\section{Analyticity properties and Mandelstam representation}
\label{sec:mandelstam}

The differential equation can be most readily integrated in the Euclidean region $u,v>0$, but this is not the only region of physical interest. To carry out their analytic continuation to other regions, it is important to know where the singularities lie.

A useful way to encode the analytic properties of an amplitude is through dispersive-type representations.
The one-loop box integral (\ref{I1exact}) is known to admit the following remarkable integral representation \cite{Mandelstam:1959bc,Davydychev:1993ut}
\begin{align}
 M^{(1)}(u,v) =& -2\int_{\Delta(\xi,\eta)}  \frac{d\xi d\eta}{(\xi+u)(\eta+v)\sqrt{1-\xi-\eta}}\,. \label{mandel_oneloop}
\end{align}
The integration region is the triangle $\Delta(\xi,\eta)=\{\xi,\eta:\xi\geq0,\eta\geq0,\xi+\eta\leq1\}$.
This representation makes the analyticity properties of the integral in the complex $u$ and $v$ planes completely manifest,
because all the singularities originate from the denominator. The integral is therefore analytic for
$(u,v)\in (\mathbb{C}\backslash [-1,0])^2$. Furthermore the discontinuity with respect to $u$ enjoys analytic properties with respect to $v$. 

Representations of this type are known as Mandelstam representations, following an early proposal by Mandelstam \cite{Mandelstam:1958xc}.\footnote{
The classic Mandelstam representation for the four-pion amplitude
contains three terms, which account for the double discontinuities in
$(s,t)$, $(s,u)$ and $(u,t)$ respectively. Planarity is the reason why there is only one term here.}  To our knowledge, the existence of the Mandelstam representation has never been conclusively proved nor disproved, not even to all orders in perturbation theory.
There is a known perturbative obstruction at one-loop, namely the presence of so-called anomalous thresholds.
These are singular branch points that are unrelated to the physical thresholds $s,t=4m^2$
and which occur when the internal and external masses in a problem violate certain inequalities. 
In the equal mass case this obstruction is absent, in agreement indeed with the existence of the formula (\ref{mandel_oneloop}).
Mandelstam proposed that if such a representation exists at one-loop for a given process, it should continue to exist at all loop orders for this process. We can test this hypothesis using the obtained differential equation.

We checked the analyticity properties of the integrals as follows.
Starting from the point $u,v=\infty$, we integrate the differential equation along $u$ with $v=\infty$. We find a single branch cut along $u\in [-1,0]$, as expected (in this interval $s>4m^2$)
and we evaluate the discontinuity across it. (In fact the $u$ dependence at $v=\infty$ can be expressed analytically in terms of harmonic polylogarithms, similarly to the Regge limit as discussed around eq.~(\ref{alphabet_Regge}), so at this stage everything is analytic.)
We then integrate the discontinuity along the $v$ direction, keeping $u$ fixed. Although the differential equation has in principle many loci of singularity, we find
(numerically) for all integrals in our three-loop basis only two branch points, located at $v=0$ and $v=-1-u$, indicating that the other singularities are spurious.
This implies that \emph{all} integrals, in our two- and three-loop families,
possess the analyticity properties implied by the Mandelstam representation.

We should note that, for these statements to be true, it is important to consider integrals with rational coefficients,
e.g. we have to divide the $g_i$'s in eq.~(\ref{newform1loop}) by the square roots which appear in their normalization.
Computing the discontinuity of the discontinuity, for $v\in [-1-u,0]$, constructively produces the integrand of the Mandelstam representation.
Up to two loops we obtained closed-form analytic formulae:
\begin{equation}
\begin{aligned}
 g_6(u,v) =& 
\buv \int_{\Delta(\xi,\eta)}  \frac{d\xi d\eta}{(\xi+u)(\eta+v)\sqrt{1-\xi-\eta}}
\\
 g_8(u,v) =&
  \buv\int_{\Delta(\xi,\eta)}  \frac{d\xi d\eta}{(\xi+u)(\eta+v)\sqrt{1-\xi-\eta}} \log\left(\frac{\eta}{\xi+\eta}\right)
\\
g_9(u,v) =&
  \int_{\Delta(\xi,\eta)}  \frac{d\xi d\eta}{(\xi+u)(\eta+v)} \log\left(\frac{1-\sqrt{1-\xi-\eta}}{1+\sqrt{1-\xi-\eta}}\right)
  \\ &+ \int_0^1 \frac{d\xi}{\xi+u}\left( 4\Li_2(\xi)+2\log \xi\log(1-\xi)-\frac{\pi^2}{6}\right)
\\
g_{10}(u,v) =& 
  \bu\buv\int_{\Delta(\xi,\eta)}  \frac{d\xi d\eta\,h_{10}(-\xi,-\eta)}{(\xi+u)(\eta+v)\sqrt{1-\xi}\sqrt{1-\xi-\eta}}
 \end{aligned}\label{doubledisc1}
 \end{equation}
where
\begin{align}
 h_{10}(u,v) =&
\Li_2\left(\frac{\buv+\bu}{\buv+1}\right)+\Li_2\left(\frac{\buv-\bu}{\buv-1}\right)-\Li_2\left(\frac{\buv+\bu}{\buv-1}\right)-\Li_2\left(\frac{\buv-\bu}{\buv+1}\right)
\nn\\ &+\log\left(\frac{1+\bu}{1-\bu}\right)\log \left(\frac{u+v}{v}\right)-\frac{\pi^2}{2}. \nn
\end{align}
All of these integrals are manifestly analytic in the cut complex plane $(u,v)\in (\mathbb{C}\backslash[-1,0])^2$.
The square roots $\bu,\buv$ factored out in front of the integrals are precisely those which appear
in the corresponding normalization factors in eq.~(\ref{newform1loop}), as just mentioned.

The remaining integrals in our two-loop basis depend on only one variable at a time and admit
the more familiar single-variable dispersion relations:
\begin{equation}
\begin{aligned}
g_2(u) =& -\bu \int_0^1 \frac{d\xi}{(\xi+u)\sqrt{1-\xi}}
 \\
g_4(u) =& \int_0^1 \frac{d\xi}{(\xi+u)}\log\left(\frac{1+\sqrt{1-\xi})}{1-\sqrt{1-\xi}}\right)
 \\
g_5(u) =& \bu\int_0^1 \frac{d\xi}{(\xi+u)\sqrt{1-\xi}}\log\left(\frac{\xi}{4(1-\xi)}\right)\,. 
\end{aligned} \label{single_disp}
\end{equation}
Also $g_1=1$, $g_3=g_2(v)$ and $g_7=g_4(v)$.

The analyticity properties which we have verified for the three-loop integrals imply that they also admit Mandelstam representations, but we did not study these explicitly.

These dispersion relations can be written in a more familiar form using the change of variable $u=4m^2/(-s)$, $\xi=4m^2/s'$:
\begin{align}
 \int_0^1 \frac{d\xi}{\xi+u+i0} F(\xi) = \int_{4m^2}^\infty \frac{s\,ds'}{s'(s-s'-i0)} F\left(\frac{4m^2}{s'}\right)\,.
\end{align}
Here we have included Feynman's $i0$ prescription, implicit in all preceding equations, but required to make sense of the integral when $s>4m^2$ or $t>4m^2$.
We recognize the right-hand-side as a once-subtracted dispersion relation. The subtraction constants happen to vanish for all integrals in eqs.~(\ref{single_disp}),
because the left-hand-sides vanish at $s=0$.  Similar comments apply to the Mandelstam representations (\ref{doubledisc1}), where
the one subtraction for $g_9$ is due to its nonvanishing value at $t=0$.

We mention that at an early stage of this work, we actually successfully \emph{guessed} an analytic expression for the two-loop integral $g_{10}$, by simply assuming that it admitted a Mandelstam representation!
The reason is that its integrand (i.e. the function $h_{10}$ above) has only transcendental weight 2,
so by assuming that all entries of its symbol are taken from the finite alphabet $(u,v,\bu,\bv,\bu+1,\bv+1,\buv+1,\bu+\buv,\bv+\buv)$ as suggested by the one-loop computation,
one arrives at a relatively compact ansatz. We could then fix all coefficients using reality considerations, symmetries, Regge limits, and numerical evaluation at one point.
However, we must warn the reader that such a procedure is far from systematic.
At three loops our ansatz would most likely have missed
the new entries of the symbol shown in eq.~(\ref{letters3loop}), and would thus have led us to a wrong answer, if any.
The differential equation approach in contrast is completely systematic and reliable.

%
%
%
%
%

\section{Checks}
\label{sec:checks}

We have performed a number of checks on our calculations. 
First of all, there are many internal consistency checks for the differential equations, 
for example the integrability conditions (\ref{integrabilityconditions}) for the matrices appearing 
in different partial derivatives.

Moreover, as discussed in appendix \ref{app:twoloop}, the integrals we computed contain simpler two-scale form-factor type integrals that were previously computed \cite{Anastasiou:2006hc}. We have performed tests against the results given in that reference.
Our results also include (a subset of) two-loop integrals occurring in top quark pair production cross sections \cite{Czakon:2008ii}.
They can be obtained from our formulas by taking an $s$- or $t$-channel discontinuity and going to certain forward limits.
In particular, in this way we checked that the singularity structure of the integrals shown in Fig.~4(b)-(d) of 
the above reference can be understood by taking forward limits of the alphabet we found here in the 
general kinematical case.
Moreover, we have done a detailed check for the cut double box integral shown in Fig.~4(b) of that reference, 
which corresponds to 
${\rm Disc}_{v}\left[g_{10}\right](u,-u)$,
up to a normalization factor.\footnote{We thank M.~Czakon for sending us results for 
individual master integrals computed in ref. \cite{Czakon:2008ii}.} 
Perhaps the easiest check can be done using the Mandelstam representation discussed in section \ref{sec:mandelstam}, cf. eq. (\ref{doubledisc1}), which gives
\begin{align}
{\rm Disc}_{v}\left[g_{10}\right](u,-u) =   \bu  \int_0^{1-u}d\xi   \frac{ \,h_{10}(-\xi,-u)}{(\xi+u)(\eta+v)\sqrt{1-\xi}\sqrt{1-\xi-u}} \,.
\end{align}
In this way, we numerically verified the result of  \cite{Czakon:2008ii} for that integral.
It is also possible to give an analytic answer for this (and other) integrals, starting from the Chen iterated integral representation. In eq. (\ref{wzparam}) we give a convenient parametrization of $u$ and $v$ for rewriting those integrals in
terms of multiple polylogarithms. Here it will suffice to remark that by specializing to the $v=-u$ case gives 
$u = 4 z(1-z^2)/(1+z^2)^2$, in which case the 
alphabet relevant for ${\rm Disc}_{v}\left[g_{10}\right](u,-u)$ reduces to $z, 1 \pm z, 1+z^2$,
i.e. a particularly simple case. 
Our results can also be used to derive similar formulas for other integrals in the same limit $v\mapsto -u$, 
including the three-loop ones, albeit using a larger alphabet. 

Very strong tests were also provided by expansions of the integrals in various physical limits that
will be discussed in detail in ref. \cite{chpaper2}. 
For example, we have successfully reproduced the known Regge limit of the three-loop ladder and tennis court integrals $g_{37}$ and $g_{38}$.
Because of the iterative structure of the solution, this also tests to some extent the whole system of
differential equations.

Finally we have performed numerical tests. The iterative structure of the differential equation allows for a simple numerical evaluation,
as detailed in section \ref{sec:N}.
In the three-loop case, by working out analytic expressions for all weight 3 $g_i$ functions in terms of $\Li_3$ functions,
and for the $t_5,t_6$ integrals in terms of dilogarithms, we actually succeeded in obtaining a \emph{one-fold} integral representation similar
to eq.~(\ref{twoloopN}), which also converge rather fast.
We then performed numerical comparisons against values obtained from FIESTA \cite{Smirnov:2008py,Smirnov:2013eza}.
While it is difficult to estimate the expected error of multidimensional integration routines, based
on gradually increasing the number of sampling points and observing stable results, we 
arrived at the following values and error estimates
\begin{align}
g^{\rm FIESTA}_{10}(4,4) = & \phantom{-} 0.0528808 \pm 0.0001\,, \\
g^{\rm FIESTA}_{37}(2,4) = & \phantom{-} 0.0764993 \pm 0.0001\,, \\
g^{\rm FIESTA}_{38}(2,4) = & -0.195813  \pm 0.001 \,. 
\end{align}
Indeed, they perfectly agree to that precision with the results obtained from our analytic answer,
\begin{align}
g_{10}(4,4) = &  \phantom{-}  0.0530386297835364296406023741052 \ldots \\
g_{37}(2,4) = &  \phantom{-} 0.0764922717271986970254859257468 \ldots \\
g_{38}(2,4) = & -0.195801160078160255184758547581\ldots
\end{align}We have also performed successful checks at other kinematical points, 
and for other integrals.

%
%
%
%
%

\section{Conclusions}
\label{sec:discussion}

We have shown in this paper how to implement the differential equations method \cite{Kotikov:1990kg,Remiddi:1997ny,Gehrmann:1999as,Argeri:2007up} for a closed subset of finite (and dual conformal) integrals. 
The restriction to $D=4$ turns out to be a significant simplification of the method of ref. \cite{Henn:2013pwa}, which was originally developed for arbitrary $D$ dimensions.  

We expect that our approach can also be applied for other classes of convergent four-dimensional integrals,
such as for example position space correlation functions.
For instance, in $\cN=4$ super Yang-Mills, expressions for the integrands of four-point correlation functions
are known to relatively high loop order \cite{Eden:2012tu}, and our method could be used to shed light on their 
functional dependence and to evaluate them analytically. Currently, the planar integrals up to three 
loops are known \cite{Drummond:2013nda}.
Another set of finite and dual conformal functions that this method could be applied to are ratio and remainder functions in $\cN=4$ super Yang-Mills \cite{Drummond:2008vq}.\footnote{In that context, we wish to mention that differential equations involving integrals at different loop orders have a natural explanation in terms of supersymmetry \cite{CaronHuot:2011ky,Bullimore:2011kg}. Also, we would expect the differential equations of \cite{Drummond:2010cz,Dixon:2011ng} to appear naturally as part of the system of differential equations described here.}.
Using a suitable four-dimensional regulator, such as for example differential renormalization \cite{Freedman:1991tk,Gracia-Bondia:2014zwa}, it seems reasonable to anticipate applicability of the method
to correlation functions in an arbitrary quantum field theory.

It is important to stress that this approach to calculating integrals is systematic.
We find it extremely satisfying that the algorithm developed in section \ref{sec:defourdim} for the two-loop problem
worked without any modification at three loops.  
However we do not have a proof that the algorithm will always succeed, and
it will be interesting to try it on other problems, and also maybe generalize it to the $D$-dimensional case.

Our final result for the double box in $D=4$ reveals a simple iterative structure, summarized in Fig.~\ref{fig:blocktriangular2loop}.
In fact the figure contains all needed information about the integral!
For example, it can be evaluated numerically almost directly, as explained in section \ref{sec:N}.
It also provides an efficient starting point for expanding around various kinematic limits, as will be pursued in a companion paper \cite{chpaper2}.
Finally, it can be automatically converted to a sum of standard special functions, for example Goncharov polylogarithms, as shown in appendix \ref{app:polylogs}.
It is tempting to call Fig.~\ref{fig:blocktriangular2loop} \emph{the answer} for the double-box; it is the most concise and useful presentation which we can imagine on this day.
The similar figure at three-loops is given in matrix form in appendix \ref{app:de}.

It is tempting to expect such an iterative structure of the differential equation to be a general feature of Feynman integrals in integer dimensions.
We expect this regardless of whether or not the integrals are expressible in terms of polylogarithm-type functions.
For example, a similar iterative structure is also seen for integrals involving elliptic-type functions (see, for example, section 5 of \cite{Bloch:2013tra}).

We comment that the method automatically finds the `alphabet' relevant to a problem. This is important since the alphabet may change from one loop order to the next.
The three-loop octagon Wilson loops of ref. \cite{Caron-Huot:2013vda} are a case in point;
this paper provides another example, where the letters (\ref{letters3loop}) would likely have been very difficult to guess.
The alphabet specifies the class of two-variable functions describing the problem.
Although the latter are rather complicated, we have demonstrated in this paper that they can be easily dealt with.
In particular, we have shown how to evaluate them numerically, and discussed their analytic structure.

In addition to the expressions for all dual conformal invariant master integrals up to three loops, we have performed
a complete calculation in dimensional regularization of the planar two-loop master integrals.
The results, given in appendix \ref{app:twoloop}, can be used in future studies of light-by-light scattering at NLO, keeping the exact mass dependence. For previous studies in various mass regimes, see refs. \cite{Bern:2001dg,Binoth:2002xg}.

The integrals we computed may also be helpful for tackling problems on the 2013 Les Houches QCD wishlist\footnote{{\tt https://phystev.in2p3.fr/wiki/2013:groups:sm:nnlowishlist}},
in particular when one wants to keep the exact dependence on quark masses.
Indeed, the three-loop family we computed contain several integrals relevant for $gg\to H$ at NNNLO.
Moreover, the process $gg\to g H$ at NNLO involves four-point functions similar to the ones computed here, albeit with one additional mass scale. 

%
%
%
%
%

\section*{\it Acknowledgements}
We would like to thank V.~Smirnov for useful discussions and help with checks.
J.M.H. is supported in part by the DOE grant DE-SC0009988.
S.C.H is supported in part by the NSF grant PHY-1314311. 
Both authors are supported by the Marvin L. Goldberger fund.

%
%
%
%
%

%
%
%
%
%

%
%
%
%
%

%
%
%
%
%

%
%
%
%
%

%
%
%
%
%

%
%
%
%
%

%
%
%
%
%

\appendix

\section{Two-loop master integrals in $D$ dimensions from differential equations}
\label{app:twoloop}

Here we report on the computation of all master integrals of the integral family (\ref{def-2loopfamily}), see Fig.~\ref{fig:massivefamily}(b), in $D$ dimensions.
Contrary to the main text, in this appendix we keep the full dependence on $\eps$.
These integrals may be used in future applications for light-by-light scattering in QED or QCD, and we also hope that
they provide a first useful step towards similar but more complicated integrals needed e.g. for $gg\to H g$ at NNLO.

\subsection{Choice of integral basis}
\label{sec:basis}

\begin{figure}[t] 
\captionsetup[subfigure]{labelformat=empty}
\begin{center}
\subfloat[$\gA_1$ ]{\includegraphics[width=0.17\textwidth]{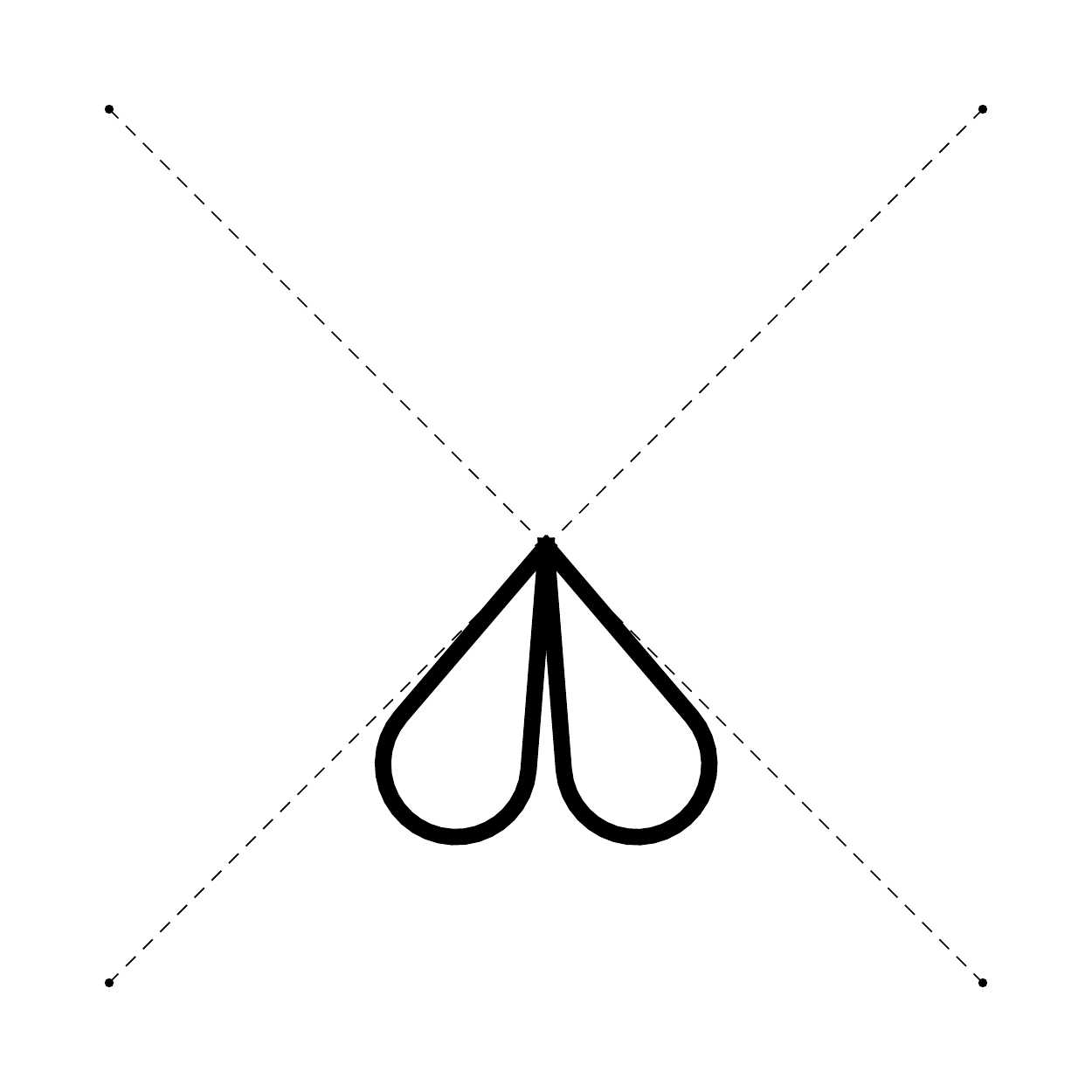}}
\subfloat[$\gA_{14} ,  \gA_{15} $]{\includegraphics[width=0.17\textwidth]{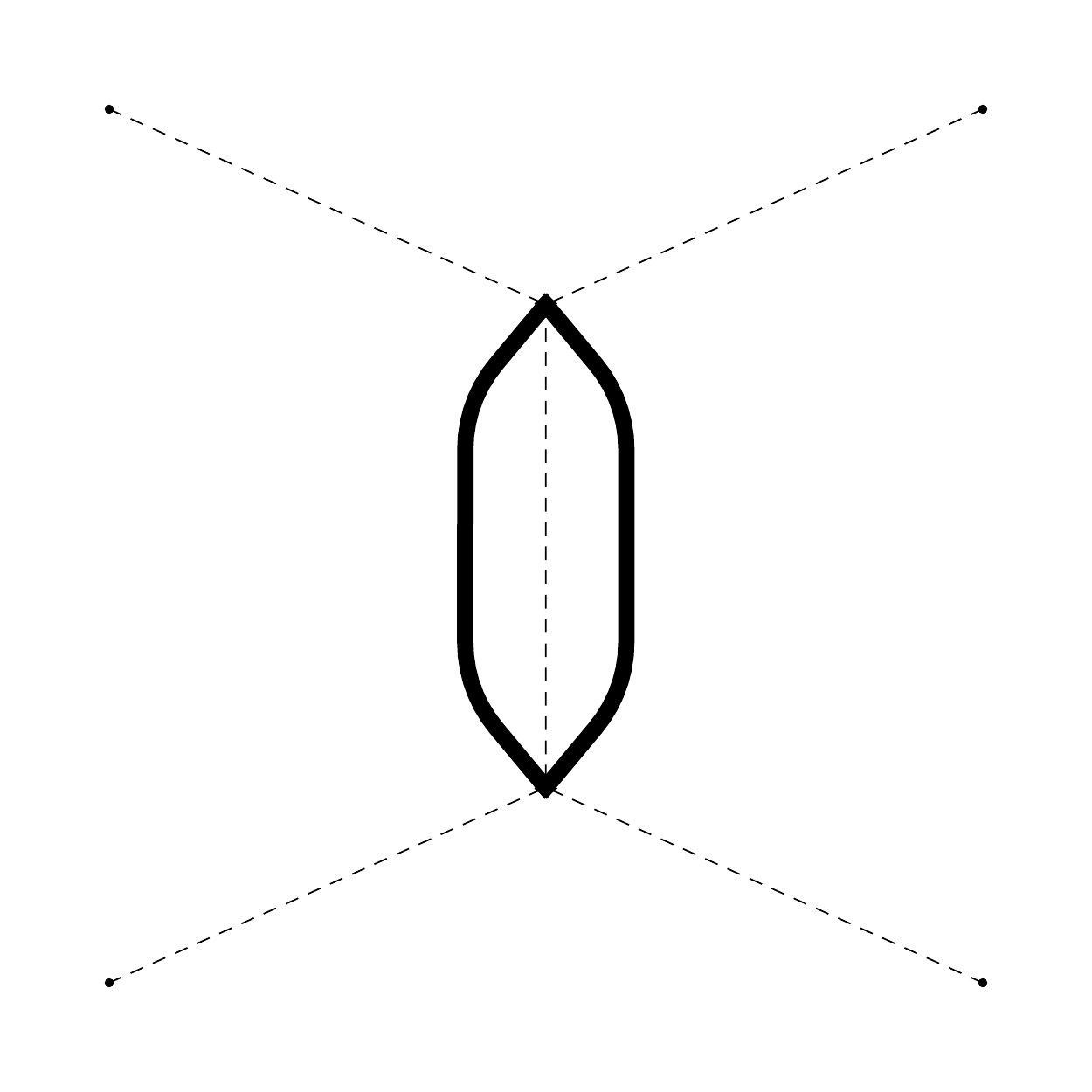}}
\subfloat[$\gA_{6}, \gA_{7} $]{\includegraphics[width=0.17\textwidth]{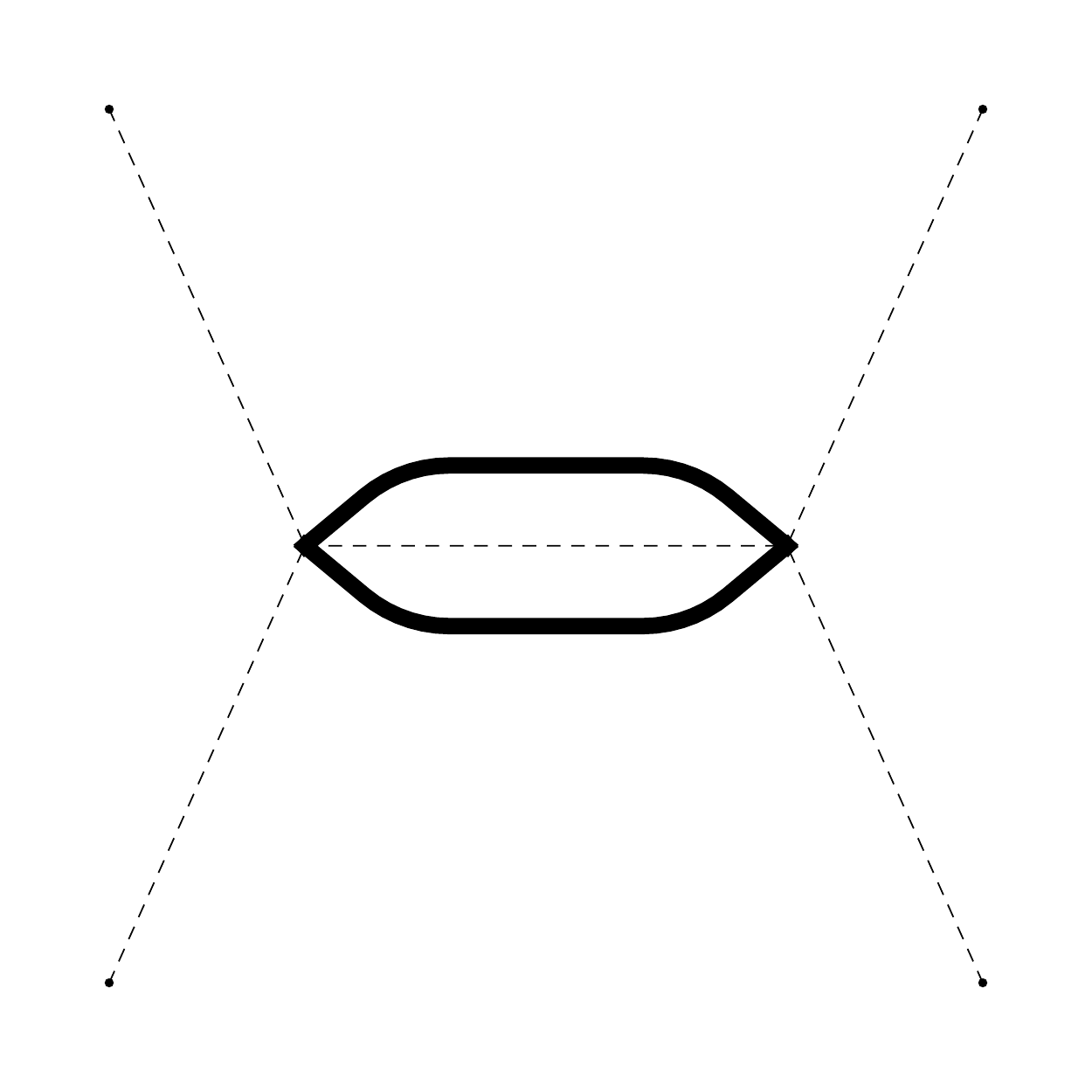}}
\subfloat[$\gA_{2}$]{\includegraphics[width=0.17\textwidth]{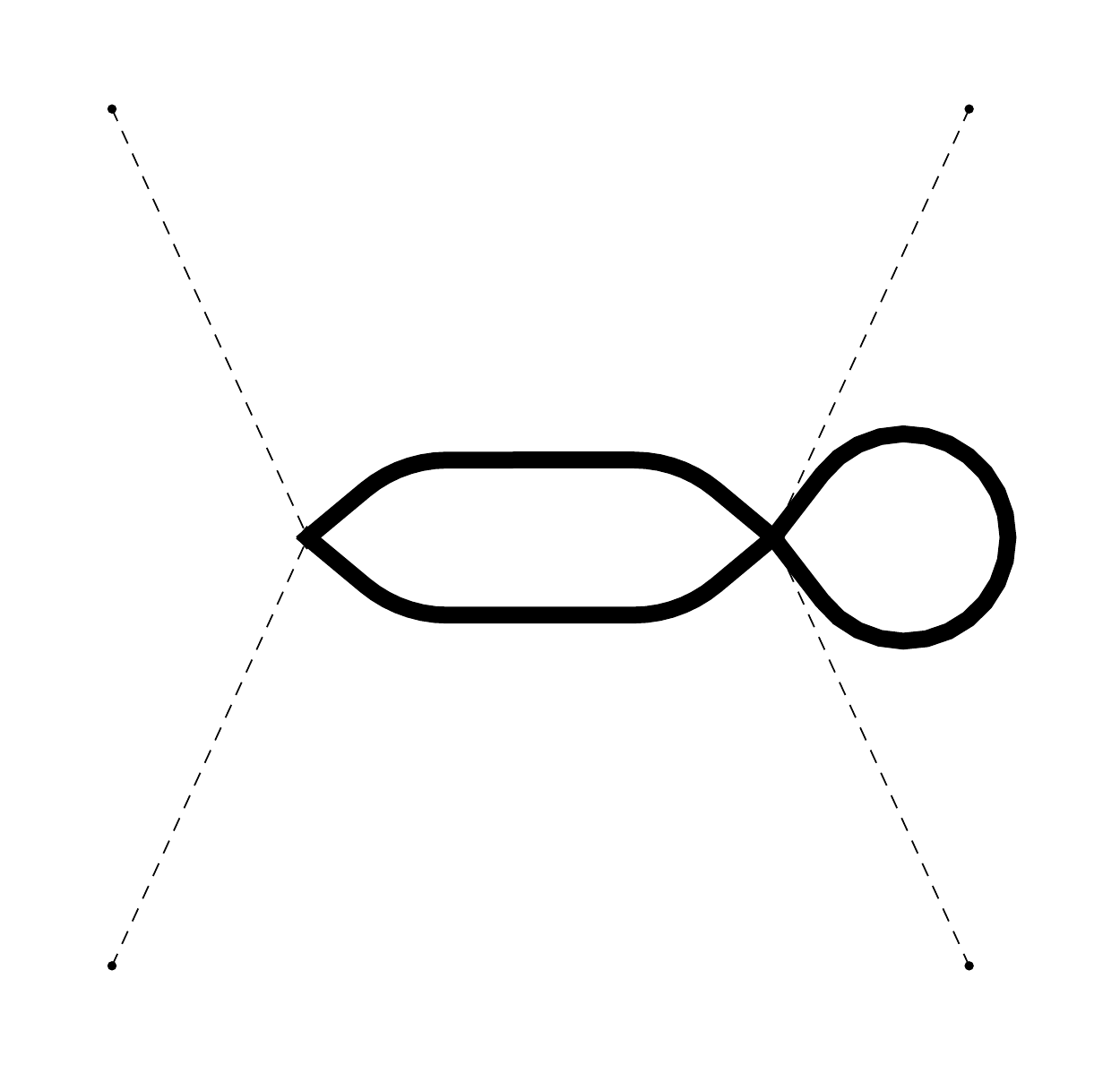}}
\newline
\subfloat[$\gA_{8}$]{\includegraphics[width=0.17\textwidth]{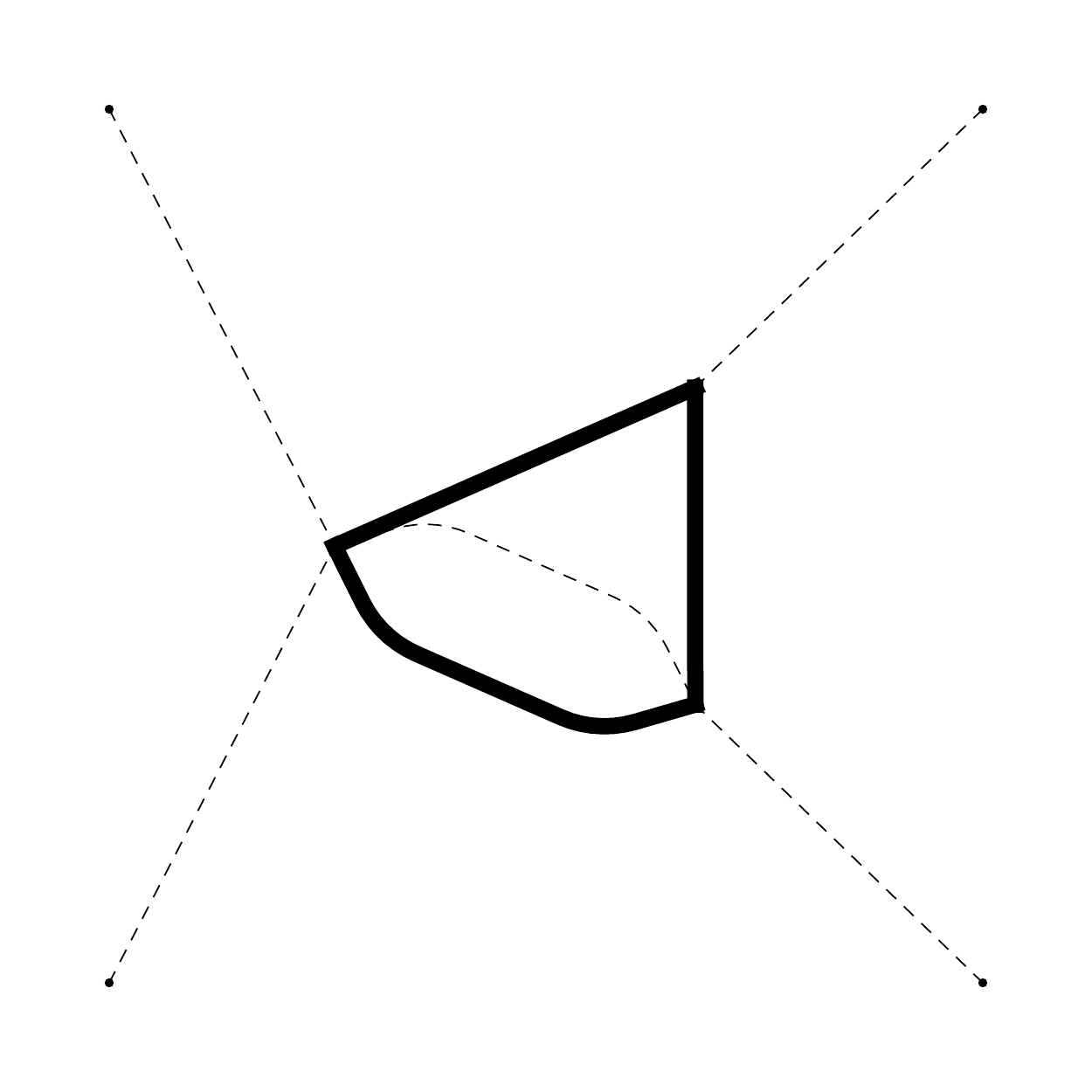}}
\subfloat[$\gA_{4}$]{\includegraphics[width=0.17\textwidth]{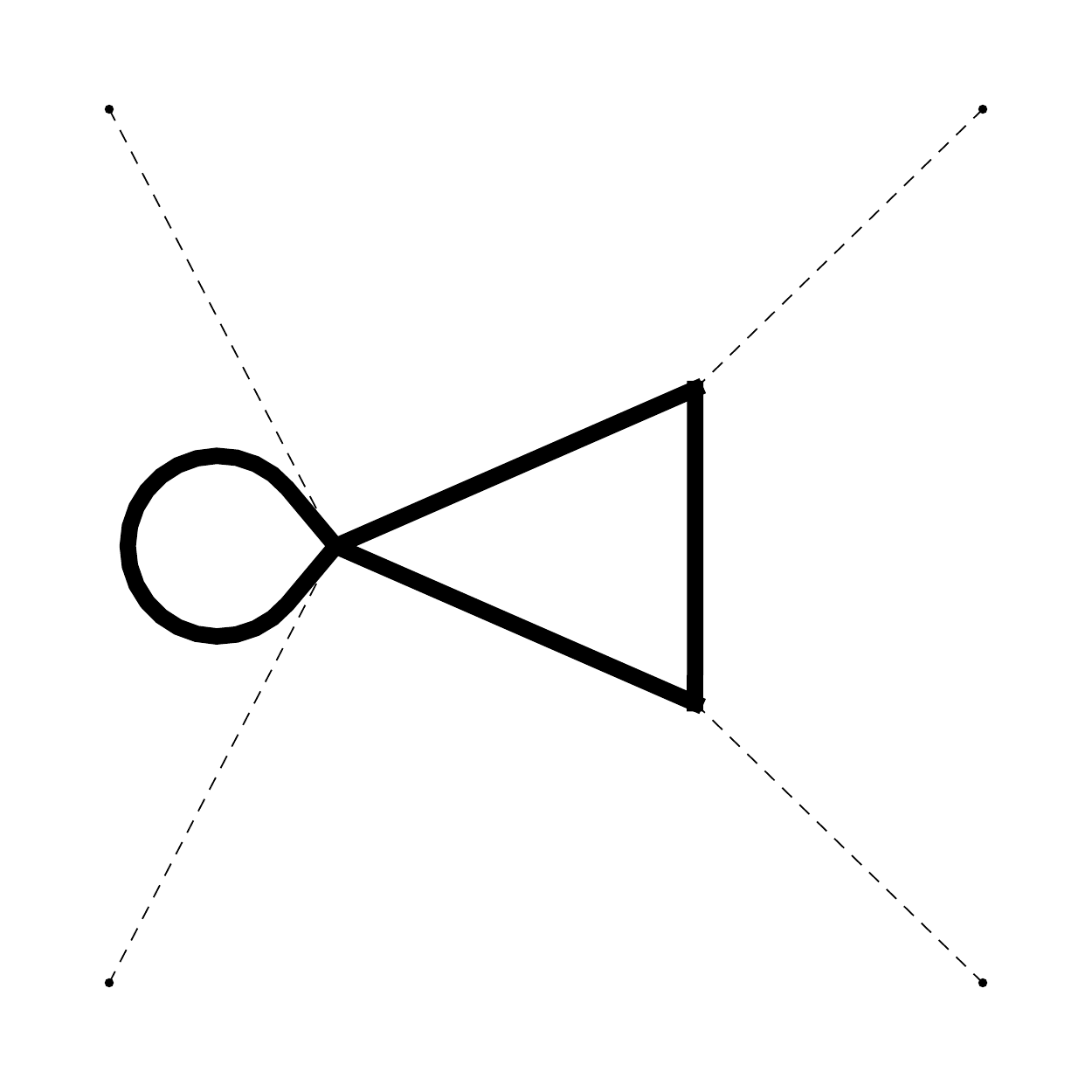}}
\subfloat[$\gA_{9},\gA_{10}$]{\includegraphics[width=0.17\textwidth]{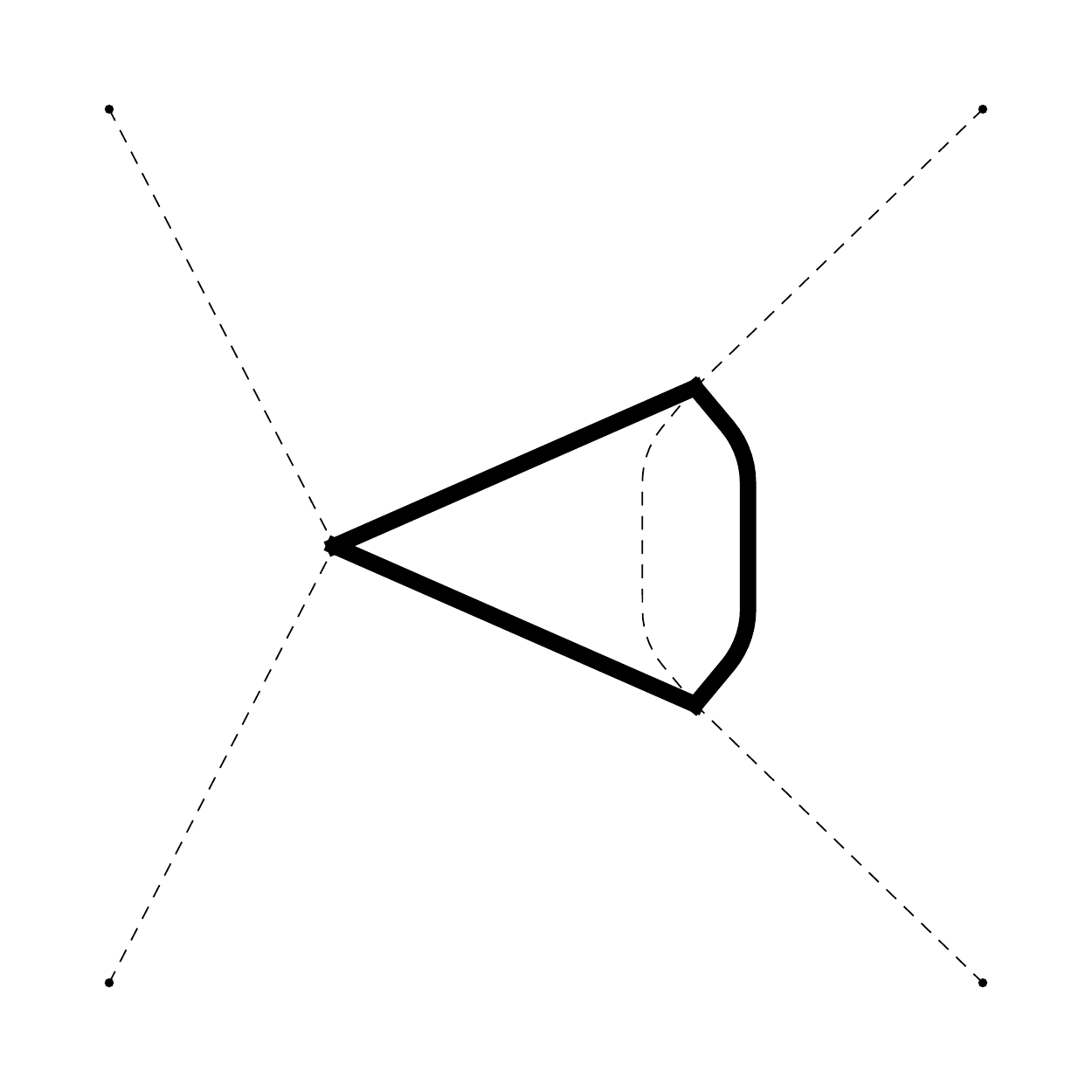}}
\subfloat[$\gA_{3}$]{\includegraphics[width=0.17\textwidth]{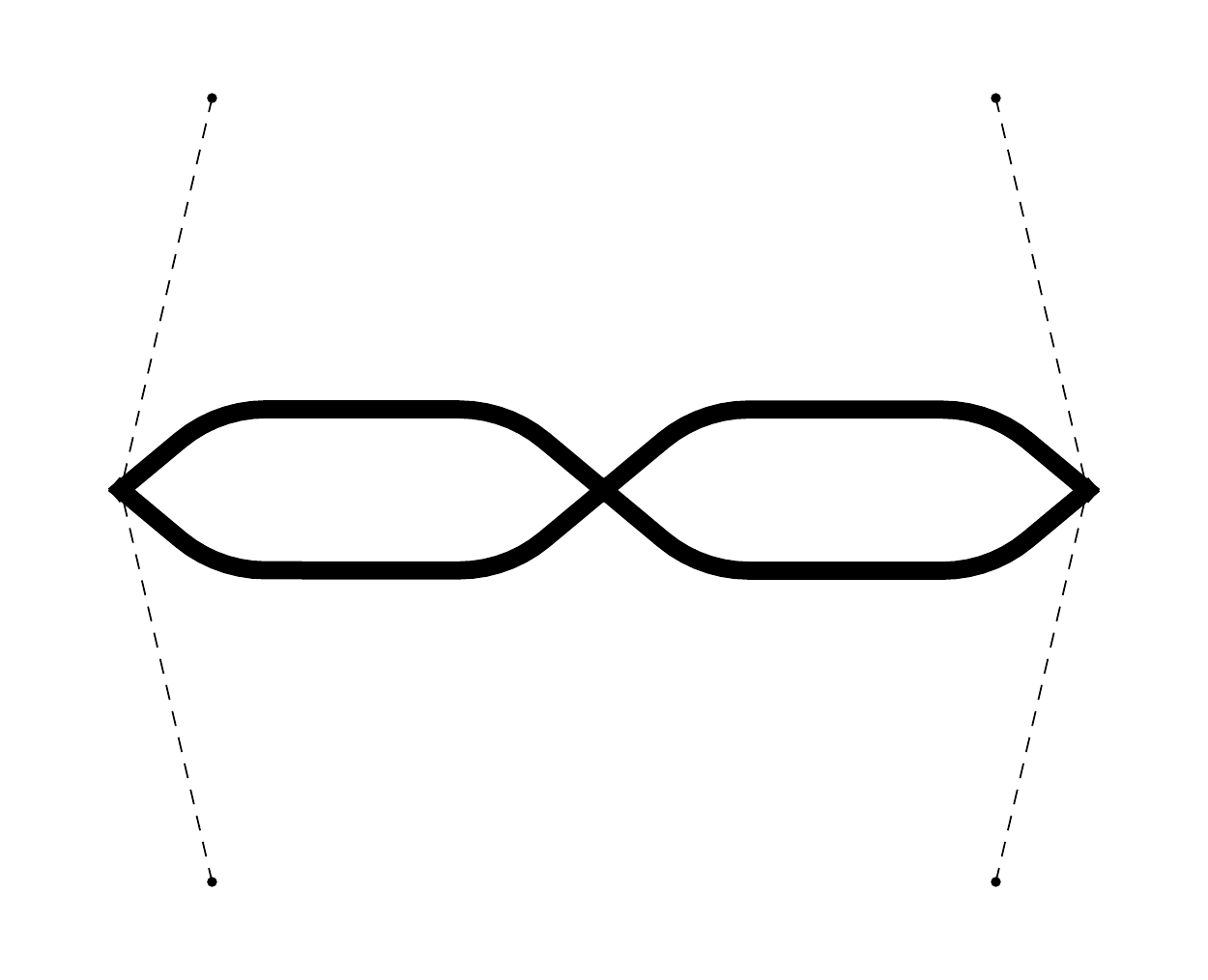}}
\subfloat[$\gA_{16}$]{\includegraphics[width=0.17\textwidth]{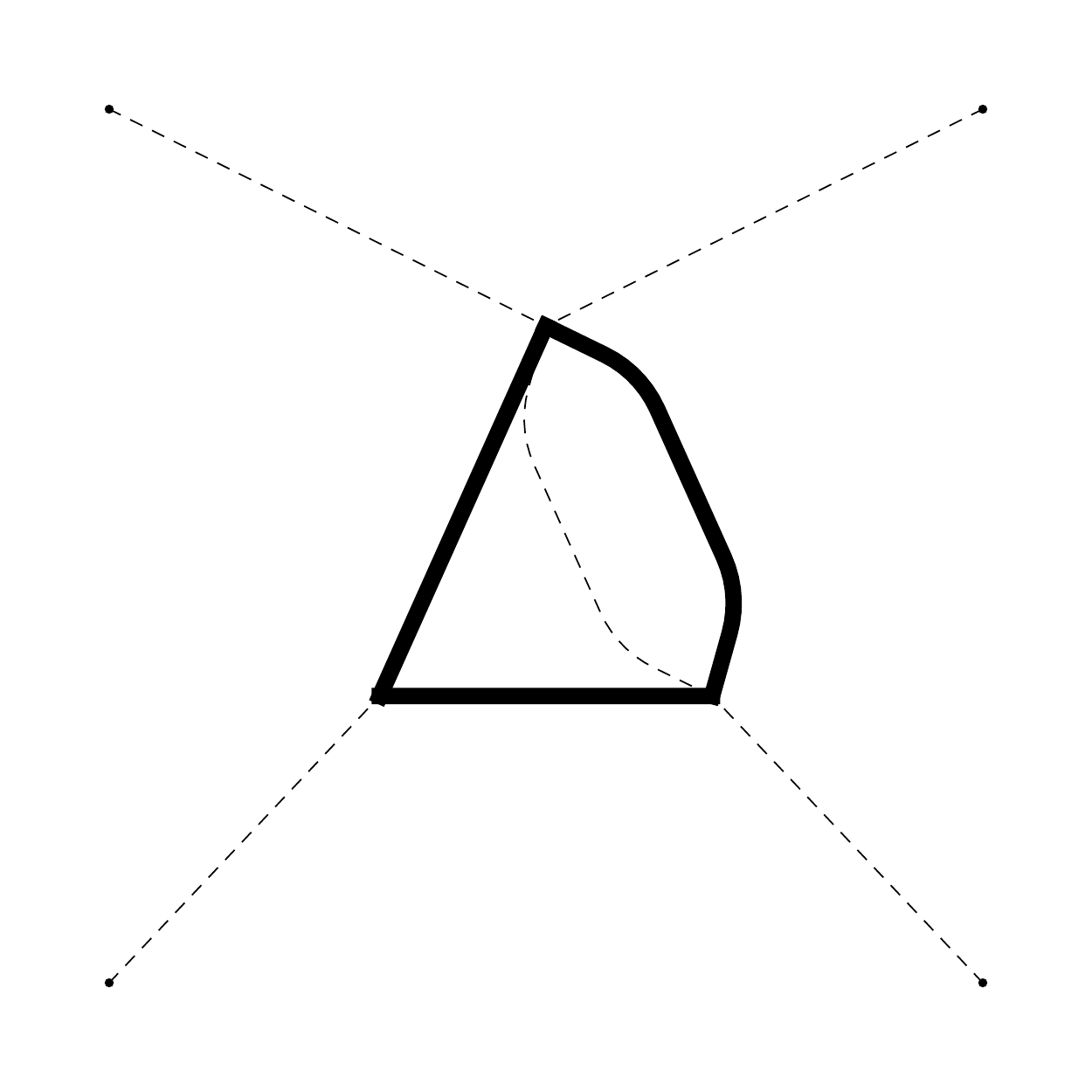}}
\newline
\subfloat[$\gA_{21},\gA_{22},\gA_{23}$]{\includegraphics[width=0.17\textwidth]{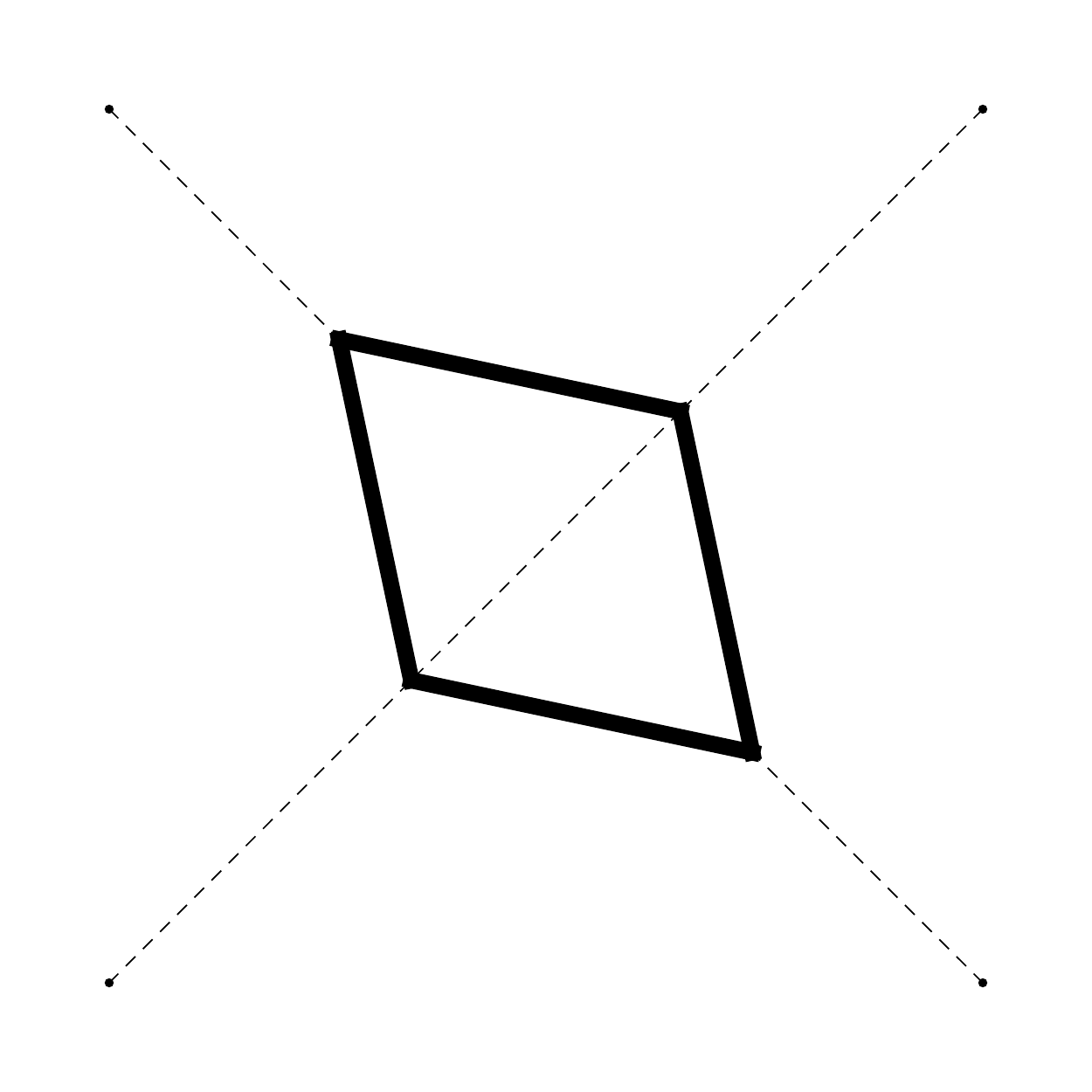}}
\subfloat[$\gA_{12}$]{\includegraphics[width=0.17\textwidth]{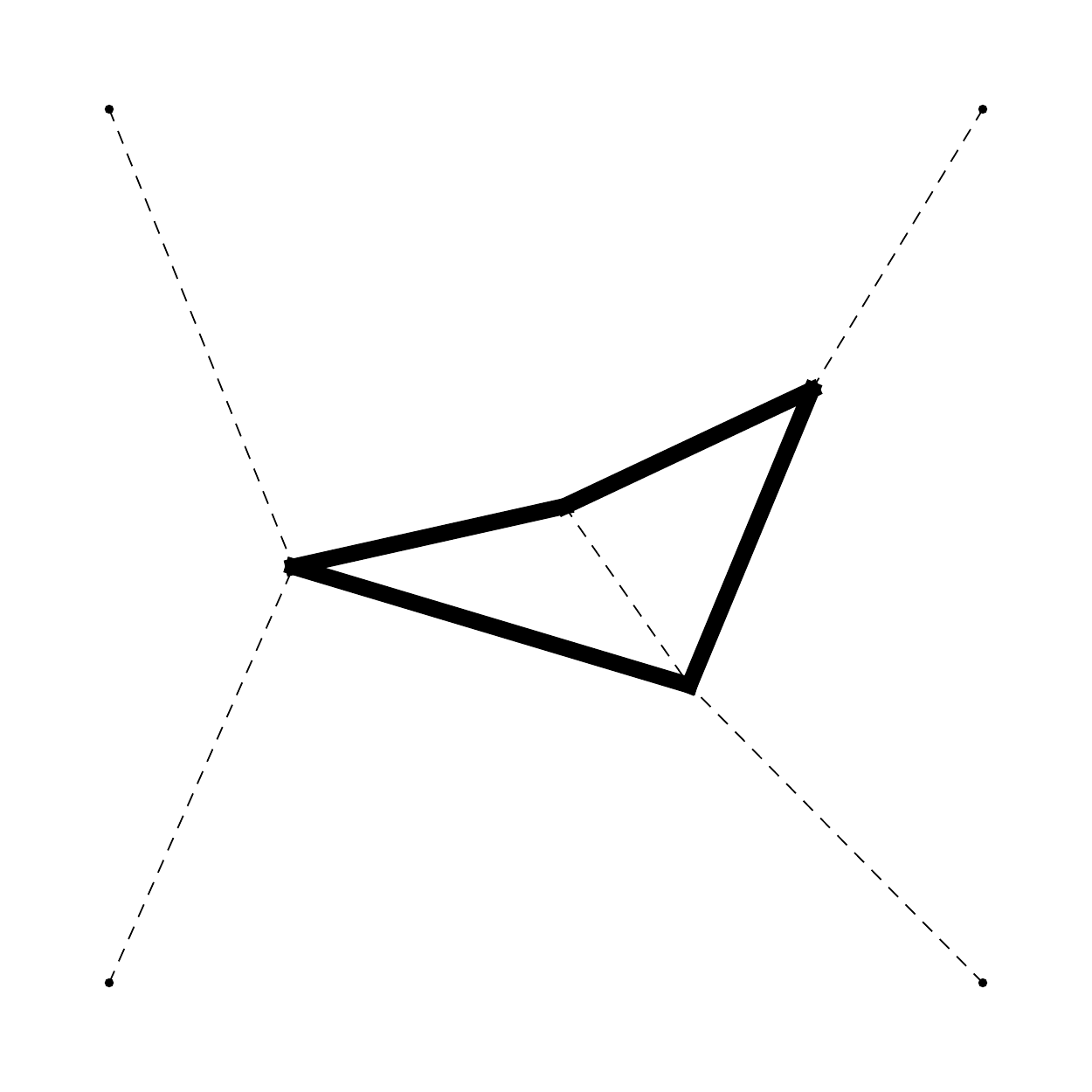}}
\subfloat[$\gA_{5}$]{\includegraphics[width=0.17\textwidth]{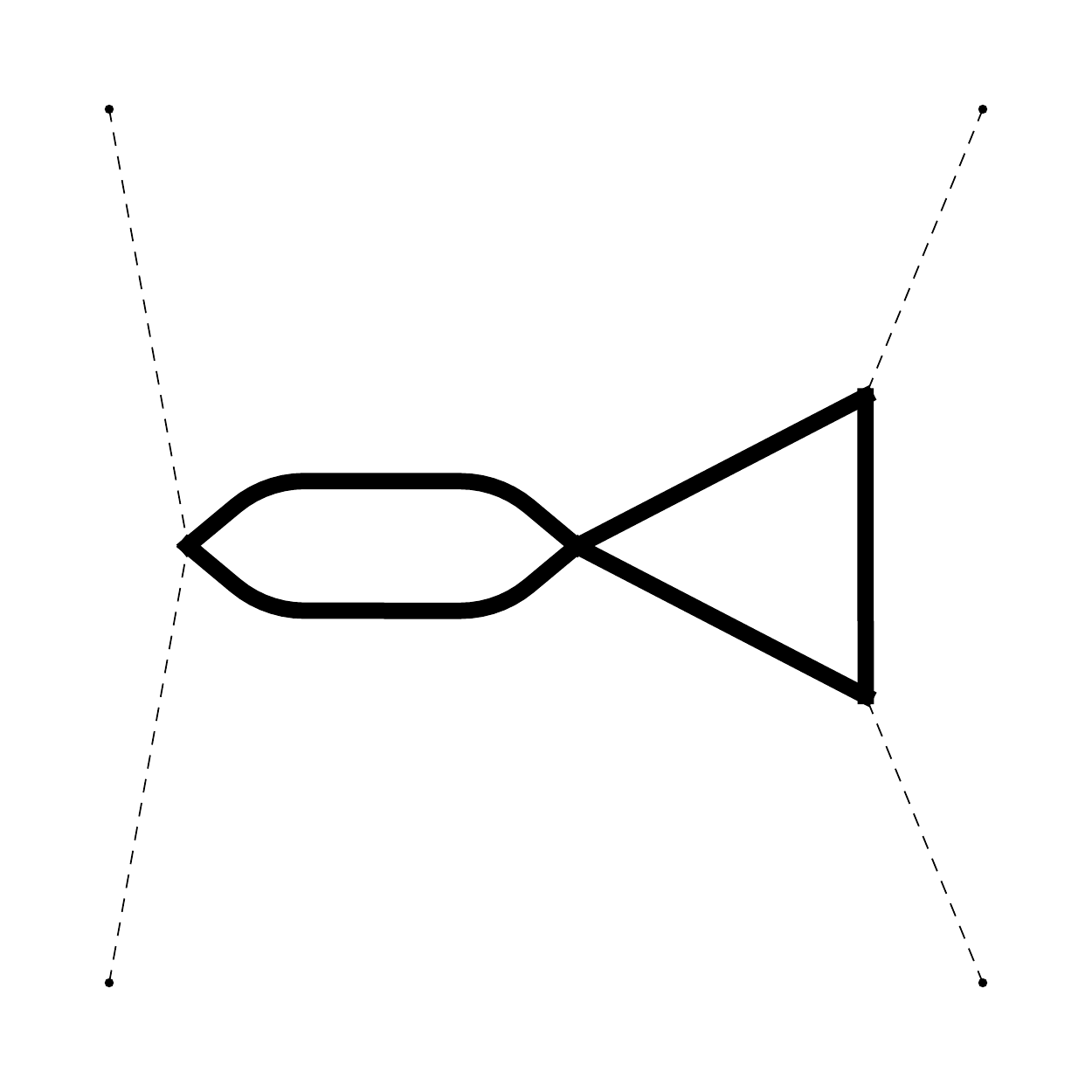}}
\subfloat[$\gA_{17}$]{\includegraphics[width=0.17\textwidth]{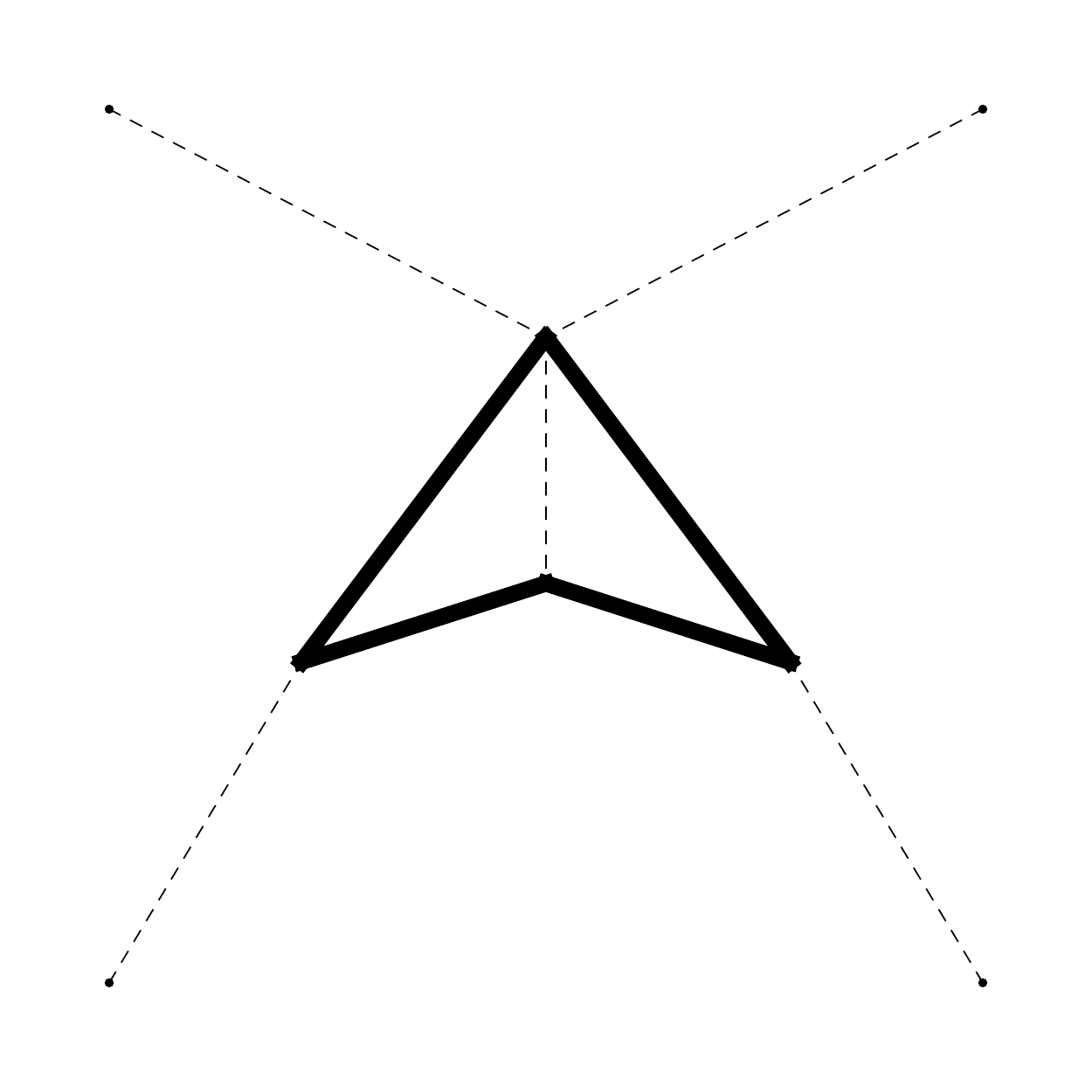}}
\subfloat[$\gA_{18},\gA_{19},\gA_{20}$]{\includegraphics[width=0.17\textwidth]{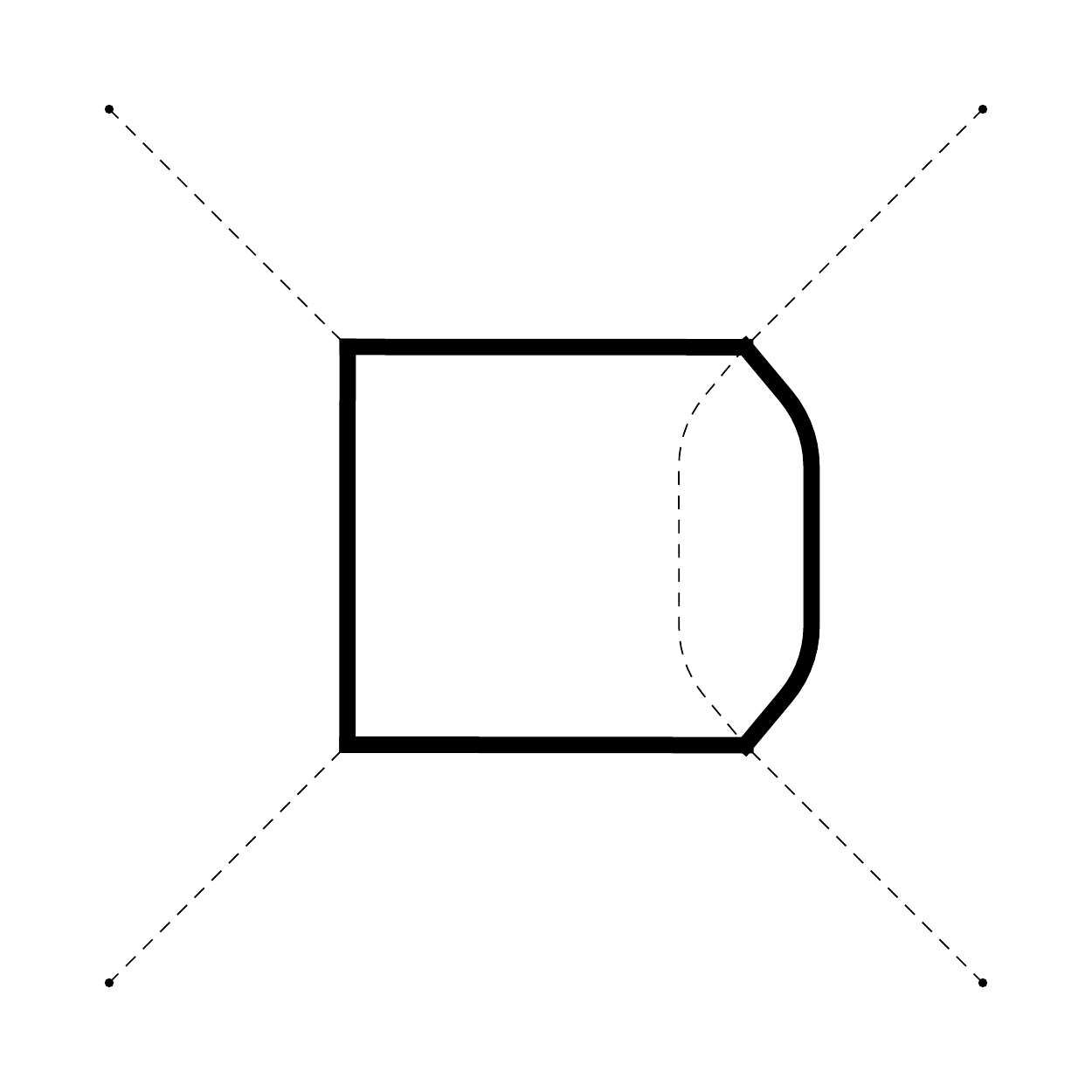}}
\newline
\subfloat[$\gA_{13}$]{\includegraphics[width=0.17\textwidth]{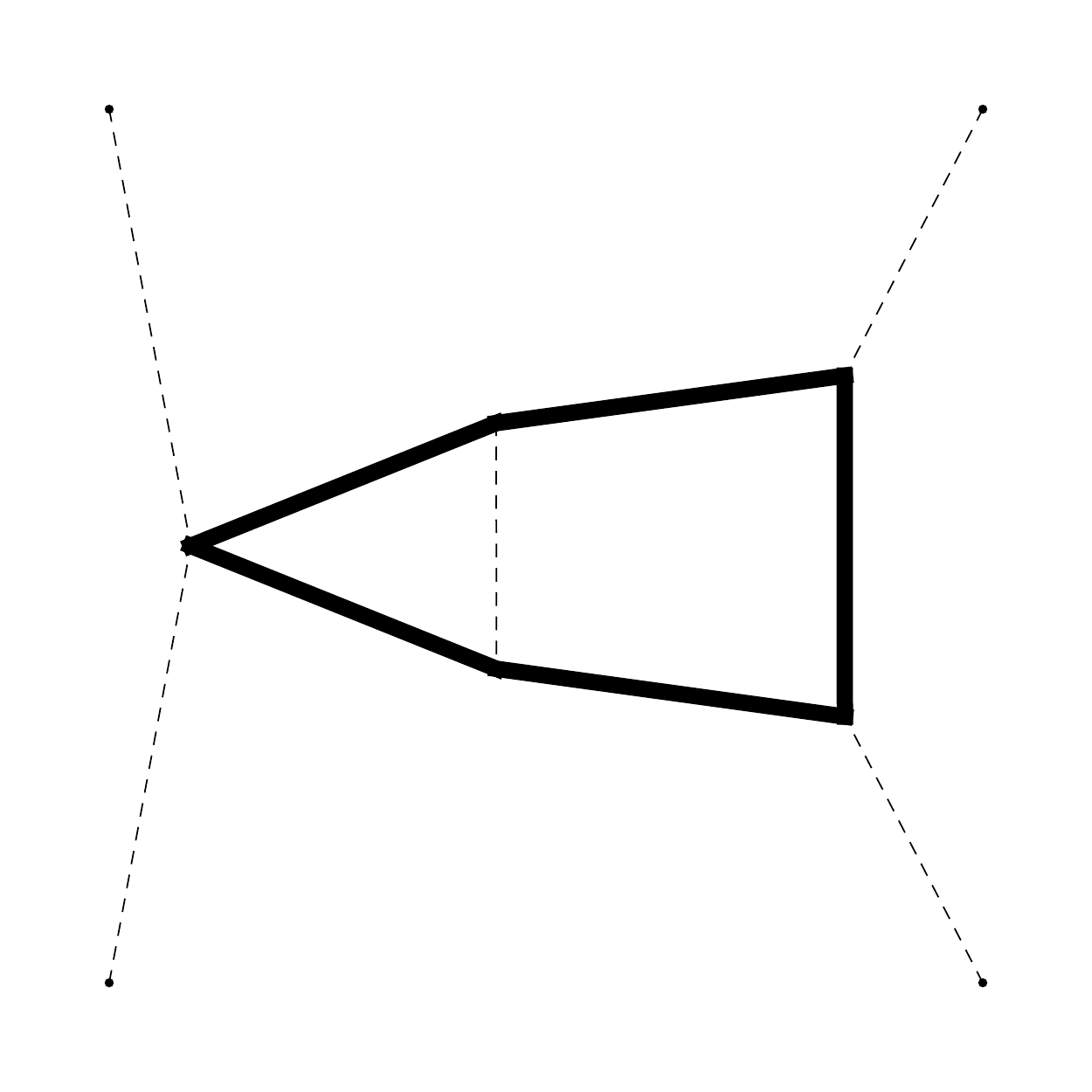}}
\subfloat[$\gA_{25}, \gA_{26}$]{\includegraphics[width=0.17\textwidth]{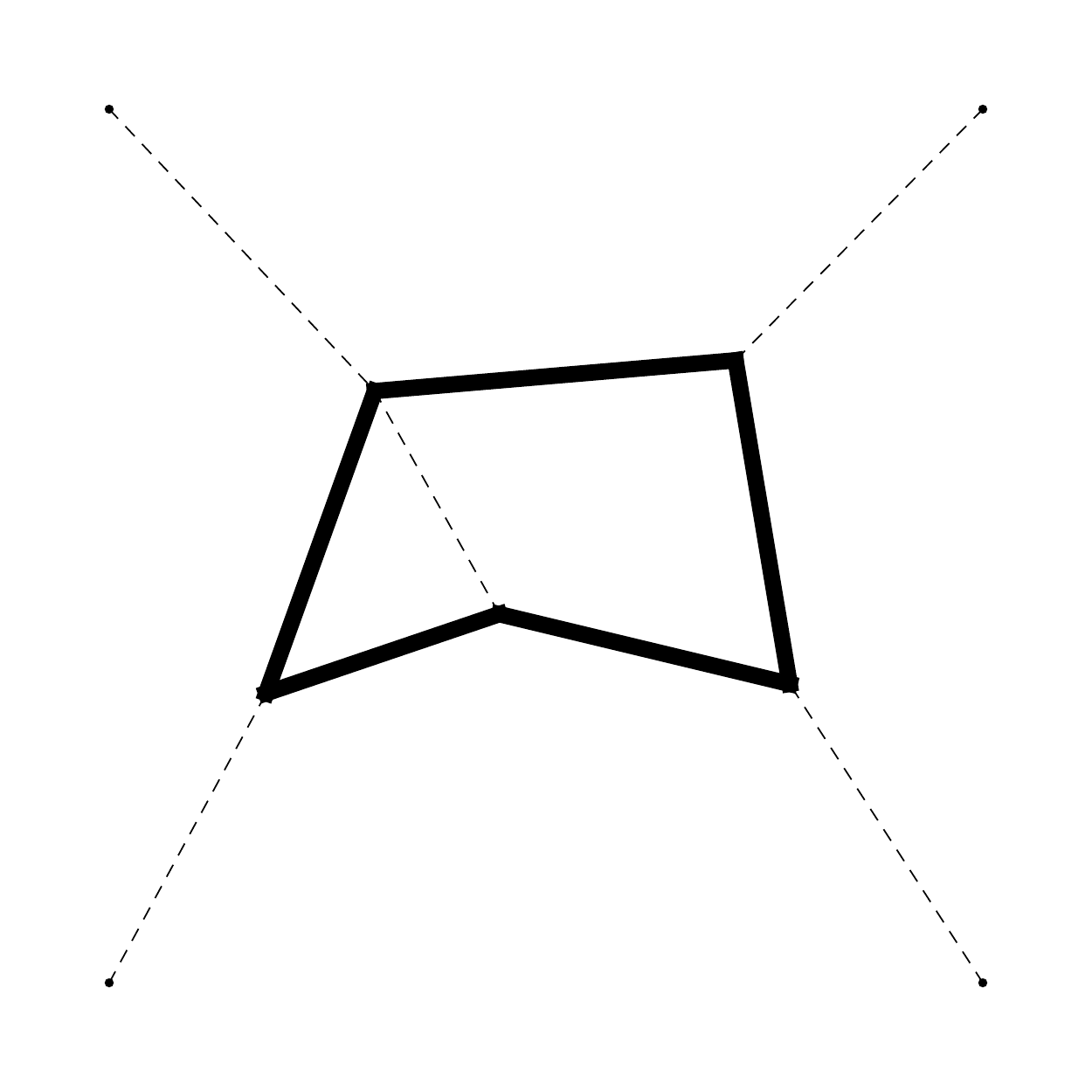}}
\subfloat[$\gA_{24}$]{\includegraphics[width=0.17\textwidth]{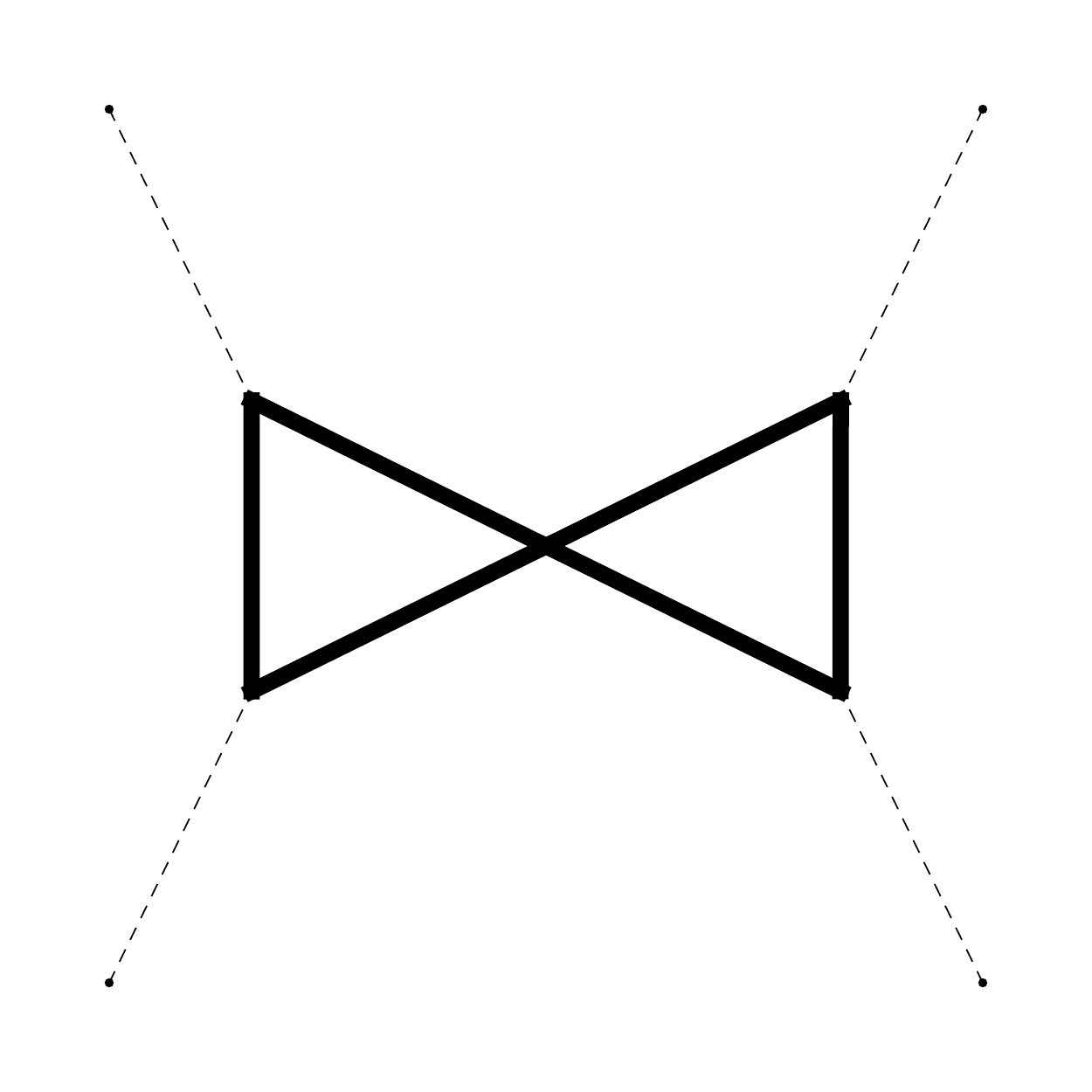}}
\subfloat[$\gA_{27},\gA_{28},\gA_{29}$]{\includegraphics[width=0.17\textwidth]{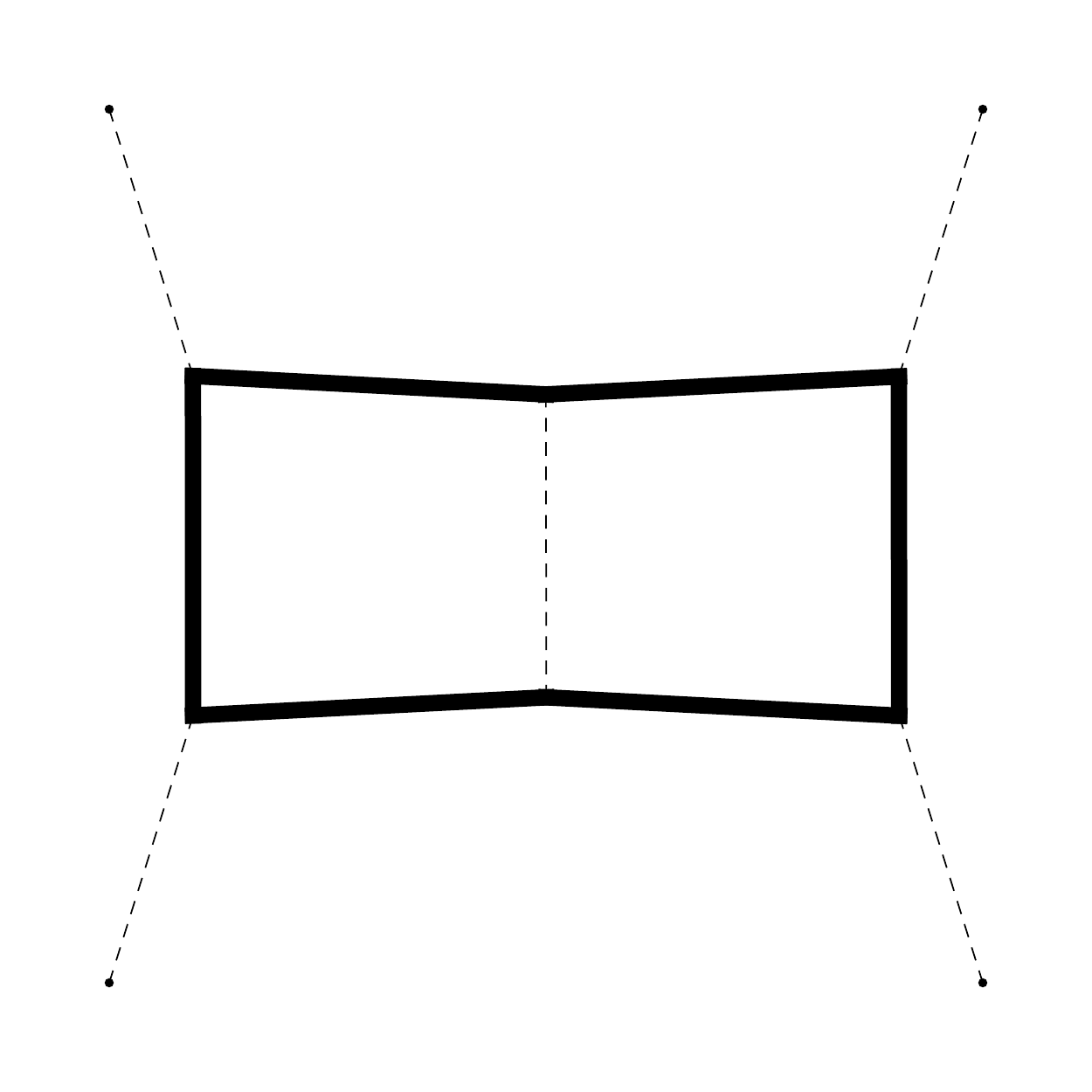}}
\caption{Master integrals the integral family, organized by the number of propagators. 
Each integral stands for a set of integrals sharing the same propagators, with possible
numerator factors not shown in the pictures.}
\label{fig:basismassive}
\end{center}
\end{figure}

We apply the method of differential equations as explained in the main text. Using the methods of ref. \cite{Henn:2013pwa}, we arrived at the following convenient basis choice, 
\begin{align}
\gA_1 =& \, c  G_{2,0,0,0,2,0,0,0,0} \,,  \label{eqbasis1} \\
\gA_2 =& c \sqrt{s \left(s-4 m^2\right)}  G_{2,0,1,0,2,0,0,0,0}  \,, \\
\gA_3 =& c s  \left(4 m^2-s\right) G_{2,0,1,0,2,0,1,0,0} \,,\\
\gA_4 =& -2 c \eps s G_{2,0,0,0,1,0,1,1,0} \,,\\
\gA_5 =& c 2 \eps s \sqrt{s \left(s-4 m^2\right)} G_{2,0,1,0,1,0,1,1,0} \,, \\
\gA_6 =& c s G_{2,0,0,0,0,0,2,0,1} \,,\\
\gA_7 =& c \frac{1}{2}   \sqrt{s
   \left(s-4 m^2\right)} \left[ 2 G_{2,0,0,0,0,0,1,0,2}+G_{2,0,0,0,0,0,2,0,1}\right] \,,\\
   \gA_8 =&c m^2 s G_{3,0,0,0,0,0,1,1,1} \,, \\
   \gA_9 =&c m^2 s G_{1,0,1,0,0,0,0,3,1} \,,\\
   \gA_{10} =& c (-2) {\eps} s G_{1,0,1,0,0,0,0,2,1} \,,\\
   \gA_{11} =&c \sqrt{s \left(s-4 m^2\right)} [m^2 \left(G_{1,0,1,0,0,0,0,3,1}+G_{2,0,1,0,0,0,0,2,1}\right)-\frac{3}{2} {\eps}
   G_{1,0,1,0,0,0,0,2,1}] \,, \\
   \gA_{12} =& 4c  {\eps}^2 s G_{1,0,1,0,0,0,1,1,1} \,,\\
   \gA_{13} =&-4c {\eps}^2 s \sqrt{s \left(s-4 m^2\right)} G_{1,0,1,0,1,0,1,1,1}  \,,\\
   \gA_{14} =&  t c G_{0,2,0,0,0,0,0,2,1} \,,\\
   \gA_{15} =& c \frac{1}{2}  \sqrt{t
   \left(t-4 m^2\right)}  \left[2 G_{0,2,0,0,0,0,0,1,2}+G_{0,2,0,0,0,0,0,2,1}\right]  \,,\\
   \gA_{16} =& c m^2 t G_{1,1,0,0,0,0,0,3,1} \,,\\
   \gA_{17}=& 4 c {\eps}^2 t G_{1,1,0,0,1,0,0,1,1} \,,\\
   \gA_{18} =& -2 c {\eps} \sqrt{s \left(m^4 s-2 m^2 t (s+2 t)+s t^2\right)} G_{1,1,1,0,0,0,0,2,1}  \,,\\
   \gA_{19}=& c \sqrt{s t \left(s t-4 m^2 (s+t)\right)} \left[ m^2 G_{1,1,1,0,0,0,0,3,1}-{\eps} G_{1,1,1,0,0,0,0,2,1}\right]\,,\\
   \gA_{20} =& -2  c {\eps} s \left[ G_{1,1,1,-1,0,0,0,2,1}-m^2 G_{1,1,1,0,0,0,0,2,1}\right] \,, \\
   \gA_{21} =& 4 c {\eps}^2 (s+t) G_{0,1,1,0,1,0,0,1,1} \,,\\
   \gA_{22} =& -2 c  {\eps} \sqrt{s t \left(s t-4 m^2 (s+t)\right)} G_{0,1,1,0,1,0,0,1,2} \,,\\
   \gA_{23} =& -2 c  {\eps} m^2 (s+t) \left[ G_{0,1,2,0,1,0,0,1,1}+G_{0,2,1,0,1,0,0,1,1}\right]  \,,\\
   \gA_{24} =& 4  c{\eps}^2 s^2 G_{1,1,1,0,1,0,1,1,0} \,,\\
   \gA_{25}=& 4 c {\eps}^2 \left[ s G_{1,1,0,0,1,-1,1,1,1}+t G_{0,1,1,0,1,0,0,1,1}\right] \,,\\
   \gA_{26}=&  4 c {\eps}^2  \sqrt{s t \left(s t-4 m^2 (s+t)\right)} G_{1,1,0,0,1,0,1,1,1} \,,\\
   \gA_{27}=&  4 c {\eps}^2 s \sqrt{t \left(s-4 m^2\right) \left(s t-4 m^2
   (s+t)\right)} G_{1,1,1,0,1,0,1,1,1}  \,,\\
   \gA_{28} =& -4c  {\eps}^2 \sqrt{s \left(s-4 m^2\right)} \left[ s G_{1,1,1,0,1,-1,1,1,1}+2 t
   G_{1,1,0,0,1,0,1,1,1}\right] \,,\\
\gA_{29} =& 4 c \eps^2 
    s \Big[ -G_{1, 0, 1, 0, 1, 0, 1, 0, 1} +
     s G_{1, 0, 1, 0, 1, 0, 1, 1, 1} -
     2 G_{1, 1, 0, 0, 1, -1, 1, 1, 1} -
     2 t G_{1, 1, 0, 0, 1, 0, 1, 1, 1} \nonumber \\
     & \quad\quad +
     G_ {1, 1, 1, -1, 1, -1, 1, 1, 1} -
     s G_{1, 1, 1, 0, 1, -1, 1, 1, 1} +
     t G_{1, 1, 1, 0, 1, 0, 1, 1, 0} \Big]\,.
\label{eqbasis29}
   \end{align}
   where the normalization factor $c$ is
   \begin{align}
c= 1/ \Gamma^2(1+\eps) \, e^{2 \eps \gamma_{E}} \, \eps^2 \, (m^2)^{2 \eps}\,.
\end{align}
The basis integrals are depicted (qualitatively) in Fig.~\ref{fig:basismassive}.
Integrals $\gA_{1}$ to $\gA_{13}$ are $s$-channel triangle integrals, while the subsequent four integrals are $t$-channel triangle integrals. These integrals were computed previously, to some order in $\eps$, in ref. \cite{Anastasiou:2006hc}. The form of the basis we have chosen has the advantage that all basis elements have uniform weight and therefore are simpler compared to that reference.
Some related integrals were also considered in ref. \cite{puhlfuerst}.

We wish to comment that, contrary to the $D=4$ case, where we could employ the systematic algorithm presented in section \ref{sec:defourdim},
in the $D$-dimensional case finding the basis where \ref{diffeqsimpleDdim} is satisfied required a fair amount of intuition, and trial and error.
Also, the resulting $\tilde{A}$-matrix is no long block-triangular (nor even triangular).

\subsection{Differential equations and analytic solution}

With the above basis choice, we find the differential equations in the form of eq. (\ref{canonicalequationssection3}),
with $\tilde{A}$ being a $29 \times 29$ matrix. It is given in electronic form as an ancillary file with the arXiv submission of this paper.
As was discussed in the main text, the differential equations can be immediately solved, to any order in $\eps$, in terms of Chen iterated integrals, cf. eq. (\ref{path_ordered}). The relevant boundary condition is 
\begin{align}
f_{i}(s=0,t=0;m^2) = \delta_{i,1} \,.
\end{align}

The {\it alphabet}, i.e. the set of entries of the matrix appearing in the differential equations turns out to be given by
\begin{align}\label{extralettersDdim}
\{ 1+u+v \,, \frac{4-v+\beta}{4-v-\beta} \,, \frac{4+v+\beta}{4+v-\beta}\,, \frac{ (4 \beta_u + \beta)(4 \beta_u+\beta_u v + \beta)}{ (4 \beta_u - \beta)(4 \beta_u+\beta_u v - \beta)} \,, \frac{ (4 \beta_{uv} + \beta)(4 \beta_{uv}-\beta_{uv} v + \beta ) }{ (4 \beta_{uv} - \beta)(4 \beta_{uv}-\beta_{uv} v - \beta ) }\,,
\end{align}
with a new square root,
\begin{align}\label{extraroot}
\beta = \sqrt{16+16 u+8 v+v^2} \,.
\end{align}
in addition to the letters present in (\ref{Amat2loop}). We see that the full $D$-dimensional solution has a more complicated structure than just the four-dimensional dual conformal integrals.

Remarkably, we find that the only new singularity introduced by (\ref{extralettersDdim}) w.r.t. the four-dimensional case is $1+u+v=0$.

\section{Generation of IBP identities using the embedding formalism}
\label{app:embedding}

Here we describe the generation of integration-by-parts identities
among dual conformal integrals.  We first briefly review the embedding
formalism for arbitrary integrals in $D$-dimensions, following closely
ref.~\cite{SimmonsDuffin:2012uy},  and then we set $D=4$ and restrict to conformal integrals.

\subsection{Loop integrals in the embedding formalism}

To each point $x^\mu$ in $D$-dimensional
spacetime we assign a $(D{+}2)$-dimensional vector $X^a$ defined up to rescaling:
\be
 X^a = \left(\begin{array}{c} X^\mu \\ X^-\\ X^+\end{array}\right)\equiv
 \left(\begin{array}{c} x^\mu \\ -x^2\\ 1\end{array}\right)\,,\quad\mbox{with}\quad
 X^a \simeq \alpha X^a \quad(\alpha\neq 0)\,. \label{embed_four}
\ee
The projective identification naturally removes one degree of freedom
from the $(D+2)$-vector $X$.
The second degree of freedom is removed by the fact that the vector
(\ref{embed_four}) is null with respect to the
$(D+2)$-dimensional metric
\be
 \cd{X}{Y} \equiv 2X_\mu X^\mu + X^+Y^- + X^- Y^+\,.
\ee
Thus eq.~(\ref{embed_four}) defines a one-to-one map between
null projective $(D{+}2)$-vectors and points in
(compactified) $D$-dimensional spacetime.
Furthermore, the dot product of the representative (\ref{embed_four})
evaluates to $\cd{X}{Y}=-(x-y)^2$, effectively linearizing the
propagators.

To handle massive propagators in momentum space, we relax the null condition
and modify $X^-$ in eq.~(\ref{embed_four}) such that $X^2=m^2$. As a concrete example, the inverse
propagators defined in eq.~(\ref{inverse_propagators}) map to the
following $(D{+}2)$-vectors:
\be
 X_1 =\left(\begin{array}{c} 0^\mu\\ m^2\\1\end{array}\right)\,,\qquad
 X_2 =\left(\begin{array}{c} -p_1^\mu\\ m^2\\1\end{array}\right)\,,\qquad
 X_3 =\left(\begin{array}{c} -(p_1+p_2)^\mu\\
  -s+  m^2\\1\end{array}\right)\,,\qquad
X_4 =\left(\begin{array}{c} p_4^\mu\\ m^2\\1\end{array}\right)\,. \label{defXDp2}
\ee
To make formulas invariant under choices of representative
(\ref{embed_four}), it is useful to introduce an additional ``infinity point''
\be
I \equiv \left(\begin{array}{c} 0^\mu\\ 1\\0\end{array}\right)\,.
\ee
Assigning the vector $Y_1\propto(k_1^\mu,-k_1^2,1)$ to the loop integration
variable, in this notation we have
\be
 \frac{1}{D_1(k_1)D_2(k_1)D_3(k_1)D_4(k_1)} =
 \frac{\cd{Y_1}{I}^4}{\cd{Y_1}{X_1}\cd{Y_1}{X_2}\cd{Y_1}{X_3}\cd{Y_1}{X_4}}\,. \label{propsDp2}
\ee
All propagator factors are now linear in the integration variable $Y_1$.

The quadratic nature of the propagators has effectively been transferred to the integration
measure: the $(D+2)$-vector $Y_1$ is to be integrated over the null cone $Y_1^2=0$,
subject to the GL(1) ``gauge symmetry'' $Y_1\simeq \alpha Y_1$. A natural
integration measure for this space is
\be
 \int_{Y_1} \equiv \int \frac{d^{D+2}Y_1
   \delta(\frac12Y_1^2)}{i\pi^{D/2}{\rm vol(GL(1))}}\,. \label{intDp2}
\ee
Here the notation $1/{\rm vol(GL(1))}$ means to insert appropriate
gauge-fixing factors and Faddeev-Popov determinant, which should
integrate to unity over each GL(1) gauge orbit. The notation is borrowed from ref.~\cite{ArkaniHamed:2010gh}.

A particularly simple gauge choice is $1/{\rm vol(GL(1))}\mapsto\delta(\cd{Y}{I}-1)$; the
Faddeev-Popov determinant is then unity.\footnote{The triviality of the  Faddeen-Popov
  determinant can be verified easily by integrating over a GL(1) orbit: $\int \frac{d\alpha}{\alpha} \delta(\cd{\alpha Y}{I}-1)=1$.}
The general solution to the $\delta$-function constraint in this gauge
takes the form (\ref{embed_four}), and in this gauge it is easy to see
that the above measure is proportional to the usual one:
\be
\int_Y \frac{1}{\cd{Y}{I}^{D}} = \int \frac{d^Dk}{i\pi^{D/2}} 
 \label{measDp2}
\ee
The left-hand-side is manifestly gauge-invariant (the GL(1) weight of
the denominator cancels that of the measure in eq.~(\ref{intDp2})),
so the identity does not depend on the gauge choice.  For example we could gauge-fix
using $1/{\rm vol(GL(1))}\mapsto \delta(\cd(Y,I')-1)$ for any $I'$,
and although the identity would be more difficult to see, it
would still be true since it is clear that nothing can depend on $I'$.

Combining the ingredients (\ref{propsDp2}) and (\ref{measDp2}) we can
write the integrals in our one-loop family as
\be
 \GA{a_1,\ldots,a_4} = \int_{Y_1}
 \frac{\cd{Y_1}{I}^{2\eps+\sum_i
     a_i-4}}{\cd{Y_1}{X_1}^{a_1}\cd{Y_1}{X_2}^{a_2}\cd{Y_1}{X_3}^{a_3}\cd{Y_1}{X_4}^{a_4}}\,. \label{one_loop_DCI}
\ee
The special simplification in the conformal case $\eps=0$ and $\sum_i
a_i=4$ is manifest: the dependence on $I$ drops out.

Generalization to higher loop planar integrals is straightforward: internal propagators
become, for example, $1/\cd{Y_1}{Y_2}$.  This will be sufficient for
our purposes. It is unclear whether the
formalism could be used for non-planar momentum space integrals, although
we note that non-planar \emph{coordinate-space} integrals would pose no problem.

\subsection{Integration by parts identities: one-loop case}

To derive integration-by-parts identities within the embedding formalism,
we start from the usual fundamental identity \cite{Chetyrkin:1981qh},
\be
 0 = \int \frac{d^{D+2}Y}{{\rm vol(GL(1))}}
\frac{\partial}{\partial Y^a}\delta(\frac12Y^2) V^a(Y)\,.  \label{IBPDp2}
\ee
This expression makes sense (e.g. is invariant under the
GL(1) gauge symmetry) provided that the vector is homogeneous:
$V^i(Y) = \alpha^{D-1}V^i(\alpha Y)$.  We stress that the derivative
\emph{does not} need to act on the gauge fixing factor ``$1/{\rm
  vol(GL(1))}$'' for this identity to be true.

To avoid taking derivatives of $\delta(\frac12Y^2)$, however, which would have
no interpretations in terms of Feynman integrals, one must require orthogonality: $Y_i V^i(Y)=0$.
This is trivially solved by $V^a\propto Y^a$, but, due to the
homogeneity property, the right-hand-side of
(\ref{IBPDp2}) can be seen to become zero trivially in this case,
so one obtains no useful identity in this case. We conclude that $V^a(Y)$ is a homogeneous $(D{+}2)$-vector orthogonal to
$Y^a$ and defined modulo $Y^a$.  This effectively reduces its number of
vector components to $D$, as expected.  This agrees with other treatments of vector fields in the
embedding formalism, see \cite{Weinberg:2010fx,Costa:2011mg} and references therein.

Let us now illustrate the rules with a simple example, involving a one-loop bubble:
\begin{align}
 0=& \int_Y \frac{\partial}{\partial Y^a} Y_b \frac{ X_1^aX_3^b -X_1^b X_3^a}{\cd{Y}{X_1}\cd{Y}{X_2}\cd{Y}{I}^{2-2\eps}} \nonumber\\
 =&  (4 m^2 -s)\GA{2,0,1,0} +2(1-\eps)\big(\GA{1,0,1,0}-\GA{2,0,0,0}\big)-4 m^2  \GA{3,0,0,0}\,.
\end{align}
The left-hand side is an allowed IBP vector because the replacement $\partial/\partial
Y^a\mapsto Y_a$ yields zero, as required. This justifies our omission
of the explicit $\delta(\frac12Y^2)$ factor, part of the
definition of the measure (\ref{intDp2}).
The right-hand side (in which we have used the symmetry
$\GA{a_1,a_2,a_3,a_4}=\GA{a_3,a_2,a_1,a_4}$)
can be easily verified to be indeed a valid integral identity.

To illustrate nontrivial identities among conformal integrals,
we consider the triangle integrals:
\begin{align}
 0&=\int_Y\frac{\partial}{\partial Y^i} Y_j \frac{ X_1^i X_3^j- X_1^j
   X_3^j}{\cd{Y}{X_1}\cd{Y}{X_2}\cd{Y}{X_3}^2} =
 2 m^2 \GA{2,1,1,0}+2 m^2 \GA{1,2,1,0}+(2 m^2 -s)\GA{1,1,2,0}+\ldots\nn\\
 0&=\int_Y\frac{\partial}{\partial Y^i} Y_j \frac{ X_2^i X_3^j- X_2^j
   X_3^j}{\cd{Y}{X_1}\cd{Y}{X_2}\cd{Y}{X_3}^2} =
 2 m^2 \GA{2,1,1,0}+2 m^2 \GA{1,2,1,0}+2 m^2 \GA{1,1,2,0} +\ldots
\end{align}
where we have set $\eps=0$, and suppressed bubble
and tadpole topologies from the right-hand-sides.  One can see that combining these
identities allows the triangles to be removed. Further reducing the
bubbles, one obtains in this way the following reductions:
\begin{align}
 \GA{1,1,2,0} = \frac{g_2-g_1}{s}=\GA{2,1,1,0}, \qquad \GA{1,2,1,0} = \frac{(s-4 m^2)g_2+4 m^2 g_1}{2s}\,,
\end{align}
where the master integrals $g_1$ and $g_2$, defined in eq.~(\ref{newform1loop}), have at
most two propagators. These identities,
which hold only when $D=4$, explain the absence of conformal triangle integrals in the
four-dimensional basis, cf. eq. (\ref{UTbasis1loop}).

\subsection{Integration by parts identities: multi-loop case}

The procedure we have just described generalizes trivially to the multi-loop
case. However, new subtleties arise in the limit $D\to 4$.

The first subtlety is that we must carefully avoid
divergent integrals. Ultraviolet convergence is
already guaranteed for dual conformal integrals, to which we will restrict our attention
when discussing $D=4$.  Infrared divergences, however, could appear
as a result of squaring the (massless!) internal propagators. 

It is important to stress that, when working without a regulator, one
should consider only integrals which are not only finite but \emph{convergent}, e.g.
separately finite in each integration region \cite{ArkaniHamed:2010gh,Weinzierl:2014iaa}. 

As a simple \emph{sufficient} condition for convergence, we have
imposed the following constraints on the integrals in our basis:
\begin{itemize}
 \item At two-loops, we require the internal propagator to occur with
   at most two powers, $1/\cd{Y_1}{Y_2}^2$.  Squared internal propagators, when they appear, must multiply numerators which vanish when $Y_1= Y_2$.
 \item At three-loops, we impose the preceding conditions separately for each
   internal propagator. Convergence
   in the region $Y_1\to Y_2\to Y_3$ is then ensured by requiring that no more than one squared propagators appears.
\end{itemize}
These conditions define a restricted set of integrals, within which
one can safely operate without regulator in $D=4$. 

A second subtlety when $D=4$ is related to contact terms. These appear through
the familiar identity
\be
 \frac{\partial}{\partial k_1^\mu} \frac{(k_1-k_2)^\mu}{(k_1-k_2)^4} = 2\pi^2 i \delta^4(k_1-k_2)\,, \label{delta_function_term}
\ee
which holds when $D$ is exactly equal to four.  As a simple way
to derive the corresponding identity in the embedding formalism,
we may the following result for logarithmically divergent divergent integrals:
\be
 \int_{Y_1,Y_2} \frac{1}{\cd{Y_1}{Y_2}^2} F(Y_1,Y_2) = -\frac{1}{\eps}
 \int_Y F(Y,Y) + \mbox{finite}\,.
\ee
Imagine, now, that we have an IBP vector that contains a squared
internal propagator (but with a numerator such that it generates only convergent
conformal integrals when $D=4$). When we evaluate the
IBP identity for $\eps\neq 0$, we get all the terms we get
naively from $D=4$, plus some $\eps$-dependent terms.
Actually the only $\eps$ dependence in the formalism comes from the derivative acting on
$\cd{Y,I}{}^{2\eps}$, producing
\begin{align}
 &\int_{Y_1,Y_2} \left(\frac{\partial}{\partial Y_1^a}\cd{Y_1}{I}^{2\eps}\right)
 \frac{\cd{Y_2}{I}^{2\eps}}{\cd{Y_1}{Y_2}^2}
 V^{(4)a}(Y_1,Y_2)  = 2\eps \int_{Y_1,Y_2} \frac{\cd{Y_1}{I}^{2\eps}\cd{Y_2}{I}^{2\eps}}{\cd{Y_1}{I}\cd{Y_1}{Y_2}^2}\cd{I,V^{(4)}(Y_1,Y_2)}
\nn\\ &=\left.-\frac{2\eps}{\eps} \int_Y
  \frac{\cd{I}{V^{(4)}(Y,Y)}}{\cd{I}{Y}}\right|_{D=4} + \mathcal{O}(\eps)\,.
\end{align}
We notice the $\epsilon/\epsilon$ cancelation, which means that this
term survives the $\eps\to 0$ limit. This additional term can be
interpreted as a contact term as in eq.~(\ref{delta_function_term}):
\be
 \frac{\partial}{\partial Y_1^a} \frac{1}{\cd{Y_1}{Y_2}^2}
 V^{(4)a}(Y_1,Y_2) = -2\delta^4(Y_1,Y_2)
 \frac{\cd{I}{V^{(4)}(Y,Y)}}{\cd{I}{Y}} + \mbox{regular terms}\,,  \label{contact_term}
\ee
where the $\delta$-function is normalized according to the measure (\ref{measDp2}): $\int_Y \delta^4(Y,Y')F(Y)= F(Y')$.
The ``regular terms'' are all those obtained by the naive
differentiation procedure in $D=4$.

The presence of the infinity point in this formula is surprising at
first sight, but it has a simple interpretation.  To see this, we can
consider, as an alternative to
dimensional regularization, regulating (\ref{IBPDp2})
by excising a small neighborhood of $Y_1=Y_2$.
The contact term will then arise as a boundary term in the integration by part identity, coming from a small
three-sphere surrounding $Y_1=Y_2$.  One can verify that this boundary term is independent of the shape
of the (small) boundary \emph{if and only if} $V^{(4)a}(Y,Y)\propto Y^a$. This
is precisely the condition for (\ref{contact_term}) to be independent
of $I$!  Only IBP vectors satisfying this condition have
unambiguous meaning without a regulator.

This completes our description of the class of integration-by-part
identities we have considered, which generates valid identities among
dual conformal integrals defined in $D=4$.
For the reader's convenience, we recapitulate the full list of constraints:
\begin{itemize}
\item The identities are generated by vectors
  $\frac{\partial}{\partial Y_k^a}V_k^a$ as in eq.~(\ref{IBPDp2}),
  where $V_k$ is orthogonal to $Y_k$ and homogeneous with respect to
  the loop variables.
\item The vector $V_k$ does not contain any cubed internal
  propagator. If it contains a squared internal propagator $1/\cd{Y_1,Y_2}{}^2$,
   its numerator becomes proportional to $Y_1^a$ when
  $Y_1=Y_2$.  Furthermore the resulting identities should involve integrals which
  obey the convergence properties listed at the beginning of this subsection.
\end{itemize}

Generating identities satisfying these rules is now a simple linear
algebra problem. In practice, we used a
computer to generate all possible vectors subject to prescribed bounds on the
various powers which can appear in the denominator (and hence in the
numerator, by the DCI constraint). Within this set the constraints become
a set of linear conditions, which can be solved by
Gaussian elimination (using, for example, Mathematica's 
{\tt NullSpace[]} routine).

We stress that we have not tried to generate the most
general possible IBP identities. The
restricted class we considered turned out to be sufficient for
reducing all two- and three- loop conformal integrals we considered
to a minimal basis!


\subsection{Derivatives with respect to external parameters}

For our restriction to conformal integrals to make sense, it is
important that we be able to express all derivative of conformal integrals,
in terms of conformal integrals.

This property is actually rather trivial within the embedding
formalism. Here we record the relevant formula. The first step is to
write the required differential operators, see e.g. (\ref{differs}), in terms of the $X_i$ variables
defined in eq.~(\ref{defXDp2}). The correct vectors are easily found if one
notes that they should preserve the normalization and on-shell
constraints $X_i^2=m^2$ and
$\cd{X_i}{X_{i{+}1}}=0$, which are satisfied by these expressions.
For example, we have
\be
 \frac{\partial}{\partial s} = 2m^2\frac{\left(\frac{t}{2m^2}-\frac{t}{s}-2\right)
   X_1^a + X_2^a + \frac{t}{s} X_3^a + X_4^a}{st-4m^2(s+t)}
 \frac{\partial}{\partial X_1^a}\,.
\ee
The derivative of an integral is obtained immediately. For example, we have
\begin{align}\label{DCIderivative}
&\frac{{st-4m^2(s+t)}}{-2m^2 a_1} \frac{\partial}{\partial s} \GA{a_1,a_2,a_3,a_4}=\\
& 
{\left(\frac{t}{2m^2}-\frac{t}{s}-2\right)\GA{a_1,a_2,a_3,a_4} +\GA{a_1{+}1,a_2{-}1,a_3,a_4}+\frac{t}{s}\GA{a_1{+}1,a_2,a_3{-}1,a_4}+\GA{a_1{+}1,a_2,a_3,a_4{-}1}}
\,. \nonumber
\end{align}
It is apparent that conformal integrals are mapped to conformal
integrals.

\subsection{Simple implementation of an integral reduction scheme}
\label{app:reduction}

Well known algorithms and public computer packages allow for the generation and
management of integration-by-parts identities, see e.g. 
\cite{Smirnov:2008iw,vonManteuffel:2012np,Lee:2012cn,Smirnov:2013dia}.  To deal
with the tables of identities generated manually as just described, we employed a private implementation of
the most basic ideas behind the public packages. We describe it here.

The task is to convert raw lists of (mostly useless) identities to
useful reductions, which express ``complicated'' integrals in terms of ``simpler'' ones.
The central tool in this task is Gaussian elimination, as also used in the Laporta algorithm \cite{Laporta:2001dd}.
Specifically, to add new identities to the system, we first reduce
them using existing ones, and then rank the surviving integrals in terms of their
``complexity''.  Gaussian elimination is then used to express the most
complicated remaining integrals in terms of simpler ones, as much as possible.
In our implementation, we kept \emph{all} identities: the total number
of identities stored in memory at a given time equals
the total number of linearly independent identities generated so far.

Of critical importance for the success of this strategy is the choice of ``complexity'' function.
We obtained the best results when we tuned it such that the algorithm strived to
reduce, in the following order of priority: the number of loops; the number of internal propagators; the powers of each internal propagator; the number of massive propagators; the powers of each propagator.
(For conformal integrals, the order of the numerator is not
an independent number.)


For indication, generating a sufficient set of reductions to compute all derivatives of the two-loop ladder and its sub-topologies,
turned out to require approximately 500 identities and about 30 seconds of CPU time on a single laptop computer
(endowed with a 2.6GHz processor and 8Go RAM).
The three-loop ladder required approximately 2000 identities, of
considerably longer complexity, but still only on the order of one hour CPU time.
The ``tennis court'' topology, on the other hand, required over 20000 identities and several days of CPU time.
We attribute the relative difficulty in this case to the rather
conservative convergence conditions we imposed on our IBP vectors; hopefully this situation could be improved by
relaxing these conditions, as discussed further in the conclusion section.

\section{One- and two-loop box integrals in terms of Goncharov polylogarithms}\label{app:polylogs}

It turns out that, for the alphabet that occurs in $D=4$ at both two- and three- loops, one can perform
a change of variables that linearizes it.  That is, the three square roots $\sqrt{1+u}$, $\sqrt{1+v}$ and $\sqrt{1+u+v}$
are simultaneously removable.
While we did not use this in the main text,
we wish to mention it here for convenience of the reader.
This allows for a straightforward expression of all $D=4$ integrals in terms of Goncharov polylogarithms.
Specifically we are going to discuss the one- and two-loop box integrals at $\eps=0$, i.e. $g_{6}$ and $g_{10}$.

The change of variable is the following:
\begin{align}\label{wzparam}
u= \frac{(1-w^2)(1-z^2)}{(w-z)^2} \,,\qquad 
v = \frac{4 w z}{(w-z)^2} \,.
\end{align}
We take $w,z$ real and
\begin{align}\label{euclideanwz}
0<w<z<1\,,
\end{align} 
which corresponds to 
the Euclidean region $s/m^2<0,t/m^2<0$, where all functions are real-valued. 
In terms of those variables, the alphabet needed up to two loops, eq.~(\ref{Amat2loop}), can be written as
\begin{align}\label{letters}
\{ z\,, 1 \pm z \,, w \,, 1 \pm w\,, w-z \,,  1-w z \,, 1+w-z+w z \,,  1-w+z+w z  \} \,.
\end{align}
(At three loops one finds additional polynomials in $z,w$.)
We remark that for (\ref{euclideanwz}), all letters (\ref{letters}) are sign-definite, and the corresponding functions
will therefore be manifestly real-valued in the Euclidean region.

Note that the above definition (\ref{wzparam} is separately invariant under the transformations
\begin{align}\label{symmetries}
w \leftrightarrow z \,,\qquad \{ w,z \} \leftrightarrow \left\{ \frac{1}{w}, \frac{1}{z} \right\}  \,,\qquad 
\{w,z\} \leftrightarrow \{-w,-z\} \,,
\end{align}
and therefore we expect $G$ of eq. (\ref{def-1loopfamily}), (\ref{def-2loopfamily}) to be invariant under these transformations.
Note that the basis integrals involve normalization factors that transform covariantly under
(\ref{symmetries}), which translates into corresponding symmetries of the basis integrals.

Note that all letters appearing in the alphabet  (\ref{letters})  are multilinear in $w,z$.
This property implies that we can always write the corresponding integral functions in terms of multiple
polylogarithms, as we will see presently.
The latter can be defined iteratively as follows, 
\begin{align}
G(a_1,\ldots a_n ; z) = \int_0^z \frac{dt}{t-a_{1}} G(a_{2}, \ldots ,a_{n}; t) \,,
\end{align}
with 
\begin{align}
G(a_1 ;z) = \int_0^z \frac{dt}{t-a_{1}}  \,, \qquad a_{1} \neq 0\,.
\end{align}
For $a_{1}=0$, we have $G(\vec{0}_{n};x) = 1/n! \log^n(x)$.
A subset of these functions for indices $0,\pm1$ are called harmonic polylogarithms (HPL) \cite{Remiddi:1999ew}, 
and will be denoted by $H(a_{1}, \ldots a_{n};z)$. \footnote{Note a conventional sign change - for each index equal to $1$, one needs to multiply by $-1$ to convert from the $G$ to $H$ notation.}

There are many ways of doing this which lead to different integral representations.
A straightforward possibility is to choose the integration contour $\mathcal{C}$ of eq. (\ref{path_ordered}) 
to consist of two segments that are straight lines along the $w$ and the $z$ direction, respectively.
In formulas, denoting base point and argument of the function by $(w_i,z_i )$ and $(w_f ,z_f )$, respectively,
we parametrize the two segments as follows,
\begin{align}
\mathcal{C}_{1}: & \quad w(\tau) = w_{i} (1-\tau) + w_{f} \tau \,, \quad z(\tau) = z_{i} \,, \\ 
\mathcal{C}_{2}: & \quad w(\tau) = w_{f}  \,, \quad  z(\tau) = z_{i} (1-\tau) + z_{f} \tau \,. 
\end{align}
It is then immediately obvious that each term in the $\epsilon$ expansion will be of the
form
\begin{align}
G(a_1, \ldots a_n;1) = \int_0^1 \frac{dt_{1}}{t_{1}-a_{1}} \int_0^{t_1}  \frac{dt_{2}}{t_{2}-a_{2}} \ldots  \int_0^{t_{n-1}}  \frac{dt_{n}}{t_{2}-a_{n}}  \,.
\end{align}
These integrals fall exactly into the definition of Goncharov polylogarithms of argument $1$.
It is easy to see that with the above definitions all $a_{i}<0$ for $0<w<z<1$, such that the
integrals are manifestly real there. 

Another possibility is to integrate the differential equations one variable at a time. This essentially corresponds
to choosing a piecewise contour first along one axis, and then along the other.
Integrating back the equations in terms of the multiple polylogarithms is trivial and is best implemented as a  computer algorithm. 
By construction, this form of the answer will depend on Goncharov polylogarithms with argument $w$ and indices drawn from $\{ 0, \pm1,  1/z, (1-z)/(1+z), (1+z)/(1-z)  \}$, and harmonic polylogartihms in $z$. This follows from eq. (\ref{letters}).
The explicit formulas below will be given using this variant.

In the $w,z$ variables, the one-loop box integral $g_{6}$ takes the form (cf. eq. (\ref{I1exact}))
\begin{align}\label{g6exact}
g_{6}=& -G_{-1,0}(w)+G_{0,-1}(w)-G_{0,1}(w)+G_{1,0}(w)+H_{-1,0}(z)-H_{0,-1}(z)-H_{0,1}(z) \nonumber \\
&+H_{1,0}(z)-G_0(
   w) H_{-1}(z)+G_{-1}(w) H_0(z)-G_1(w) H_0(z)-G_0(w) H_1(z) \,.
\end{align}
This particular case only requires Goncharov polylogarithms with indices $0,\pm 1$, so they could be replaced by HPL. The boundary condition, vanishing of $g_{6}$ at $z=w$, can be immediately verified.

The two-loop box integral $g_{10}$ is given in a similar way,
\begin{align}
g_{10}=&G_{-1,0} H_{-1,-1}+G_{-1,\frac{z-1}{z+1}} H_{-1,-1}-G_{-1,-\frac{z+1}{z-1}} H_{-1,-1}+2
   G_{0,-1} H_{-1,-1}+G_{0,0} H_{-1,-1} \nn \\ & \hspace{-0.7cm} 
   -2 G_{0,\frac{1}{z}}
   H_{-1,-1}-G_{0,\frac{z-1}{z+1}} H_{-1,-1}-G_{0,-\frac{z+1}{z-1}} H_{-1,-1}-G_{1,0}
   H_{-1,-1}-G_{1,\frac{z-1}{z+1}} H_{-1,-1} \nn\\ & \hspace{-0.7cm} 
     +G_{1,-\frac{z+1}{z-1}} H_{-1,-1}+G_{-1,0}
   H_{-1,0}-G_{-1,\frac{z-1}{z+1}} H_{-1,0}-G_{-1,-\frac{z+1}{z-1}}
   H_{-1,0}+G_{0,\frac{z-1}{z+1}} H_{-1,0} \nn\\ & \hspace{-0.7cm} 
     -G_{0,-\frac{z+1}{z-1}} H_{-1,0}-G_{1,0}
   H_{-1,0}+G_{1,\frac{z-1}{z+1}} H_{-1,0}+G_{1,-\frac{z+1}{z-1}} H_{-1,0}+G_{-1,0}
   H_{-1,1}+G_{-1,\frac{z-1}{z+1}} H_{-1,1} \nn\\ & \hspace{-0.7cm} 
     -G_{-1,-\frac{z+1}{z-1}} H_{-1,1}+2 G_{0,-1}
   H_{-1,1}+G_{0,0} H_{-1,1}-2 G_{0,\frac{1}{z}} H_{-1,1}-G_{0,\frac{z-1}{z+1}}
   H_{-1,1}-G_{0,-\frac{z+1}{z-1}} H_{-1,1}\nn\\ & \hspace{-0.7cm} 
-G_{1,0} H_{-1,1}    -G_{1,\frac{z-1}{z+1}}
   H_{-1,1}+G_{1,-\frac{z+1}{z-1}} H_{-1,1}-G_{-1,0} H_{0,-1}+G_{-1,\frac{z-1}{z+1}}
   H_{0,-1}+G_{-1,-\frac{z+1}{z-1}} H_{0,-1}\nn\\ & \hspace{-0.7cm} 
-G_{0,\frac{z-1}{z+1}}
   H_{0,-1}+G_{0,-\frac{z+1}{z-1}} H_{0,-1}+G_{1,0} H_{0,-1}-G_{1,\frac{z-1}{z+1}}
   H_{0,-1}-G_{1,-\frac{z+1}{z-1}} H_{0,-1}+G_{-1,\frac{z-1}{z+1}}
   H_{0,0}\nn\\ & \hspace{-0.7cm} 
-G_{-1,-\frac{z+1}{z-1}} H_{0,0}+G_{0,-1} H_{0,0}+G_{0,1}
   H_{0,0}-G_{0,\frac{z-1}{z+1}} H_{0,0}-G_{0,-\frac{z+1}{z-1}}
   H_{0,0}-G_{1,\frac{z-1}{z+1}} H_{0,0}\nn\\ & \hspace{-0.7cm} 
+G_{1,-\frac{z+1}{z-1}} H_{0,0}-G_{-1,0}
   H_{0,1}+G_{-1,\frac{z-1}{z+1}} H_{0,1}+G_{-1,-\frac{z+1}{z-1}}
   H_{0,1}-G_{0,\frac{z-1}{z+1}} H_{0,1}+G_{0,-\frac{z+1}{z-1}} H_{0,1}\nn \\ & \hspace{-0.7cm} 
+G_{1,0}
   H_{0,1}-G_{1,\frac{z-1}{z+1}} H_{0,1}-G_{1,-\frac{z+1}{z-1}} H_{0,1}-G_{-1,0}
   H_{1,-1}+G_{-1,\frac{z-1}{z+1}} H_{1,-1}-G_{-1,-\frac{z+1}{z-1}} H_{1,-1}\nn \\ & \hspace{-0.7cm} 
+G_{0,0}
   H_{1,-1}+2 G_{0,1} H_{1,-1}-2 G_{0,\frac{1}{z}} H_{1,-1}-G_{0,\frac{z-1}{z+1}}
   H_{1,-1}-G_{0,-\frac{z+1}{z-1}} H_{1,-1}+G_{1,0} H_{1,-1}\nn \\ & \hspace{-0.7cm} 
-G_{1,\frac{z-1}{z+1}}
   H_{1,-1}+G_{1,-\frac{z+1}{z-1}} H_{1,-1}+G_{-1,0} H_{1,0}-G_{-1,\frac{z-1}{z+1}}
   H_{1,0}-G_{-1,-\frac{z+1}{z-1}} H_{1,0}+G_{0,\frac{z-1}{z+1}}
   H_{1,0}\nn\\ & \hspace{-0.7cm} 
-G_{0,-\frac{z+1}{z-1}} H_{1,0}-G_{1,0} H_{1,0}+G_{1,\frac{z-1}{z+1}}
   H_{1,0}+G_{1,-\frac{z+1}{z-1}} H_{1,0}-G_{-1,0} H_{1,1}+G_{-1,\frac{z-1}{z+1}}
   H_{1,1}\nn\\ & \hspace{-0.7cm} 
-G_{-1,-\frac{z+1}{z-1}} H_{1,1}+G_{0,0} H_{1,1}+2 G_{0,1} H_{1,1}-2
   G_{0,\frac{1}{z}} H_{1,1}-G_{0,\frac{z-1}{z+1}} H_{1,1}-G_{0,-\frac{z+1}{z-1}}
   H_{1,1}\nn\\ & \hspace{-0.7cm} 
+G_{1,0} H_{1,1}-G_{1,\frac{z-1}{z+1}} H_{1,1}+G_{1,-\frac{z+1}{z-1}}
   H_{1,1}+H_{-1} G_{-1,0,-1}+H_0 G_{-1,0,-1}+H_1 G_{-1,0,-1}\nn\\ & \hspace{-0.7cm} 
-H_{-1} G_{-1,0,0}-H_1
   G_{-1,0,0}+H_{-1} G_{-1,0,1}-H_0 G_{-1,0,1}+H_1 G_{-1,0,1}-2 H_{-1}
   G_{-1,0,\frac{1}{z}}\nn\\ & \hspace{-0.7cm} 
-2 H_1 G_{-1,0,\frac{1}{z}}-H_{-1} G_{-1,\frac{z-1}{z+1},-1}-H_0
   G_{-1,\frac{z-1}{z+1},-1}-H_1 G_{-1,\frac{z-1}{z+1},-1}+H_{-1}
   G_{-1,\frac{z-1}{z+1},0}\nn\\ & \hspace{-0.7cm} 
-H_0 G_{-1,\frac{z-1}{z+1},0}+H_1
   G_{-1,\frac{z-1}{z+1},0}+H_{-1} G_{-1,\frac{z-1}{z+1},1}+H_0
   G_{-1,\frac{z-1}{z+1},1}+H_1 G_{-1,\frac{z-1}{z+1},1}\nn\\ & \hspace{-0.7cm} 
+H_{-1}
   G_{-1,-\frac{z+1}{z-1},-1}-H_0 G_{-1,-\frac{z+1}{z-1},-1}+H_1
   G_{-1,-\frac{z+1}{z-1},-1}+H_{-1} G_{-1,-\frac{z+1}{z-1},0}+H_0
   G_{-1,-\frac{z+1}{z-1},0}\nn\\ & \hspace{-0.7cm} 
+H_1 G_{-1,-\frac{z+1}{z-1},0}-H_{-1}
   G_{-1,-\frac{z+1}{z-1},1}+H_0 G_{-1,-\frac{z+1}{z-1},1}-H_1
   G_{-1,-\frac{z+1}{z-1},1}-H_0 G_{0,-1,0}\nn\\ & \hspace{-0.7cm} 
+2 H_{-1} G_{0,-1,1}+2 H_1 G_{0,-1,1}-2
   H_{-1} G_{0,-1,\frac{1}{z}}-2 H_1 G_{0,-1,\frac{1}{z}}-H_{-1} G_{0,0,-1}-H_1
   G_{0,0,-1}\nn\\ & \hspace{-0.7cm} 
+H_{-1} G_{0,0,1}+H_1 G_{0,0,1}-2 H_{-1} G_{0,1,-1}-2 H_1 G_{0,1,-1}-H_0
   G_{0,1,0}+2 H_{-1} G_{0,1,\frac{1}{z}}+2 H_1 G_{0,1,\frac{1}{z}}\nn\\ & \hspace{-0.7cm} 
+2 H_{-1}
   G_{0,\frac{1}{z},-1}+2 H_1 G_{0,\frac{1}{z},-1}-2 H_{-1} G_{0,\frac{1}{z},1}-2 H_1
   G_{0,\frac{1}{z},1}+H_{-1} G_{0,\frac{z-1}{z+1},-1}+H_0 G_{0,\frac{z-1}{z+1},-1}\nn\\ & \hspace{-0.7cm} 
+H_1
   G_{0,\frac{z-1}{z+1},-1}-H_{-1} G_{0,\frac{z-1}{z+1},0}+H_0
   G_{0,\frac{z-1}{z+1},0}-H_1 G_{0,\frac{z-1}{z+1},0}-H_{-1}
   G_{0,\frac{z-1}{z+1},1}-H_0 G_{0,\frac{z-1}{z+1},1}\nn\\ & \hspace{-0.7cm} 
-H_1
   G_{0,\frac{z-1}{z+1},1}+H_{-1} G_{0,-\frac{z+1}{z-1},-1}-H_0
   G_{0,-\frac{z+1}{z-1},-1}+H_1 G_{0,-\frac{z+1}{z-1},-1}+H_{-1}
   G_{0,-\frac{z+1}{z-1},0}\nn\\ & \hspace{-0.7cm} 
+H_0 G_{0,-\frac{z+1}{z-1},0}+H_1
   G_{0,-\frac{z+1}{z-1},0}-H_{-1} G_{0,-\frac{z+1}{z-1},1}+H_0
   G_{0,-\frac{z+1}{z-1},1}-H_1 G_{0,-\frac{z+1}{z-1},1}-H_{-1} G_{1,0,-1}\nn\\ & \hspace{-0.7cm} 
-H_0
   G_{1,0,-1}-H_1 G_{1,0,-1}+H_{-1} G_{1,0,0}+H_1 G_{1,0,0}-H_{-1} G_{1,0,1}+H_0
   G_{1,0,1}-H_1 G_{1,0,1}\nn\\ & \hspace{-0.7cm} 
+2 H_{-1} G_{1,0,\frac{1}{z}}+2 H_1
   G_{1,0,\frac{1}{z}}+H_{-1} G_{1,\frac{z-1}{z+1},-1}+H_0 G_{1,\frac{z-1}{z+1},-1}+H_1
   G_{1,\frac{z-1}{z+1},-1}-H_{-1} G_{1,\frac{z-1}{z+1},0}\nn\\ & \hspace{-0.7cm} 
+H_0
   G_{1,\frac{z-1}{z+1},0}-H_1 G_{1,\frac{z-1}{z+1},0}-H_{-1}
   G_{1,\frac{z-1}{z+1},1}-H_0 G_{1,\frac{z-1}{z+1},1}-H_1
   G_{1,\frac{z-1}{z+1},1}-H_{-1} G_{1,-\frac{z+1}{z-1},-1}\nn\\ & \hspace{-0.7cm} 
+H_0
   G_{1,-\frac{z+1}{z-1},-1}-H_1 G_{1,-\frac{z+1}{z-1},-1}-H_{-1}
   G_{1,-\frac{z+1}{z-1},0}-H_0 G_{1,-\frac{z+1}{z-1},0}-H_1
   G_{1,-\frac{z+1}{z-1},0}\nn\\ & \hspace{-0.7cm} 
+H_{-1} G_{1,-\frac{z+1}{z-1},1}-H_0
   G_{1,-\frac{z+1}{z-1},1}+H_1 G_{1,-\frac{z+1}{z-1},1}-G_{-1} H_{-1,-1,-1}-G_0
   H_{-1,-1,-1}\nn\\ & \hspace{-0.7cm} 
+G_1 H_{-1,-1,-1}-G_{-1} H_{-1,-1,0}-G_0 H_{-1,-1,0}+G_1
   H_{-1,-1,0}-G_{-1} H_{-1,-1,1}-G_0 H_{-1,-1,1}\nn\\ & \hspace{-0.7cm} 
+G_1 H_{-1,-1,1}+G_{-1}
   H_{-1,0,-1}+G_0 H_{-1,0,-1}-G_1 H_{-1,0,-1}-G_{-1} H_{-1,0,0}+G_1 H_{-1,0,0}\nn\\ & \hspace{-0.7cm} 
+G_{-1}
   H_{-1,0,1}+G_0 H_{-1,0,1}-G_1 H_{-1,0,1}-G_{-1} H_{-1,1,-1}+G_0 H_{-1,1,-1}+G_1
   H_{-1,1,-1}\nn\\ & \hspace{-0.7cm} 
-G_{-1} H_{-1,1,0}-G_0 H_{-1,1,0}+G_1 H_{-1,1,0}-G_{-1} H_{-1,1,1}+G_0
   H_{-1,1,1}+G_1 H_{-1,1,1}\nn\\ & \hspace{-0.7cm} 
+G_{-1} H_{0,-1,-1}+G_0 H_{0,-1,-1}-G_1 H_{0,-1,-1}+G_{-1}
   H_{0,-1,0}-G_1 H_{0,-1,0}+G_{-1} H_{0,-1,1}\nn\\ & \hspace{-0.7cm} 
+G_0 H_{0,-1,1}-G_1 H_{0,-1,1}-G_{-1}
   H_{0,0,-1}+G_1 H_{0,0,-1}-G_{-1} H_{0,0,1}+G_1 H_{0,0,1}-G_{-1} H_{0,1,-1}\nn\\ & \hspace{-0.7cm} 
+G_0
   H_{0,1,-1}+G_1 H_{0,1,-1}+G_{-1} H_{0,1,0}-G_1 H_{0,1,0}-G_{-1} H_{0,1,1}+G_0
   H_{0,1,1}+G_1 H_{0,1,1}\nn\\ & \hspace{-0.7cm} 
-G_{-1} H_{1,-1,-1}-G_0 H_{1,-1,-1}+G_1 H_{1,-1,-1}+G_{-1}
   H_{1,-1,0}-G_0 H_{1,-1,0}-G_1 H_{1,-1,0}\nn\\ & \hspace{-0.7cm} 
-G_{-1} H_{1,-1,1}-G_0 H_{1,-1,1}+G_1
   H_{1,-1,1}-G_{-1} H_{1,0,-1}+G_0 H_{1,0,-1}+G_1 H_{1,0,-1}-G_{-1} H_{1,0,0}\nn\\ & \hspace{-0.7cm} 
+G_1
   H_{1,0,0}-G_{-1} H_{1,0,1}+G_0 H_{1,0,1}+G_1 H_{1,0,1}-G_{-1} H_{1,1,-1}+G_0
   H_{1,1,-1}+G_1 H_{1,1,-1}\nn\\ & \hspace{-0.7cm} 
+G_{-1} H_{1,1,0}-G_0 H_{1,1,0}-G_1 H_{1,1,0}-G_{-1}
   H_{1,1,1}+G_0 H_{1,1,1}+G_1
   H_{1,1,1}-G_{-1,0,-1,-1}\nn\\ & \hspace{-0.7cm} 
-G_{-1,0,-1,0}+G_{-1,0,-1,1}+G_{-1,0,0,-1}-G_{-1,0,0,1}-G_{-
   1,0,1,-1}+G_{-1,0,1,0}+G_{-1,0,1,1}\nn\\ & \hspace{-0.7cm} 
+2 G_{-1,0,\frac{1}{z},-1}-2
   G_{-1,0,\frac{1}{z},1}+G_{-1,\frac{z-1}{z+1},-1,-1}+G_{-1,\frac{z-1}{z+1},-1,0}-G_{-
   1,\frac{z-1}{z+1},-1,1}-G_{-1,\frac{z-1}{z+1},0,-1}\nn\\ & \hspace{-0.7cm} 
+G_{-1,\frac{z-1}{z+1},0,0}+G_{-1
   ,\frac{z-1}{z+1},0,1}-G_{-1,\frac{z-1}{z+1},1,-1}-G_{-1,\frac{z-1}{z+1},1,0}+G_{-1,\frac{z-1}{z+1},1,1}-G_{-1,-\frac{z+1}{z-1},-1,-1}\nn\\ & \hspace{-0.7cm} 
+G_{-1,-\frac{z+1}{z-1},-1,0}+G_{-1
   ,-\frac{z+1}{z-1},-1,1}-G_{-1,-\frac{z+1}{z-1},0,-1}-G_{-1,-\frac{z+1}{z-1},0,0}+G_{
   -1,-\frac{z+1}{z-1},0,1}\nn\\ & \hspace{-0.7cm} 
+G_{-1,-\frac{z+1}{z-1},1,-1}-G_{-1,-\frac{z+1}{z-1},1,0}-G_
   {-1,-\frac{z+1}{z-1},1,1}+G_{0,-1,0,0}-2 G_{0,-1,1,-1}+2 G_{0,-1,1,1}\nn\\ & \hspace{-0.7cm} 
+2
   G_{0,-1,\frac{1}{z},-1}-2
   G_{0,-1,\frac{1}{z},1}+G_{0,0,-1,-1}-G_{0,0,-1,1}-G_{0,0,1,-1}+G_{0,0,1,1}+2
   G_{0,1,-1,-1}\nn\\ & \hspace{-0.7cm} 
-2 G_{0,1,-1,1}+G_{0,1,0,0}-2 G_{0,1,\frac{1}{z},-1}+2
   G_{0,1,\frac{1}{z},1}-2 G_{0,\frac{1}{z},-1,-1}+2 G_{0,\frac{1}{z},-1,1}+2
   G_{0,\frac{1}{z},1,-1}\nn\\ & \hspace{-0.7cm} 
-2
   G_{0,\frac{1}{z},1,1}-G_{0,\frac{z-1}{z+1},-1,-1}-G_{0,\frac{z-1}{z+1},-1,0}+G_{0,\frac{z-1}{z+1},-1,1}+G_{0,\frac{z-1}{z+1},0,-1}-G_{0,\frac{z-1}{z+1},0,0}-G_{0,\frac{
   z-1}{z+1},0,1}\nn\\ & \hspace{-0.7cm} 
+G_{0,\frac{z-1}{z+1},1,-1}+G_{0,\frac{z-1}{z+1},1,0}-G_{0,\frac{z-1}{
   z+1},1,1}-G_{0,-\frac{z+1}{z-1},-1,-1}+G_{0,-\frac{z+1}{z-1},-1,0}+G_{0,-\frac{z+1}{
   z-1},-1,1}\nn\\ & \hspace{-0.7cm} 
-G_{0,-\frac{z+1}{z-1},0,-1}-G_{0,-\frac{z+1}{z-1},0,0}+G_{0,-\frac{z+1}{z
   -1},0,1}+G_{0,-\frac{z+1}{z-1},1,-1}-G_{0,-\frac{z+1}{z-1},1,0}-G_{0,-\frac{z+1}{z-1
   },1,1}\nn\\ & \hspace{-0.7cm} 
+G_{1,0,-1,-1}+G_{1,0,-1,0}-G_{1,0,-1,1}-G_{1,0,0,-1}+G_{1,0,0,1}+G_{1,0,1,-1}
   -G_{1,0,1,0}-G_{1,0,1,1}\nn\\ & \hspace{-0.7cm} 
-2 G_{1,0,\frac{1}{z},-1}+2
   G_{1,0,\frac{1}{z},1}-G_{1,\frac{z-1}{z+1},-1,-1}-G_{1,\frac{z-1}{z+1},-1,0}+G_{1,\frac{z-1}{z+1},-1,1}+G_{1,\frac{z-1}{z+1},0,-1}-G_{1,\frac{z-1}{z+1},0,0}\nn\\ & \hspace{-0.7cm} 
-G_{1,\frac{
   z-1}{z+1},0,1}+G_{1,\frac{z-1}{z+1},1,-1}+G_{1,\frac{z-1}{z+1},1,0}-G_{1,\frac{z-1}{
   z+1},1,1}+G_{1,-\frac{z+1}{z-1},-1,-1}-G_{1,-\frac{z+1}{z-1},-1,0}\nn\\ & \hspace{-0.7cm} 
-G_{1,-\frac{z+1}{
   z-1},-1,1}+G_{1,-\frac{z+1}{z-1},0,-1}+G_{1,-\frac{z+1}{z-1},0,0}-G_{1,-\frac{z+1}{z
   -1},0,1}-G_{1,-\frac{z+1}{z-1},1,-1}+G_{1,-\frac{z+1}{z-1},1,0}\nn\\ & \hspace{-0.7cm} 
+G_{1,-\frac{z+1}{z-1
   },1,1}+H_{-1,-1,-1,-1}+H_{-1,-1,-1,0}+H_{-1,-1,-1,1}-H_{-1,-1,0,-1}+H_{-1,-1,0,0}\nn\\ & \hspace{-0.7cm} 
-H_
   {-1,-1,0,1}+H_{-1,-1,1,-1}+H_{-1,-1,1,0}+H_{-1,-1,1,1}-H_{-1,0,-1,-1}-H_{-1,0,-1,0}-
   H_{-1,0,-1,1}\nn\\ & \hspace{-0.7cm} 
+H_{-1,0,0,-1}+H_{-1,0,0,1}+H_{-1,0,1,-1}-H_{-1,0,1,0}+H_{-1,0,1,1}+H_{
   -1,1,-1,-1}-H_{-1,1,-1,0}\nn\\ & \hspace{-0.7cm} 
+H_{-1,1,-1,1}+H_{-1,1,0,-1}+H_{-1,1,0,0}+H_{-1,1,0,1}+H_{-
   1,1,1,-1}-H_{-1,1,1,0}+H_{-1,1,1,1}\nn\\ & \hspace{-0.7cm} 
-H_{0,-1,-1,-1}-H_{0,-1,-1,0}-H_{0,-1,-1,1}+H_{0,
   -1,0,-1}+H_{0,-1,0,1}+H_{0,-1,1,-1}-H_{0,-1,1,0}\nn\\ & \hspace{-0.7cm} 
+H_{0,-1,1,1}+H_{0,0,-1,-1}+H_{0,0,-
   1,1}+H_{0,0,1,-1}+H_{0,0,1,1}-H_{0,1,-1,-1}-H_{0,1,-1,0}-H_{0,1,-1,1}\nn\\ & \hspace{-0.7cm} 
+H_{0,1,0,-1}+H
   _{0,1,0,1}+H_{0,1,1,-1}-H_{0,1,1,0}+H_{0,1,1,1}+H_{1,-1,-1,-1}+H_{1,-1,-1,0}+H_{1,-1
   ,-1,1}\nn\\ & \hspace{-0.7cm} 
-H_{1,-1,0,-1}+H_{1,-1,0,0}-H_{1,-1,0,1}+H_{1,-1,1,-1}+H_{1,-1,1,0}+H_{1,-1,1,
   1}-H_{1,0,-1,-1}\nn\\ & \hspace{-0.7cm} 
-H_{1,0,-1,0}-H_{1,0,-1,1}+H_{1,0,0,-1}+H_{1,0,0,1}+H_{1,0,1,-1}-H_{
   1,0,1,0}+H_{1,0,1,1}+H_{1,1,-1,-1}\nn\\ & \hspace{-0.7cm} 
-H_{1,1,-1,0}+H_{1,1,-1,1}+H_{1,1,0,-1}+H_{1,1,0,0
   }+H_{1,1,0,1}+H_{1,1,1,-1}-H_{1,1,1,0}+H_{1,1,1,1} \,.
\end{align}
Here the arguments of the functions are omitted. For Goncharov polylogarithms $G$ the argument is $w$, while
for the harmonic polylogarithms $H$ it is $z$. 
We also included this formulas as an ancillary text file for convenience.
The above expression has $412$ terms and can probably be written in a more compact way -- the most compact way is probably the Chen integral representation, where it has only $10$ terms, as is evident from 
Fig.~\ref{fig:blocktriangular2loop}.

We have verified this formula numerally \cite{Bauer:2000cp,Vollinga:2004sn} for various random kinematical points in the Euclidean region against numerical results provided by FIESTA \cite{Smirnov:2008py,Smirnov:2013eza}, as well as by cross-checks against eq.~(\ref{twoloopN}) and the Mandelstam representation (\ref{doubledisc1}).

Before closing this section, we wish to mention that in the full $D$-dimensional case,
where the new square root (\ref{extraroot}) appears,
the question of finding a parametrization of the kinematic variables that removes square roots can be formulated as a diophantine problem. 

%
%
%
%
%

\section{Differential equation at three loops}
\label{app:de}

In this appendix we record the full form of the differential equation at three-loops, for the DCI integrals in four dimensions.
Out of the 48 master integrals, we show results only for the 38 of them which are related to the integrals $g_{37}, g_{38}$, defined in eqs. (\ref{defint37}) and (\ref{defint38}), which control the 3-loop
light-by-light amplitude in $\mathcal{N}=4$.
This set includes the 10 two-loop integrals $g_{1}\ldots g_{10}$ as well as 4 other related by $s\leftrightarrow t$ symmetry,
plus 24 genuine 3-loop integrals which are defined in an ancillary that file also contains the $A$-matrix reproduced here.

\def\oddbu{[\bu,1]}
\def\oddbv{[\bv,1]}
\def\oddbuv{[\buv,1]}
\def\oddbubuv{[\buv,\bu]}
\def\oddbvbuv{[\buv,\bv]}
\def\oddubu{\{\kappa_1\}}
\def\oddvbv{\{\kappa_2\}}
\def\dlog#1{\{{#1}\}}
Introducing the abbreviations,
\begin{align}
 \{a\} = d\log(a),\quad
 [a,b] \equiv d\log\left(\frac{a-b}{a+b}\right), \quad
 \kappa_1\equiv \frac{2{+}u-2\buv}{2{+}u+2\buv}\,,\quad
 \kappa_2\equiv \frac{2{+}v-2\buv}{2{+}v+2\buv}\,,
\end{align}
our result for the complete three-loop hierarchy can be written:
\begin{eqnarray}
 d\left(\begin{array}{c}g_{2}\\g_{3}\end{array}\right) &=& \left(\begin{array}{c}
  \oddbu \\
  \oddbv
\end{array}\right) g_1
\nn\\
 d\left(\begin{array}{c}g_{4}\\g_{5}\\g_{6}\\g_{7}\\g_{11}\end{array}\right) &=& \left(\begin{array}{cc}
  \oddbu & 0 \\
  \dlog{\frac{u+1}{u}} & 0 \\
  \oddbvbuv & \oddbubuv \\
  0 & \oddbv \\
  0 & \dlog{\frac{v+1}{v}}
\end{array}\right)
\left(\begin{array}{c}g_{2}\\g_{3}\end{array}\right)
\nn\\
 d\left(\begin{array}{c}g_{8}\\g_{9}\\g_{12}\\g_{13}\\g_{15}\\g_{16}\\g_{17}\\g_{18}\\g_{19}\\g_{20}\\g_{21}\\g_{22}\end{array}\right)
 &=&
  \left(\begin{array}{ccccc}
  \oddbuv & \oddbvbuv & \dlog{\frac{v}{u+v}} & \oddbuv & 0 \\
  \dlog{\frac{u v}{(u+1) (u+v)}} & \oddbu & \oddbuv & \dlog{\frac{u}{u+v}} & 0 \\
  \oddbuv & 0 & \dlog{\frac{u}{u+v}} & \oddbuv & \oddbubuv \\
  \dlog{\frac{v}{u+v}} & 0 & \oddbuv & \dlog{\frac{u v}{(v+1) (u+v)}} & \oddbv \\
  \oddbu & 0 & 0 & 0 & 0 \\
  2\dlog{\frac{u+1}{u}} & \oddbu & 0 & 0 & 0 \\
  \dlog{\frac{u+1}{u}} & \oddbu & 0 & 0 & 0 \\
  \dlog{u} & \oddbu & 0 & 0 & 0 \\
  0 & 0 & 0 & \dlog{v} & 0 \\
  0 & 0 & 0 & \oddbv & 0 \\
  0 & 0 & 0 & \dlog{v} & \oddbv \\
  0 & 4\dlog{\frac{u}{u+1}} & \oddbvbuv & \oddbu & 0
\end{array}\right)
 \left(\begin{array}{c}g_{4}\\g_{5}\\g_{6}\\g_{7}\\g_{11}\end{array}\right)
\nn\\
 d\left(\begin{array}{c}g_{32}\\g_{33}\\g_{34}\\g_{35}\\g_{36}\end{array}\right) 
 &=& \begin{array}{l}
 \left(\begin{array}{cccc}
  0 & 0 & 0 & \oddbvbuv \\
  3\dlog{u+1} - 4\dlog{u} & 0 & 0 & 0 \\
  0 & 0 & 0 & 0 \\
  0 & -4\oddbv & 2\oddbubuv & 0 \\
  0 & 0 & \oddbv & 0
\end{array}\right.\\
\hspace{1.5cm} \left.\begin{array}{ccccccc}
  0 & \oddbu & 0 & 0 & 0 & 0 & 0 \\
  \oddbu & \oddbvbuv & 0 & 0 & 0 & 0 & 0 \\
  0 & 0 & 0 & -\oddbv & 0 & \oddbubuv & 0 \\
  0 & 0 & 2\dlog{\frac{u+v}{u}} & 2\dlog{\frac{u+v}{u}} - \dlog{v} & \dlog{v} & 4\oddbuv & 2\oddbuv \\
  0 & 0 & -\oddbuv & -\oddbuv & 0 & 2\dlog{\frac{u}{u+v}} & \dlog{\frac{v}{u+v}}
\end{array}\right)\end{array}
 \left(\begin{array}{c}g_{10}\\g_{14}\\g_{23}\\g_{24}\\g_{25}\\g_{26}\\g_{27}\\g_{28}\\g_{29}\\g_{30}\\g_{31}\end{array}\right)
\nn\\
 d\left(\begin{array}{c}g_{37}\\g_{38}\end{array}\right)
 &=& \left(\begin{array}{ccccc}
  \oddbvbuv & \oddbu & 0 & 0 & 0 \\
  0 & 0 & 4\dlog{\frac{v+1}{v}} & \oddbv & 2\oddbubuv
\end{array}\right)\left(\begin{array}{c}g_{32}\\g_{33}\\g_{34}\\g_{35}\\g_{36}\end{array}\right)
\end{eqnarray}

\begin{sideways}\parbox{\textheight}{\begin{eqnarray} 
 d\left(\begin{array}{c}g_{10}\\g_{14}\\g_{23}\\g_{24}\\g_{25}\\g_{26}\\g_{27}\\g_{28}\\g_{29}\\g_{30}\\g_{31}\end{array}\right)
 &=& 
 \left(\begin{array}{ccccccc}
  \oddbu & \oddbvbuv & 0 & 0 & 0 & 0 & 0 \\
  0 & 0 & \oddbv & \oddbubuv & 0 & 0 & 0 \\
  0 & 2\oddbv & 0 & -2\oddbv & 0 & 4\oddbv & -6\oddbv \\
  \dlog{\frac{u^2}{u^2-4 v}} & \oddbuv + \oddubu & 0 & 0 & 3\oddbvbuv & \oddbuv & -\oddbuv + \oddubu \\
  4\dlog{\frac{u v}{u^2-4 v}} & 4\oddubu + \oddbuv & 0 & 0 & 5\oddbvbuv & \oddbuv & 4\oddubu - \oddbuv \\
  2\oddubu & 2\dlog{\frac{v}{u^2-4 v}} + \dlog{\frac{u v}{u+v}} & 0 & 0 & 5\oddbu & \dlog{\frac{u v}{(u+1) (u+v)}} & 2\dlog{\frac{u^2}{u^2-4 v}} + \dlog{\frac{v (u+v)}{u (u+1)}} \\
  \dlog{\frac{u (u+v)}{u^2-4 v}} & \oddbuv + \oddubu & \dlog{\frac{v}{u+v}} & 0 & 2\oddbvbuv & 2\oddbuv & -3\oddbuv + \oddubu \\
  \dlog{\frac{u}{u+v}} & \oddvbv & \dlog{\frac{v (u+v)}{v^2-4 u}} & 0 & 0 & \frac{5}{2}\oddvbv - \oddbuv & 2\oddbuv - 4\oddvbv \\
  \dlog{\frac{u^2-4 v}{u (u+v)}} & 2\oddbuv - \oddubu + \oddvbv & \dlog{\frac{v (u+v)}{v^2-4 u}} & 0 & 0 & \frac{5}{2}\oddvbv + \oddbuv & -4\oddvbv - \oddubu \\
  -\oddbuv & \frac{1}{2}\dlog{\frac{v^2-4 u}{v^2}} & -\frac{1}{2}\oddvbv + \oddbuv & 0 & 0 & \dlog{\frac{v}{u+v}} + \frac{5}{4}\dlog{\frac{v^2-4 u}{v^2}} & 2\dlog{\frac{v (u+v)}{v^2-4 u}} \\
  2\oddbuv - \oddubu & \dlog{\frac{u^2-4 v}{u^2}} & -2\oddbuv & 0 & -2\oddbu & 2\dlog{\frac{u+v}{u}} & \dlog{\frac{u^2 \left(u^2-4 v\right)}{(u+v)^4}}
\end{array}\right.\nn\\&&
 \hspace{1.5cm}\left.\begin{array}{ccccc}
  0 & 0 & 0 & 0 & 0 \\
  0 & 0 & 0 & 0 & 0 \\
  0 & -\frac{7}{2}\oddbv & -2\dlog{v+1} + \frac{7}{2}\dlog{v} & 2\oddbv & 0 \\
  -\oddbuv & \frac{1}{2}\oddubu - \oddbuv & 0 & 0 & 0 \\
  -\oddbuv & 2\oddubu - \oddbuv & 0 & 0 & -\oddbvbuv \\
  \dlog{\frac{(u+1) (u+v)}{u v}} & \dlog{\frac{u (u+v)}{u^2-4 v}} & 0 & 0 & -\oddbu \\
  -\oddbuv & -2\oddbuv + \frac{1}{2}\oddubu & \oddbubuv & \oddbuv & 0 \\
  \frac{1}{2}\oddvbv & -2\oddvbv + \oddbuv & \oddbubuv & 2\oddvbv - \oddbuv & 0 \\
  -2\oddbuv + \frac{1}{2}\oddvbv & -2\oddvbv - \frac{1}{2}\oddubu - \oddbuv & 4\oddbubuv & 2\oddvbv - \oddbuv & 0 \\
  \frac{1}{4}\dlog{\frac{v^2-4 u}{v^2}} & \dlog{\frac{v (u+v)}{v^2-4 u}} & -\oddbv & \dlog{\frac{v^2-4 u}{v (u+v)}} & 0 \\
  0 & \frac{1}{2}\dlog{\frac{u^2 \left(u^2-4 v\right)}{(u+v)^4}} & 0 & 2\dlog{\frac{u+v}{u}} & 0
\end{array}\right)
 \left(\begin{array}{c}g_{8}\\g_{9}\\g_{12}\\g_{13}\\g_{15}\\g_{16}\\g_{17}\\g_{18}\\g_{19}\\g_{20}\\g_{21}\\g_{22}\end{array}\right)
\end{eqnarray}}\end{sideways}

\bibliographystyle{JHEP}

\bibliography{bibfile}

\end{document}